%% file: specialIssue.tex
\pdfoutput=1
\documentclass[final,1p,times]{elsarticle}
	\usepackage{./myMac}	
	\usepackage{mathtools} 
	\usepackage{mathtools}

	\DeclareMathOperator{\LOG}{\ensuremath{\operatorname{log
		\mathllap{\raisebox{1.35ex}{\rule{0.95em}{0.08ex}}\hspace{0.025em}}}}}

	\newcommand{\softcompare}{{\tt SoftCompare}}

	\newcommand{\equiV}[1][\vareps]{\stackrel{#1}{\sim}}

	\newcommand{\MAX}{\max_1}
	\newcommand{\smax}{s_{\max}}
	\newcommand{\rad}{{\rm rad}}

	\newcommand{\bisect}{{\sc Bisect}}
	\newcommand{\newton}{{\sc Newton}}
	\newcommand{\qout}{Q_{\sc out}}
	
	\newcommand{\qdis}{Q_{\sc dis}}
	\newcommand{\wtT}{\wt{T}}
	\newcommand{\wtTG}{\wt{T}^G}
	\newcommand{\wtTGk}{\wt{T}^G_k}
	\newcommand{\splittt}{\ensuremath{\mathtt{Split}}}
	\newcommand{\subdivTree}{{\cal T}_{subdiv}}
	\newcommand{\compTree}{{\cal T}_{comp}}
	\newcommand{\compTreeHat}{\wh{{\cal T}}_{comp}}
	\newcommand{\gdisc}{\operatorname{GenDisc}}
	\newcommand{\lcoeff}{\operatorname{lcf}}

 \newcommand{\longShort}[2]{#2}

 \newcommand{\blemB}{\begin{lemmaB}}
 \newcommand{\elemB}{\end{lemmaB}}
 \newcommand{\bthmB}{\begin{theoremB}}
 \newcommand{\ethmB}{\end{theoremB}}
 \newtheorem{lemmaB}{\sc Lemma}
 \newtheorem{theoremB}[lemmaB]{\sc Theorem}

 \newcommand{\blemBl}[1]{\begin{lemmaB} \label{lem:#1}}
 \newcommand{\bthmBl}[1]{\begin{theoremB} \label{thm:#1}}
 \newcommand{\blemC}{\begin{lemmaC}}
 \newcommand{\elemC}{\end{lemmaC}}
 \newcommand{\bthmC}{\begin{theoremC}}
 \newcommand{\ethmC}{\end{theoremC}}
 \newcommand{\bcorC}{\begin{corollaryC}}
 \newcommand{\ecorC}{\end{corollaryC}}
 \newtheorem{lemmaC}{\sc Lemma}
 \newtheorem{theoremC}[lemmaC]{\sc Theorem}
 \newtheorem{corollaryC}[lemmaC]{\sc Corollary}

 \newcommand{\blemCl}[1]{\begin{lemmaC} \label{lem:#1}}
 \newcommand{\bthmCl}[1]{\begin{theoremC} \label{thm:#1}}
	\excludecomment{s1}
	\excludecomment{s2}
	\excludecomment{s3}
	\excludecomment{s4}
	\excludecomment{s5}
	\excludecomment{s6}
	\excludecomment{s7}
	\excludecomment{s8}
	\excludecomment{s9}
	\excludecomment{bib}
 \includecomment{s1} 
 \includecomment{s2} 
 \includecomment{s4}  
 \includecomment{bib}
 \includecomment{s5} 
 \includecomment{s6}  
 \includecomment{s7}  
 \includecomment{s8}  

\RequirePackage{lineno}	

\begin{document}


%
%
\begin{frontmatter}

\title{
  Complexity Analysis of Root Clustering \\
    for a Complex Polynomial\footnote{An extended abstract appeared in the proceedings of ISSAC 2016~\citep{becker+4:cluster:16}}
}
    \author {Ruben~Becker}
    \address{MPI for Informatics,
          Saarbr\"ucken Graduate School of Computer Science,
	  Saarbr\"{u}cken, Germany}
    \ead{ruben@mpi-inf.mpg.de}
	\author {Michael~Sagraloff}
    \address{MPI for Informatics, Saarbr\"{u}cken, Germany}
	\ead{msagralo@mpi-inf.mpg.de}
	\author {Vikram~Sharma}
	\address{Institute of Mathematical Sciences, Chennai, India
    }
	\ead{vikram@imsc.res.in}
    \author{Juan Xu\fnref{fn1}}
	\address{Beihang University, Beijing, China}
	\ead{xujuan0505@126.com}
    \author{Chee Yap\fnref{fn2}}
	\address{Courant Institute of Mathematical Sciences, New York, USA}
	\ead{yap@cs.nyu.edu}

    \fntext[fn1]{
		Supported by
		China Scholarship Council, No.20150602005.
	}
    \fntext[fn2]{
	 	Supported by NSF Grants \#CCF-1423228 
	  	and \#CCF-1564132. 		
	    }
\begin{abstract}
    Let $F(z)$ be an arbitrary complex polynomial.
    We introduce the \dt{local root clustering problem}, to compute
    a set of natural $\vareps$-clusters of roots of $F(z)$ in some box region $B_0$
    in the complex plane.  This may be viewed as an extension of the classical
    root isolation problem.
    Our contribution is two-fold: we provide an efficient certified
    subdivision algorithm for this problem,
    and we provide a bit-complexity analysis based on the
    local geometry of the root clusters.

Our computational model assumes that
arbitrarily good approximations of the coefficients of
$F$ are provided by means of an oracle at the cost of reading the coefficients.
Our algorithmic techniques come from a companion
	paper \citep{becker+3:cisolate:18} and are based on the Pellet test, Graeffe and
Newton iterations, and are independent of Sch\"onhage's splitting circle
method.
Our algorithm is relatively simple and promises to be efficient in practice.

\end{abstract}
\end{frontmatter}
\ignore{
%
%

\ccsdesc[500]{Computer systems organization~Embedded systems}
\ccsdesc[300]{Computer systems organization~Redundancy}
\ccsdesc{Computer systems organization~Robotics}
\ccsdesc[100]{Networks~Network reliability}

%
%

%
%
\printccsdesc


\keywords{ACM proceedings; \LaTeX; text tagging}
}%

\begin{s1}
\section{Introduction} 
	The problem of computing the roots of a univariate
	polynomial $F$ has a venerable history that dates back to antiquity.
	With the advent of modern computing, the subject
	received several newfound aspects;
	in particular, the introduction of algorithmic rigor and complexity
	analysis has been extremely fruitful
	(see \cite{mcnamee:roots:bk2}).
	This development is traced to
	Sch\"onhage's 1982 landmark paper,
	``{\em Fundamental Theorem of Algebra
	in Terms of Computational Complexity}''
	\citep{schonhage:fundamental}.
	\longShort{    \footnote{
       Coincidentally,
       Smale wrote his influential paper
	   \cite{smale:fundamental-thm:81}
	   on the ``{\em Fundamental Theorem and Complexity Theory}''
       around the same time.
	   Smale's work opened up a completely different line of work
	   that is outside the current scope.
	   In \cite[p.~48]{schonhage:fundamental}
	   Sch\"onhage noted that despite the similarity in their
	   titles, there is no overlap;
	   for instance, Smale's approach cannot handle multiple roots
	   which is a major concern of Sch\"onhage, and of this paper.
		}}{
	}
	Algorithms in this tradition are usually described as ``exact and
	efficient''.  Sch\"onhage considers the problem of
	approximate polynomial factorization, that is, the computation of
	approximations $\tilde{z}_i$
	of the roots $z_i$ of $F$ such that
	$\|F-\wt{F}\|_{1}<2^{-b}\cdot\|F\|_{1}$, where
	$\wt{F}(z):=\lcoeff(F)\cdot\prod_{i=1}^n (z-\wtz_i)$
	and $b$ is a given positive integer.
	The sharpest result for this problem is from
	\cite[Theorem 2.1.1]{pan:poly-roots:02} and
	\cite[p.196]{pan:history-progress:97}.
	Hereafter, we refer to the underlying algorithm in this theorem as
	``Pan's algorithm''.
	Under some mild assumption on $F$ (i.e., $|z_i|\le 1$ and
	$b\ge n\log n$), Pan's algorithm uses only $\tilde{O}(n\log b)$
	arithmetic operations with a precision bounded by $\tilde{O}(b)$, and
	thus $\tilde{O}(nb)$ bit operations.
	This result further implies that the
	complexity of approximating all $z_i$'s to any specified $b/n$ bits,
	with $b>n\log n$,
	is also $\wtO(nb)$ (see \cite[Corollary 2.1.2]{pan:poly-roots:02}).
	Here, $\wtO$ means we ignore logarithmic factors
	in the displayed parameters.  
	In a model of computation, where it is assumed that the coefficients of
	$F$ are complex numbers for which approximations
	are given up to a demanded precision, the above bound is tight (up to
	poly-logarithmic factors) for polynomial factorization
	as well as for root approximation.

	The preceding paragraph is concerned with
	\emph{root approximation},
	i.e., computing $\wtz_i$ such that $|\wtz_i-z_i|\le \vareps$ for
	specified $\vareps>0$. Our main focus is the
	stronger problem of \emph{root isolation},
	i.e., computing $(\wtz_i,r_i)$ such that $r_i\le \vareps$
	and the discs $\Delta(\wtz_i,r_i)$ centered at $\wtz_i$ of
	radius $r_i$ are pairwise disjoint and contains $z_i$.
	A central focus in exact and efficient root approximation
	research has been to determine the complexity of
	\emph{isolating} all the roots of an \emph{integer} polynomial $F(z)$
	of degree $n$ with $L$-bit coefficients.
	We call this the \dt{benchmark problem} in
	\cite{sagraloff-yap:ceval:11}
	since this case is the main theoretical tool for comparing root
	isolation algorithms.
	Although this paper addresses complex root isolation, we will also
	refer to the related \dt{real benchmark problem}
	which concerns real roots for integer polynomials.

	Root isolation can be reduced to
	root approximation.  Sch\"onhage showed that, for a
	square-free polynomial $F$, it
	suffices to choose a $b$ of size $\Omega(n(\log n+L))$ to ensure that
	the approximations $\wtz_i$ are isolated with $2\vareps$ taken as
	the root separation bound of $F$.  Together with Pan's result on
	approximate polynomial factorization, this yields a
	complexity of $\wtO(n^2L)$ for the benchmark problem.
	Interestingly, the latter bound was not explicitly stated until
	recently
	(see \cite[Theorem 3.1]{emiris-pan-tsigaridas:handbk:14}).

	\cite{mehlhorn-sagraloff-wang:15} extend the latter
	result to (not necessarily square-free) polynomials $F$
	with arbitrary complex coefficients
	for which the number of distinct roots is given as an additional input.
	That is, Pan's algorithm is used
	as a blackbox with successively increasing precision $b$ to isolate the
	roots of $F$. For the benchmark problem, this yields
	the bound $\wtO(n^3+n^2L)$; however, the actual cost adapts to the
	geometry of the roots, and for most input polynomials, the
	complexity is considerably lower than the worst case bound.

	We further remark that it seems likely that the bound $\wtO(n^2L)$ is
	also near-optimal for the benchmark problem
	because it is generally believed that
	Pan's algorithm is near-optimal for the problem of approximately
	factorizing a polynomial with complex
	coefficients. However, rigorous arguments for such claims are missing.
	\ignore{
	, they give an algorithm to isolate all roots by one can adaptively
	choose the precision $b$ to yield a
	bit complexity
		$
		\wtO(n^2(n+L))
		$
	for an arbitrary input $F(z)$ that is not necessarily square-free
	(this result requires knowing the number $m$ of distinct roots, which
	can be computed within this complexity).

	But to use this result, we must first make $F(z)$ square-free.
	Theoreticians feel justified to do this because it does not
	affect degrade the benchmark complexity; but in practice, this is
	undesirable
	since it invokes an unrelated algorithm and since the
	typical $F(z)$ is square-free.
	\ignore{ 
	  Michael argued against my ``wasted effort remark'' (OK I will remove
	  my remark, even though what I state is literally true -- I am
	  not making a complexity claim).  But Michael claims that
	  ``in practice, this is done via a Las Vegas algorithm based
	  on modular computation and Chinese Remainder, which has low cost of
	  $O(nL)$''.    I agree that the worst case complexity
	  is low (but these computations seems non-adaptive).
	  Can you tell me where this is practiced?
	  Complexity theorists in pursuit of ``record worst case
	  bounds'' has no qualms.  But outside of crypto applications,
	  I cannot imagine any numerical community willing to adopt this.
	  Anyway, to settle this issue, it would be nice
	  to see some experimental results.
	  Perhaps Michael's work with Rouillier can provide this?
	}
	So it is of interest to have root isolation algorithms that are
	not conditional on the square-freeness of $F(z)$.
	This motivates the second improvement:
	\cite{mehlhorn-sagraloff-wang:15}
	showed that by using Pan's algorithm
	as a blackbox, one can adaptively choose the precision $b$ to yield a
	bit complexity
	\ignore{
    \beql{bench}
      \wtO(n^2(n+L))
    \eeql
	}%
		$
		\wtO(n^2(n+L))
		$
	for an arbitrary input $F(z)$ that is not necessarily square-free
	(this result requires knowing the number $m$ of distinct roots, which
	can be computed within this complexity).
	Note that \cite[p.~722]{pan:poly-roots:02} refers to the bounds
	$\wtO(n^2L)$ (or $\wtO(n^2(n+L))$) as ``near-optimal'' for the benchmark problem.
	It is an informal but useful terminology.

	}
	
	Until recently, it had been widely assumed that near-optimal bounds
	need the kind of
	``muscular'' divide and conquer techniques such as
	the splitting circle method of Sch\"onhage (which underlies
	most of the previous fast algorithms in the complexity literature).
	These algorithms are far from practical (see below).
	So, also the bound $\wtO(n^2(n+L))$ achieved by
	\cite{mehlhorn-sagraloff-wang:15}
	is mainly of theoretical interest as the algorithm uses Pan's method as
	a blackbox.
	\ignore{
 	Currently, the best implemented
	algorithms for root isolation with guaranteed correctness
	are based on subdivision methods (see
	\cite{sagraloff-yap:ceval:11} for a brief survey).
        }
	Instead of these near-optimal algorithms, practitioners interested
	in a priori root isolation invariably rely on subdivision methods.
	The classical example is real root isolation based on Sturm sequences
	(1829).  For complex roots, Weyl (1924) introduced the quadtree method
	for 
	\ignore{
	Subdivision algorithms need one or more predicates for testing
	subdivision
	boxes; failure of the predicates would force the box to split, forming a
	subdivision tree. 
        }%
    Two types of subdivision algorithms are actively investigated currently: the
	\dt{Descartes Method}
	\citep{collins-akritas:76,lane-riesenfeld:81,rouillier-zimmermann:roots:04,schonhage:fundamental,sagraloff:newton-descartes:12,mehlhorn-sagraloff:real-roots:16}
	and the \dt{Evaluation Method}
	\citep{burr-krahmer-yap:continuousAmort:09,burr-krahmer:sqfree:12,sharma-yap:near-optimal:12,becker:thesis:12,sagraloff-yap:ceval:11,kamath-voiculescu-yap:study:11,pan:weyl:00}.
	%
	See \cite{sagraloff-yap:ceval:11} for a comparison of
	Descartes and Evaluation (or Bolzano) methods.

	The development of certain tools, such as the Mahler-Davenport root
	bounds
	\citep{davenport:85,du-sharma-yap:sturm:07}, have been useful in deriving
	tight bounds on the subdivision tree size for certain subdivision
	algorithms
        \citep{eigenwillig-sharma-yap:descartes:06,burr-krahmer:sqfree:12,sharma-yap:near-optimal:12}.
	Moreover, most of these analyses
	can be unified under the ``continuous amortization'' framework
	\citep{burr-krahmer-yap:continuousAmort:09,burr:contAmortization:16}
	which can even incorporate bit-complexity.
	These algorithms only use bisection in their subdivision,
	which seems destined to lag behind the above ``near optimal bounds''
	by a factor of $n$.
	To overcome this, we need to combine Newton iteration with bisection,
        an old idea that goes back to Dekker and Brent
	in the 1960s.  
	\cite{pan:weyl:00} showed that theoretically, the
	near optimal bounds can be achieved with subdivision methods.
	In recent years, a formulation of Newton iteration
	due to \cite{abbott:qir} and 
	\cite{sagraloff:newton-descartes:12} has proven especially useful.
	This has been adapted to achieve the recent near-optimal
	algorithms of 
    \cite{sagraloff:newton-descartes:12} and 
    \cite{mehlhorn-sagraloff:real-roots:16}
	for real roots, and \cite{becker+3:cisolate:18} for complex roots.
	\longShort{One key contribution of this paper is to expand the scope of
	these techniques.}{}

	{\bf The Root Clustering Problem.}
	In this paper, we are interested in root clustering.
	The requirements of root clustering represents
	a simultaneous strengthening of
	root approximation (i.e., the output discs must be disjoint)
	and weakening of root isolation (i.e., the output discs
	can have more than one root).  Hereafter, ``root finding'' refers
	generally
	to any of the tasks of approximating, isolating or clustering roots.

	For an analytic function $F:\CC\to\CC$ and a complex disc
	$\Delta\ib\CC$,
	let $\calZ(\Delta;F)$ denote the multiset of roots of $F$ in $\Delta$
	and $\#(\Delta;F)$ counts the size of this multiset.
	We write $\calZ(\Delta)$ and $\#(\Delta)$ since $F$ is usually
	supplied by the context.
	Any non-empty set of roots of the form $\calZ(\Delta)$ is called a
	\dt{cluster}.
	The disc $\Delta$ is called an \dt{isolator} for $F$ if $\#(\Delta)=$
	$\#(3\Delta)>0$.
	Here, $k\Delta=k\cdot \Delta$ denotes the centrally scaled version of
	$\Delta$
	by a factor $k\ge 0$.
	The set $\calZ(\Delta)$ is called a \dt{natural cluster} when
	$\Delta$ is an isolator.  A set of $n$ roots could contain
	$\Theta(n^3)$ clusters, but at most $2n-1$ of these are natural.
	This follows from the fact that any two natural clusters
	are either disjoint or have a containment relationship.
	The benchmark problem is a global problem because it concerns
	{\em all} roots of the polynomial $F(z)$; we now address local
	problems where we are interested in finding only {\em some} roots of
	$F(z)$.
	For instance, \cite{yakoubsohn:cluster:00} gave a method to
	test if Newton iteration from a given point will converge to a cluster.
	In \cite{yap-sagraloff-sharma:cluster:13}, we introduced
	the following \dt{local root clustering problem}:
	{\em given
	  $F(z)$, a box $B_0\ib \CC$ and $\vareps>0$,
	to compute a set $\set{(\Delta_i,m_i): i\in I}$
	where the $\Delta_i$'s are pairwise disjoint isolators,
	each of radius $\le \vareps$ and $m_i =\#(\Delta_i)\ge 1$, such that}

		$$
		\calZ(B_0) ~\ib~ \bigcup\nolimits_{i\in I}\calZ(\Delta_i) ~\ib~ \calZ(2B_0).
        $$
	We call the set
	$S =\set{\Delta_i: i\in I}$ (omitting the $m_i$'s)
	a \dt{solution} for the local
	root clustering instance $(F(z), B_0,\vareps)$.
	The roots in $2B_0\setminus B_0$ are said to be \dt{adventitious}
	because we are really only interested in roots in $B_0$.
	Suppose $S$ and $\wh{S}$ are both solutions for an instance
	$(F(z),B_0,\vareps)$.  If $S\ib\wh{S}$, then we call $\wh{S}$
	an augmentation of $S$.  Thus any $\Delta \in\wh{S}\setminus S$
	contains only adventitious roots.

	We solved the local root clustering problem
	in \cite{yap-sagraloff-sharma:cluster:13} for any
	analytic function $F$, provided an upper on $\#(2B_0)$ is known,
	but no complexity analysis was given.
	Let us see why our formulation is reasonable.
	It is easy to modify our algorithm so that the adventitious
	roots in the output are contained in
	$(1+\delta)B_0$ for any fixed $\delta>0$.
	We choose $\delta=1$ for convenience.
	Some $\delta>0$ is necessary because in our
	computational model where only approximate
	coefficients of $F$ are available, we cannot decide
    the implicit ``Zero Problem'' (see \cite{yap:praise:09}) necessary
	to decide if the input has a root on the boundary of $B_0$,
	or to decide whether $\Delta$ contains a root of multiplicity $k>1$.
	Thus, root clustering is the best one can hope for.
		\ignore{
	The advantage of our model (and its main motivation)
	is that we are now able to
	design ``unconditional'' algorithms for analytic roots
	\citep{yap-sagraloff-sharma:cluster:13}.
	For instance, Sch\"onhage
	\cite[p.~5]{schonhage:fundamental}
	noted that algorithms for root isolation
	(see references [17,18,34] therein)
	that are only ``conditionally correct'' because they use
	contour tests which fail when roots lie on the contours.
        	}%
    %
\subsection{Main Result}
	In this paper, we describe a local root clustering
	algorithm and provide an analysis of its bit-complexity.
	Standard complexity bounds for root isolation
	are based on \dt{synthetic parameters}
	such as degree $n$ and bitsize $L$ of the input polynomial.
	But our computational model for $F(z)$ has no notion of bit size.
	Moreover, to address ``local'' complexity of roots,
	we must invoke \dt{geometric parameters} such as root separation
	\citep{sagraloff:newton-descartes:12,mehlhorn-sagraloff:real-roots:16}.
	We will now introduce new geometric parameters arising from
	cluster considerations.

	Assume $F(z)$ has $m$ distinct complex roots
		$z_1\dd z_m$
	where each $z_j$ has multiplicity $n_j\ge 1$,
	thus $n=\sum_{j=1}^m n_j$ is the degree of $F(z)$.
	Let the magnitude of the
	leading coefficient of $F$ be $\ge 1/4$,
	and the maximum coefficient magnitude $\|F\|_\infty$
	be bounded by $2^{\tau_F}$ for some $\tau_F$.

	\longShort{
	This constraint on leading coefficient of $F$ is analogous
	to making the leading coefficient equal to $1$; but in our
	model of approximate coefficients, we use a lower bound of $1/4$.
	  }{}Let $k$ be the number of roots counted with multiplicities in
	  $2B_0$.
	An input instance $(F(z), B_0,\vareps)$ is called \dt{normal} if
	$k\ge 1$ and
	$\vareps\le\min\set{1,\frac{w_0}{96n}}$ with $w_0$ the width of $B_0$.
	For any set $U\ib\CC$,
	let $\LOG(U) \as \max(1,\log\sup(|z|: z\in U))$.

    Our algorithm outputs a set of discs, each one contains a
    natural cluster. We provide a bit complexity bound of the
    algorithm in terms of the output.

	\bgenDIY{Theorem A}{
	  Let $S$ be the solution computed by our algorithm for
	  a normal instance $(F(z),B_0,\vareps)$.
	  Then there is an augmentation
        $\wh{S}=\set{D_i: i\in I}$ of $S$ such that
	  the bit complexity of the algorithm is
		{
	  \beql{TotalCost}
		\wtO\Big(n^2\LOG(B_0)+ n \sum\nolimits_{D\in \wh{S}} L_D\Big)
      \eeql}
      with
      {
	     \beql{L_D}
	      \begin{aligned}
		      L_D = \wtO \Big( \tau_F
	                 +n\cdot\LOG(\xi_D)
	                 +k_D\cdot(k+\LOG(\varepsilon^{-1})) 
			       + \LOG (\prod\nolimits_{z_j\notin D}
                      |\xi_D-z_j|^{-n_j})\Big)
	      \end{aligned}
	   \eeql}where
	   $k_D=\#(D)$, and $\xi_D$ is an arbitrary root in $D$.
       Moreover, an $L_D^*$-bit approximation of the coefficients of $F$
       is required with
          $L_D^*\as\max_{D\in\wh{S}}L_D$.
	}
	\\The solution $\wh{S}$ in this theorem is called
	an \dt{augmented solution} for input $(F(z),B_0,\vareps)$.
	Each natural $\vareps$-cluster $D \in \wh{S}$ is an isolator of radius
	$\le \vareps$.
    From \refeQ{TotalCost}, we deduce:

    \bgenDIY{\sc Corollary to Theorem A}{
     \ \\The bit complexity of the algorithm is bounded by
     {
         \beql{complexity}
         \wtO\Big(n^2(\tau_F + k+m)
	  		    + nk\LOG(\vareps^{-1})
			    + n\LOG|\gdisc(F)|^{-1}
			    \Big)
        \eeql}
    where
        \begin{equation*}\label{gdisc}
        \gdisc(F) \as {\lcoeff(F)}^{m}
		\prod_{1 \le i< j\le m} \paren{\alpha_i - \alpha_j}^{n_i+n_j}.
        \end{equation*}
      In case $F$ is an integer polynomial, this bound becomes
      {
	\beql{complexity2}
		\wtO\Big(n^2(\tau_F + k +m)
			    + nk\LOG(\vareps^{-1}) \Big).
	\eeql}
	}
	The bound \refeQ{complexity2} is the sum of two terms:
	the first is essentially the near-optimal root bound,
	the second is linear in $k$, $n$ and $\LOG(\vareps^{-1})$.
	This suggests that Theorem A is quite sharp.

	{\bf On strong $\vareps$-clusters.}
    Actually, the natural $\vareps$-clusters in the $\wh{S}$ have some
    intrinsic property captured by the following definition.
	Two roots $z,z'$ of $F$ are \dt{$\vareps$-equivalent},
	written $z\equiV z'$,
	if there exists a disk $\Delta=\Delta(r,m)$ containing $z$ and $z'$
	such that
	$r\le \frac{\varepsilon}{12}$
	and $\#(\Delta)=\#(114\cdot\Delta)$.
	Clearly $\Delta$ is an isolator; from this, we see
	that $\vareps$-equivalence is an equivalence relationship.
	We define a \dt{strong $\vareps$-cluster} to be
	any such $\vareps$-equivalence class.
	Unlike natural clusters, any two strong
	$\vareps$-clusters must be disjoint.

    \bgenDIY{Theorem B}{ \ \\
      Each
	  natural cluster $D\in\wh{S}$ is a union of strong
	  $\vareps$-clusters.
	\vspace{4pt}
	}
	This implies that our algorithm will never split
	any strong $\vareps$-cluster.  It might appear
	surprising that our ``soft'' techniques
	can avoid accidentally splitting a strong $\vareps$-cluster.

	\longShort{
    With Theorem B, we can also give a bit complexity bound
    in terms of the set of strong $\vareps$-clusters.
	
	\bgenDIY{Lemma C}{\ \\
    The bit complexity of the algorithm can be bound by \refeQ{TotalCost}
    where $\wh{S}$ is replaced by the set of strong
    $\vareps$-clusters contained in $2B_0$.
    }
    This bound is not as good as \refeQ{TotalCost}, nevertheless,
    it is
    intrinsic and does not depend on the output of the algorithm.
	}{}
\subsection{What is New}
    Our algorithm and analysis is noteworthy for its wide applicability:
    (1) We do not require square-free polynomials. This is important
    because we cannot compute the square-free part of
    $F(z)$ in our computational model where the coefficients
    of $F(z)$ are only arbitrarily approximated.
	Most of the recent fast subdivision
	algorithms for real roots
	\citep{sagraloff:newton-descartes:12,mehlhorn-sagraloff:real-roots:16}
	require square-free polynomials.
    (2) We address the local root problem
	and provide a complexity analysis
	based on the local geometry of roots.
	Many practical applications (e.g., computational geometry)
	can exploit locality. The companion paper \citep{becker+3:cisolate:18}
    also gives a local analysis. However, it is under the condition that
    the initial box is not too large or is centered at the origin, and
    an additional preprocessing step is needed for the latter case.
    But our result does not depend on any assumptions on $B_0$ nor
    require any preprocessing.
    (3) Our complexity bound is based on cluster geometry instead of individual
    roots.  To see its benefits, recall that the bit complexity in
    \cite{becker+3:cisolate:18}
     involves a term $\LOG \sigma(z_i)^{-1}$ where $\sigma(z_i)$ is the
     distance to the nearest root of $F(z)$.  If $z_i$ is a multiple root,
     $\sigma(z_i)=0$.  If square-freeness is not
    assumed, we must replace $\sigma(z_i)$ by the distance $\sigma^*(z_i)$ to
    the closest root $\neq z_i$ (so $\sigma^*(z_i)>0$).
    But in fact, our bound in \refeQ{TotalCost} involves
    $T_D\as \LOG\prod_{z_j\notin D} |\xi_i-z_j|^{-n_j}$ which depends only on the inverse distance
    from a root within a cluster $D$ to the other roots outside of $D$,
    which is smaller than $\LOG \sigma^*(z_i)^{-1}$. So the
    closeness of roots within $D$ has no consequence on $T_D$.

    Why can't we just run the
    algorithm in \cite{becker+3:cisolate:18}
	by changing the stopping
	criteria so that it terminates as soon as a component $C$
	is verified to be a natural $\vareps$-cluster?
    Yes, indeed one can. But our previous method of charging the work associated
    with a box $B$ to a root $\phi(B)$ may now cause a cluster of multiplicity
    $k$ to be charged a total of $\Omega(k)$ times, instead of $\wtO(1)$ times.
    Cf.~Lemma 11 below where $\phi(B)$ is directly charged to a cluster.
\end{s1}
\begin{s2}
\subsection{Practical Significance}

	Our algorithm is not only theoretically efficient, but
	has many potential applications.  Local root
	isolation is useful in applications where
	the roots of interest lie in a known locality,
	and this local complexity can be much
	smaller than that of finding all roots.
	\longShort{
	  Many problems of computational geometry (e.g.,
	  locating Voronoi vertices in a non-linear diagrams),
	we often have a good idea of the region $B_0$ where the Voronoi vertex
	lies.  The actual number of roots in $B_0$ may be very small
	compared to the total number of roots.
        }{}
	From this perspective, focusing on
	the benchmark problem is misleading for such applications.

	We believe our algorithm is practical, and plan to implement it.
	Many recent subdivision algorithms were implemented, with
	promising results: 
	\cite{rouillier-zimmermann:roots:04}
	engineered a very efficient
	Descartes method algorithm which is widely used in the
	Computer Algebra community, through Maple.  The {\tt CEVAL}
	algorithm in \cite{sagraloff-yap:ceval:11} was implemented in
	\cite{kamath:thesis,kamath-voiculescu-yap:study:11}.
	\cite{becker:thesis:12} gave a Maple implementation of the
	{\tt REVAL} algorithm
	for isolating real roots of a square-free real polynomial.
	Most recently, 
	\cite{kobel-rouillier-sagraloff:for-real:16}
	implemented the {\tt ANewDsc} algorithm
	from \cite{mehlhorn-sagraloff:real-roots:16}, showing
	its all round superiority; it especially shines against
	known algorithms when roots are clustered.

	Although there are  several fast divide-and-conquer algorithms
	\citep{renegar:approximate-poly-zeros,neff-reif:complex-roots:96,kirrinnis:poly-factor-newton:98},
    there are no implementations of these methods.
	\cite{pan:poly-roots:02} notes in p.~703:
	``{\em Our algorithms are quite involved, and their implementation
	  would require a non-trivial work, incorporating numerous known
	  implementation techniques and tricks.}''
	Further in p.~705:
	``{\em since Sch\"onhage (1982b) already has 72 pages
	and Kirrinnis (1998) has 67 pages,
	this ruled out a self-contained presentation of our root-finding
	algorithm}''.
	\ignore{footnote:
	  In \cite[p.~49]{schonhage:fundamental}, it is
	  noted that Kirrinnis's Habilitationsschrift at Universit\"at Bonn
	  (1999) is about 200 pages and is ``hopefully getting published soon''.
	}%
	Our paper
	\citep{becker+3:cisolate:18} is self-contained with over
	50 pages, and explicit precision bounds for all
	numerical primitives: we use asymptotic bounds
	only in complexity analysis (since it has no consequence for
	implementations) but not in computational primitives.

	\ignore{ 
	Root approximation is a very practical problem that
	is used in many areas of applied science (see
	\cite{mcnamee:roots:bk1,mcnamee:roots:bk2}).
	So which root solvers do practitioners actually use?
	We describe three distinct computing domains, each
	with its own unique computing concerns.
	For each domain, we mention the implemented algorithm
	that is widely considered ``best in its domain''.
	\benum[(A)]
	\item
	  In numerical analysis, algorithms are optimized to exploit
	fixed-precision floating point arithmetic; here
	Jenkins-Traub \cite{jenkins-traub:roots:70}
	is popular.  The algorithm is stable, globally convergent and
	used in commercial packages.
	There is no a priori guarantee of correctness;
	a posteriori error analysis (depending on condition numbers)
	might confirm correctness.
	\item
	For stronger guarantees, many applications require multiprecision
	arithmetic.  In this domain, the MPSOLVE library of Bini and Fiorentino
	\cite{bini-fiorentino:design:00,bini:mpsolve:96}
	(with a recent update \cite{bini-robol:solving:14})
	is widely recognized as the best in this class.
	It is based on the Aberth-Erlich Iteration \cite{aberth-erlich:roots:73}
	whose global convergence is, unfortunately, unknown.
	Aberth-Erlich, along with Durand-Kerner \cite{durand-kerner:roots:66},
	falls under the class of ``Weierstrass Methods'' that
	can be viewed as simultaneous path following
	(see \cite{batra:effective:10}).
	\item
	In Computer Algebra, algorithms with a priori guarantee of correctness
	is a pre-requisite.  Here the Descartes method %
	from Collins and Akritas \cite{collins-akritas:76} (in the power basis form)
	or Lane and Riesenfeld \cite{lane-riesenfeld:81} (in the Bernstein basis form)
	is regarded as among the best.  A highly engineered package from
	Rouillier-Zimmerman \cite{rouillier-zimmermann:roots:04}
	is available in Maple and widely used in exact computation.
	\eenum

	Clearly our new algorithm competes in
	in the exact domain (C), but we believe it is also competitive in
	the multiprecision domain (B).  We mention three favorable
	properties of our algorithm which
	might be obscured by an extreme focus on the benchmark problem:
	\\ 1.
	Our algorithm does not need any preprocessing of the input polynomial.
	Even though such preprocessing has no impact on the benchmark complexity,
	this is an important practical consideration.
	As mentioned, we do not require square-free input polynomials;
	instead, the algorithm achieves the bound \refeQ{bench}
	in general, but automatically achieves $\wtO(n^2L)$ in the square-free case
	(see \cite{mehlhorn-sagraloff:real-roots:16}).  
	Another transformation we avoid, but commonly assumed
	in the divide-and-conquer algorithms is to
	transform the polynomial so that its roots
	lie in the unit disc. Pan \cite[p.~702-3]{pan:poly-roots:02}
	noted the practical difficulties of this transformation.
	\\ 2.
	Root isolation is actually
	only one aspect of root approximation; but it is the critical
	one because after a root has been isolated,
	the refinement of the root to any desired precision can be
	achieve using Newton-type iterations (e.g., \cite{kerber-sagraloff:qir:15}).
	Moreover, in subdivision methods such as our algorithm, this refinement process can be
	automatically built\footnote{
	  	So the criticism of introducing a separate
		square-free polynomial extraction algorithm into Pan's algorithm
		does not apply here.
	}
	into the root isolation algorithm.
	\\ 3.
	The benchmark problem solves what might be called the \dt{global version}
	of root isolation, i.e., finding all roots of a polynomial.  But
	many applications are interested in some subset of the roots only,
	and it is often possible to compute a
	simple bounded region (like a disc or a box region)
	in which to search for roots.  Subdivision algorithms can naturally solve
	this \dt{local version} of root isolation.
	Algorithms optimized for the global problem might be adapted as easily.
	Note that the current best algorithm in (B) solves the global version.
      }%

	\ignore{Pan'02, page 702-703:
		Our algorithms are quite involved, and their implementation would require a
		non- trivial work, incorporating numerous known implementation techniques and
		tricks (Bini and Fiorentino, 2000; Fortune, 2001; Bini and Pan, to appear). We
		do not touch this vast domain here and just briefly comment on the precision of
		computing.

		Our algorithms involve the shifts of the variable (or equivalently of the
		origin), its scaling, and approximation of the root radii, that is, the
		distances of the unknown roots from a selected complex point. These techniques
		have low arithmetic and Boolean cost and are customary for reducing the study
		to the canonical cases, say where all roots lie in the unit disc {x : |x| ≤ 1}
		(see Renegar, 1987; Pan, 1996; Kirrinnis, 1998). On the other hand, using these
		techniques requires precision of computation of the order of n or n log n bits,
		which creates an implementation problem for larger n. Although the Chi- nese
		remainder algorithm and Sch ̈onhage–Strassen’s algorithm (Sch ̈onhage and
		Strassen, 1971) overcome this problem in principle (at least at the asymptotic
		complexity level), the problem is still substantial for numerical
		implementation, which is most efficient using the single or double IEEE
		precision. Thus the current champions in practical numerical root-finding rely
		on Jenkins–Traub’s, modified Laguerre’s and modified Newton’s algo- rithms,
		variations of the Weierstrass method, and the QR algorithm for the companion
		matrix (McNamee, 1993, 1997; Fortune, 2001; Pan, 2002a; Bini and Pan, to
		appear). We note, however, that our splitting approach has all the potential to
		be effective for lower precision computation of polynomial factorization (see
		Malajovich and Zubelli, 1997, Bini et al., 2002) and that the shift-free
		variations of our algorithms (say in the beginning of Section 2.3) have the
		same promise (in both cases substantial implementation work would be required).
	}%

	\ignore{COMPLEXITY CONSIDERATIONS:
		TAKEN from Mehlhorn/Sagraloff -- discussion after Theorem 2:

		For polynomials with integer coefficients, the bound can be stated more simply.

		Theorem 2. For a square-free polynomial P ∈ Z[x] with integer coefficients of absolute value 2τ or less, the algorithm ANewDsc computes isolating intervals for all real roots of P with O ̃(n3 + n2τ) bit operations. If P has only k non-vanishing coefficients, the bound improves to O ̃(n2(k + τ)).

		For general real polynomials, the bit complexity of the algorithm ANewDsc matches the bit complexity of the best algorithm known ([26]). For polynomials with integer coefficients, the bit complexity of the best algorithm known ([12, Theorem 3.1]) is O ̃(n2τ), however, for the price of using Ω(n2τ) bit operations for every input.8 Both algorithms are based on Pan’s approximate factorization algorithm [30], which is quite complex, and always compute all complex roots.

		Our algorithm is simpler and has the additional advantage that it can be used to isolate the real roots in a given interval instead of isolating all roots. Moreover, the complexities stated in the theorems above are worst-case complexities, and we expect a better behavior for many instances. We have some theoretical evidence for this statement. For sparse integer polynomials with only k non-vanishing coefficients, the complexity bound reduces from O ̃(n2(n + τ)) to O ̃(n2(k + τ))
		(Theorem 2). Also, if we restrict the search for roots in an interval I0, then only the roots contained in the one-circle region ∆(I0) of I0 have to be considered in the complexity bound (1) in Theorem 1. More precisely, the first summand n3 can be replaced by n2 · m, where m denotes the number of roots contained in ∆(I0), and the last summand 􏰀ni=1 logM(P′(zi)−1) can be replaced by 􏰀i:zi∈∆(I0) logM(P′(zi)−1). We can also bound the size of the subdivision tree in terms of the number of sign changes in the coefficient sequence of the input polynomial (Theorem 27). In particular, if P is a sparse integer polynomial, e.g., a Mignotte polynomial, with
		only (log(nτ))O(1) non-vanishing coefficients, our algorithm generates a tree of size (log(nτ))O(1). Our algorithm generates a tree of size (log(nτ))O(1), whereas bisection methods, such as the classical Descartes method, generate a tree of size Ω(nτ), and the continued fraction method [6] generates a tree of size Ω(n).
	}%
\end{s2}
\begin{s3}
\subsection{Some Background}
	We now provide the necessary background for the techniques
	in our algorithm.  But first we mention
	are several lines of related work for which we can only offer
	brief pointers:
	Sagraloff \cite{sagraloff:sparse:14}
	has achieved near-optimal bounds for the
	real benchmark problem for sparse integer polynomial.
	One can generalize integer polynomials to real polynomials
	whose coefficients are given as a bit stream.
	Most of the algorithms here use the
	Descartes Method (e.g., \cite{mehlhorn-sagraloff:real-roots:16}).
	For the Evaluation Method, a bit stream algorithm
	for real root isolation has been achieved in
	Becker's thesis \cite{becker:thesis:12}.

	There are two main algorithmic ingredients in our algorithm:
	the first is Pellet's test
	\cite{marden:bk}.
	In root isolation, this test was
	recently introduced in our algorithm for clustering
	analytic roots \cite{yap-sagraloff-sharma:cluster:13}.
	Let $T_k(D) = T_k(D;F)$ denote
	the $k$th order Pellet test which provides a criteria for the complex disc $D$
	to contain exactly $k$ roots (counted with multiplicity)
	of a complex analytic function $F$.
	When $k=0$, $T_0$ is precisely the $C_0$ predicate (exclusion predicate)
	used in several of our subdivision algorithms for isolating roots,
	curves and surfaces
	(e.g., \cite{sagraloff-yap:ceval:11,cxy}).
	Instead of $T_1$, our previous root isolation algorithms
	\cite{sagraloff-yap:ceval:11}
	used the $C_1$ predicate, viewed as $1$-dimensional form of 
	the normal variation predicate of Plantinga-Vegter
	\cite{pv}.  Note that $C_1$ is weaker than $T_1$ and only ensures
	that that is at most one root; so we need some other method to
	confirm that there is at least one root.

	The second ingredient is Newton iteration cleverly combined with
	subdivision.  In root isolation, such a combination is an old idea
	going back to Dekker (1967) and Brent (1973)
	\cite[p.~49]{brent:minimization}.  Recently, Abbot \cite{abbott:qir}
	introduced an adaptive form of the Newton-Bisection called
	QIR (quadratic interval refinement): this method requires
	that the iteration remembers some ``speed parameter'' $\sigma$ (a
	positive integer).
	When a Newton step is successful, we increment $\sigma$.
	If not, we perform bisection and decrement $\sigma$ (but never allowing it to
	go below some constant).  This idea had been previously been
	exploited in real root isolation
	\cite{sagraloff:newton-descartes:12},
	leading to the near-optimal bound \refeQ{bench}
	for the real benchmark problem.
	One twist in the Newton iteration is that we must
	apply the order $k$ Newton iterator $z' = z- k F(z)/F'(z)$
	when there is a cluster of $k$ roots.
	Such clusters are associated with a connected set of
	subdivision boxes which fail the $T_0$ test.
	Pellet's test enable us to confirm the presence of such clusters;
	success of a Newton is measured in terms the new $z'$
	being the center of the same cluster of $k$ roots with radius
	smaller by a factor of $2^{2^\sigma}$.
	If case of failure, we subdivide each box into four,
	and re-compute the clusters.   This make cause the original cluster of $k$
	roots to split into two or more smaller clusters.
	For the complexity analysis, one must show that there are no
	``long chains'' of failures in which the multiplicity $k$ is not reduced.
\end{s3}
\begin{s4}
\section{Preliminary}
	We review the basic tools from \cite{becker+3:cisolate:18}.
	%
	%
	The coefficients of $F$ are viewed as an oracle from
	which we can request approximations to any desired absolute precision.
	Approximate complex numbers are represented by a pair of
	dyadic numbers, where the set of
	dyadic numbers (or BigFloats) may be denoted
	$\ZZ[\half]\as \set{n2^m: n,m\in\ZZ}$.
	We formalize\footnote{
	  This is essentially the ``bit-stream model'',
	  but the term is unfortunate because it suggests that we
	  are getting successive bits of an infinite binary
	  representation of a real number.
	  We know from Computable Analysis that this representation
	  of real numbers is not robust.
	}
	this as follows:
	a complex number $z\in\CC$ is an \dt{oracular number}
	if it is represented by an \dt{oracle function}
	$\wtz:\NN\to \ZZ[\half]$ with some $\tau\ge 0$
	such that for all $L\in\NN$,
	$|\wtz(L)-z|\le 2^{-L}$ and $\wtz(L)$ has $O(\tau+L)$ bits.
	The oracular number is said to be \dt{$\tau$-regular} in this case.
	In our computational model, the algorithm is charged the cost
	to read these $O(\tau+L)$ bits.
	This cost model is reasonable when $z$ is
	an algebraic number because in this case,
	$\wtz(L)$ can be computed in time $\wtO(\tau+L)$ on a Turing machine.
	Following \citep{becker+3:cisolate:18,yap-sagraloff-sharma:cluster:13},
	we can construct a procedure $\softcompare(z_\ell,z_r)$
	that takes two non-negative real
	oracular numbers $z_\ell$ and $z_r$ with $z_\ell +z_r>0$,
	that returns a value in $\set{+1,0,-1}$
	such that if $\softcompare(z_\ell,z_r)$ returns $0$ then
	$\frac{2}{3}z_\ell<z_r<\frac{3}{2}z_\ell$; otherwise
	$\softcompare(z_\ell,z_r)$ returns $\sign(z_\ell-z_r)\in\set{+1,-1}$.
	Note that $\softcompare$ is non-deterministic
	since its output depends on the underlying oracular functions used.

	\blemT{\protect{see \cite[Lemma 4]{becker+3:cisolate:18}}
 	and \cite{yap-sagraloff-sharma:cluster:13}}{softcompare}

	\ \\ In evaluating $\softcompare(z_\ell, z_r)$:
	\ \\(a)
	The absolute precision requested from
	the oracular numbers $z_\ell$ and $z_r$
	is at most $$L= 2(\LOG(\max(z_\ell,z_r)^{-1})+4).$$
	(b)
	The time complexity of the evaluation is $\wtO(\tau+L)$
	where $z_\ell, z_r$ are $\tau$-regular.
	\elemT
	%

	The critical predicate for our algorithm is a test from
	Pellet (1881)
	(see \cite{marden:bk}).
	Let $\Delta =\Delta(m,r)$ denote a disc with radius $r>0$ centered at
	$m\in\CC$.
	For $k=0,1\dd n$ and $K\ge 1$,
	define the \dt{Pellet test} $T_k(\Delta,K) =T_k(\Delta, K;F)$ as
	the predicate
		\begin{equation*}\label{pellet}
		|F_k(m)| r^k > K\cdot \sum_{i=0, i\neq k}^n |F_i(m)|r^i
		\end{equation*}
	Here $F_i(m)$ is defined as the Taylor coefficient
	$\frac{F^{(i)}(m)}{i!}$.
	Call the test $T_k(\Delta,K)$ a \dt{success} if
	the predicate holds; else a \dt{failure}.
	Pellet's theorem says that for $K\ge 1$, a success implies $\#(\Delta)=k$.
    	Following \citep{yap-sagraloff-sharma:cluster:13,becker+3:cisolate:18},
	we define the ``soft version'' of Pellet test
	$\wtT_k(\Delta)$ 
	to mean that $\softcompare(z_\ell,z_r)>0$ where $z_\ell=|F_k(m)| r^k$
	and $z_r= \sum_{i=0, i\neq k}^n |F_i(m)|r^i$.
	%
	%
	We need to derive quantitative information in case
	the soft Pellet test fails.  Contra-positively, what quantitative
	information ensures that the soft Pellet test will succeed?
	Roughly, it is that $\#(\Delta)=\#(r\Delta)=k$ for a suitably
	large $r>1$, as captured by the following theorem:

	\bthml{theorem 2}
		\ \\Let $k$ be an integer with $0\le k\le n=\deg(F)$ and $K\ge 1$.
		Let $c_1=7kK$, and
			$\lambda_1=3K(n-k)\cdot \max\set{1,4k(n-k)}$.
		\\If
			$\#(\Delta) = \#(c_1\lambda_1\Delta)=k,$
		then
			$$T_k(c_1\Delta,K,F)\quad \text{holds.}$$
	\ethml

	The factor $c_1\lambda_1$ is $O(n^4)$ in this theorem, an
    	improvement from
	$O(n^5)$ in \cite{becker+3:cisolate:18}.
	A proof is given in Appendix A.
	In application, we choose $K=\frac{3}{2}$ and thus
		$c_1\cdot \lambda_1 \le (7Kn)\cdot (12Kn^3) = 189n^4$.
		The preceding theorem implies that
	  	if $\#(\Delta)=\#(189n^4\Delta)$
		then $T_k(\frac{21}{2}n\Delta,\frac{3}{2},F)$ holds.
		This translates into the main form for our application:

	\bgenDIY{{\sc Corollary}}{
	  \ \\ If $k=\#(\efrac{11}n\Delta)=\#(18n^3\Delta)$
		then $T_k(\Delta,\frac{3}{2};F)$ holds.
	}

	In other words, under the hypothesis of
	this Corollary, $\wtT_k(\Delta)$ succeeds.
	%
	We need one final extension: instead of applying
	$\wtT_k(\Delta)$ directly on $F$, we apply $\wtT_k(\Delta(0,1))$ to
	the $N$th Graeffe iterations of $F_\Delta(z)\as F(m+rz)$.
	Here, $\Delta=\Delta(m,r)$ and
	$N=\ceil{\log(1+\log n)}+4 =O(\log\log n)$.
	The result is called the \dt{Graeffe-Pellet test}, denoted
	$\wtTG_k(\Delta)=\wtTG_k(\Delta;F)$.
	As in \cite{becker+3:cisolate:18}
	we combine $\wtTG_k(\Delta)$ for all $k=0,1\dd n$ to obtain
		$$\wtTG_*(\Delta)$$
		which returns the unique $k\in\set{0\dd n}$
	such that $\wtTG_k(\Delta)$ succeeds,
	or else returns $-1$.
	We say that the test $\wtTG_*(\Delta)$
	\dt{succeeds} iff $T^G_*(\Delta,K)\ge 0$.

	\ignore{
		\blemT{Hard Pellet-Graeffe Test}{hard}
		Let $\rho_1=\frac{2\sqrt{2}}{3}\simeq 0.943$ and $\rho_2=\frac{4}{3}$.
		\\ (a) If $T^G_k(\Delta,K)$ succeeds for some $K\ge 1$ then
			$\#(\Delta)=k$.
		\\ (b) If $T^G_k(\Delta,3/2)$ fails for every $k\in\set{0\dd n}$
		then $\#(\rho_2\Delta)> \#(\rho_1\Delta)$.
		\elemT

		(without the $K$ parameter) in which the comparison is performed
		on two subexpressions $E_\ell:E_r$ of the underlying polynomial.
		Again, let
			$\wtTG_*(\Delta)$
		return the $k$ such that $\wtTG_k(\Delta)$ succeeds, or returns $-1$
		if there is no such $k$.
		In implementation, we would perform
		$\wtTG_k$ for $k=0,1\dd $ until we get a success, but
		in our algorithm, we often have an upper bound on the number
		of roots in $\Delta$
		that is better than the trivial bound of $n$.  In that case
		we can return $-1$ as soon as $k$ exceed that upper bound.

		If $\Delta=\Delta(m,r)$, let $F_\Delta(z) \as F(m+rz)$.
		\blem
		\ \\(a) If $T_k(\Delta,3/2)$ succeeds, then $\wtT_k(\Delta)$ succeeds.
		\ \\(b) $\wtT_k(\Delta)$ needs an
		$\wtO(n\LOG(r)+\tau_F+L(\Delta,F))$-bit approximation of $F$ where
			$$L(\Delta,F)\as 2\cdot(4+\LOG(\|F_\Delta\|_{\infty}^{-1})).$$
		\ \\(c) The cost of computing $\wtT_*(\Delta)$ for $k=0\dd n$ is
		$\wtO(n(n\LOG(r)+\tau_F+L(\Delta,F)))$.
		\elem
	The soft version of \refLem{hard} is:
        }%

	The key property of $\wtTG_i(\Delta)$ is
	\cite[Lemma 6]{becker+3:cisolate:18}:

	\blemT{Soft Graeffe-Pellet Test}{soft}
	\ \\ Let $\rho_1=\frac{2\sqrt{2}}{3}\simeq 0.943$ and $\rho_2=\frac{4}{3}$.
	\ \\(a) If $\wtTG_k(\Delta)$ succeeds then $\#(\Delta)=k$.
	\ \\(b) If $\wtTG_*(\Delta)$ fails then $\#(\rho_2\Delta)>\#(\rho_1\Delta)$.
	\elemT

	The bit complexity of the combined test $\wtTG_*(\Delta)$
	is asymptotically the same as any individual test
	\cite[Lemma 7]{becker+3:cisolate:18}:
	\blem
	Let
		$$L(\Delta,F)\as 2\cdot(4+\LOG(\|F_\Delta\|_{\infty}^{-1})).$$
	(a) To evaluate $\wtTG_k(\Delta)$, it is sufficient
		to have an $M$-bit approximation of
		each coefficient of $F$ where
		$M = \wtO(n\LOG(m,r)+\tau_F+L(\Delta,F))$.
	\\(b) The total bit-complexity of computing $\wtTG_*(\Delta)$
		is $\wtO(n M).$
	\elem
\subsection{Box Subdivision}
	Let $A,B\ib\CC$.  Their \dt{separation} is $\Sep(A,B) \as$
	$\inf\{|a-b|: a\in A, b\in B\}$,
	and $\rad(A)$, the \dt{radius} of $A$, is the smallest radius
	of a disc containing $A$.
	Also, $\partial A$ denotes the boundary of $A$.

	We use the terminology of subdivision trees (quadtrees)
    (see \cite{becker+3:cisolate:18}).  All boxes are closed subsets
	of $\CC$ with square shape and axes-aligned.
	Let $B(m,w')$ denote the axes-aligned box centered at $m$
	of width $w(B)\as w'$.
	As for discs, if $k\ge 0$ and $B=B(m,w')$,
	then $kB$ denotes the box $B(m,kw')$.
	The smallest covering disc of $B(m,w')$ is $\Delta(m,\efrac{\sqrt{2}}w')$.
	If $B=B(m,w')$ then we define
	$\Delta(B)$ as the disc $\Delta(m,\frac{3}{4}w')$.
	\longShort{This definition will be useful in the
	application of \refLem{soft} above.
      }{}Thus $\Delta(m,\efrac{\sqrt{2}}w')$ is properly contained in $\Delta(B)$.
	Any collection $\calS$ of boxes is called a (box) \dt{subdivision}
	if the interior of any two boxes in $\calS$ are disjoint.
	The union $\bigcup \calS$ of these boxes is called the \dt{support} of $S$.
	Two boxes $B,B'$ are \dt{adjacent} if $B\cup B'$ is a connected set, equivalently, $B\cap B'\neq\es$.
	A subdivision $\calS$ is said to be \dt{connected} if its support is connected.
	A \dt{component} $C$ is the support of some connected subdivision
	$\calS$, i.e., $C= \bigcup \calS$.

	The \dt{split} operation on a box $B$ creates
	a subdivision $\splittt(B)=\set{B_1\dd B_4}$
	of $B$ comprising four congruent subboxes.
	Each $B_i$ is a \dt{child} of $B$, denoted $B\to B_i$.
	Therefore, starting from any box $B_0$, we may split
	$B_0$ and recursively split zero or more of its children.
	After a finite number of such splits, we obtain a \dt{subdivision tree}
	rooted at $B_0$, denoted $\subdivTree(B_0)$.
	\ignore{
	Any box that can potentially appear in $\subdivTree(B_0)$
	is called\footnote{
	  Aligned boxes are necessarily axes-aligned but the converse is not true.
	}
	an \dt{aligned box} (relative to $B_0$.
	The significance of aligned boxes is that if $B$ is
	represented by dyadic numbers, then all aligned boxes remain
	representable by dyadic numbers.
      }%

	The \dt{exclusion test} for a box $B(m,w')$ is
	$\wtTG_0(\Delta(m,\frac{3w'}{4})) =\wtTG_0(\Delta(B))$.
	We say that $B(m,w')$ is \dt{excluded} if this test succeeds,
	and \dt{included} if it fails.
	The key fact we use is a consequence of \refLem{soft} for
	the test $\wtTG_0(\Delta)$:
	\bcorl{exclusion}
	Consider any box $B = B(m,w')$.
	\\ (a) If $B$ is excluded, then
	$\#(\Delta(m,\frac{3w'}{4}))=0$, so $\#(B)=0$.
	\\ (b) If $B$ is included, then
	$\#(\Delta(m,w'))>0$, so $\#(2B)>0$.
	\ecorl
	%
\subsection{Component Tree}
	In traditional subdivision algorithms, we focus on
    the complexity
	analysis on the subdivision tree $\subdivTree(B_0)$.
	But for our algorithm, it is more natural to work with a tree
	whose nodes are
	higher level entities called components above.

	Typical of subdivision algorithms, our algorithm consists
	of several while loops, but for now,
    we only consider the main loop.
	This loop is controlled by the \dt{active queue} $Q_1$.
	At the start of each loop iteration,
	there is a set of included boxes. The maximally connected
	sets in the union of these boxes constitute our (current) components.
    And the boxes in the subdivision of a component $C$
    are called the \dt{constituent boxes} of $C$.
	\ignore{	These components may have been discarded (they are
	``adventitious''
	as discussed below), or stored in the \dt{output queue}
	$\qout$, or remain active in $Q_1$.
	}%
	While $Q_1$ is non-empty, we remove a component $C$ from $Q_1$
	for processing.   There are 3 dispositions for $C$:
	We try to put $C$ to the \dt{output queue} $\qout$. Failing this, we try a \dt{Newton Step}.
	If successful, it produces a single
	new component $C'\ibp C$ which is placed in $Q_1$.
	If Newton Step fails, we apply a \dt{Bisection Step}.
	In this step,
	we split each constituent box of $C$, and apply the
	exclusion test to each of its four children.
	The set of included children are again organized into maximally
	connected sets $C_1\dd C_t$ ($t\ge 1$).
	Each subcomponent $C_i$ is either placed in $Q_1$ or $\qdis$,
    depending on whether $C_i$ intersects the initial box $B_0$.
	The components in $\qdis$ are viewed as \dt{discarded} because we
	do not process them further (but our analysis need to ensure
	that other components are sufficiently separated from them in the
	main loop).
	We will use the notation $C\to C'$ or $C\to C_i$ to indicate
	the parent-child relationship.  The \dt{component tree} is
	defined by this parent-child relationship, and denoted $\compTree$.
	In \cite{becker+3:cisolate:18}, the root of the component tree
	is $B_0$; we take $\frac{5}{4}B_0$ as the root to
    address boundary issues.
\ignore{this is sufficient for global root isolation, but for
	local root isolation, matters are simplified with
	$\frac{5}{4}B_0$ as root.}
    So we write $\compTree=\compTree(\frac{5}{4}B_0)$
	to indicate that $\frac{5}{4}B_0$ is the root.
	The leaves of $\compTree$ are either discarded (adventitious)
	or output.

	For efficiency, the set of boxes in the subdivision of a component $C$
	must maintain links to adjacent boxes within the subdivision;
	such links are easy to maintain
	because all the boxes in a component have the same width.
\section{Component Properties}
	Before providing details about the algorithm,
	we discuss some critical data associated with each component $C$.
	Such data is subscripted by $C$.
\longShort{
	E.g., $\calS_C$ represents the subdivision of $C$.}
{}We also describe some qualitative properties so that
	the algorithm can be intuitively understood.
	\refFig{component} may be an aid in the following description.

	    	\begin{figure}[htb]
	    	  \begin{center}
		   \scalebox{0.3}{
	    	     \input{./component.tex}}
	    	   \caption{Three components $C_1, C_2, C_3$:
	blue dots indicate roots of $F$,
	pink boxes are constituent boxes,
	and the non-pink parts of each $B_C$ is colored cyan.
	Only $C_3$ is confined.
	}
	    	   \label{fig:component}
	    	  \end{center}
	    	\end{figure}
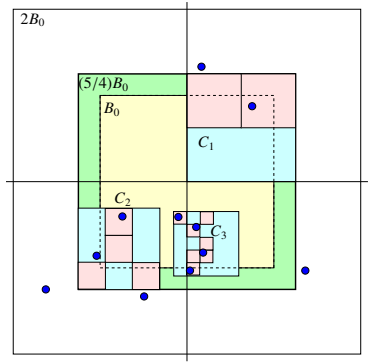 	

	\benum[(C1)]\setlength{\itemsep}{-2pt}
	\item
\longShort{
		The primary data associated with a component $C$
		is the connected subdivision $\calS_C$
		comprised of included boxes.
	  	The boxes in $\calS_C$ are called \dt{constituent boxes} of
		$C$, and they all 
		share a common width, $w_C$.
		The \dt{component size} $|\calS_C|$ is 
	  	the number of boxes.
}{All the constituent boxes of a component share a common width,
denoted by $w_C$.}
	\item
		Our algorithm never discards any box $B$
		if $B$ contains a root in $B_0$; it follows that
		all the roots in $B_0$ are contained in
		$\bigcup_{C} C$ where $C$ ranges over components in
		$Q_0\cup Q_1\cup \qout$ (at any moment during our
		algorithm).
	\item
\longShort{
		Since the initial component is $\frac{5}{4}B_0$, all the
		components are subsets of $\frac{5}{4}B_0$.
}{}
		Recall that a zero $\zeta$ of $F(z)$ in
		$2B_0\setminus B_0$ is called adventitious.
		A component $C$ is \dt{adventitious} if $C\cap B_0$
		is empty\longShort{. In this case, it follows from (C3) that
		then all the roots in $C^+$ are
		adventitious, and therefore we can discard $C$
		(i.e., store in $\qdis$).
}{ (placed in $\qdis$).}
\longShort{Let $\partial(\frac{5}{4}B_0)$ be called the \dt{critical boundary}.
}{}We say a component $C$ is \dt{confined}
	  if $C\cap \partial(\frac{5}{4}B_0)$ is empty; otherwise it is non-confined. \refFig{confined} shows these different kinds of components.
      Note that after the preprocessing step,
		all components are confined.
\ignore{
and for such
		we have $\#(C)=\#(C^+)$ (see Appendix B).
		Moreover, all the zeros in $C$ are separated from
		zeros outside $C$ by at least $w_C$.
}
	\item
	  	If $C, C'$ are distinct active components, then
		their separation $\Sep(C,C')$ is at least $\max\set{w_C, w_{C'}}$.
		If $C$ is an adventitious component, then
        $\Sep(C, B_0)\ge w_C$.
        If $C$ is a confined component, then
        $\Sep(C,\partial(\frac{5}{4}B_0))\ge w_C$.
    \item
    Let $C^+$ be the \dt{extended component} defined as the set
       $\bigcup_{B\in\calS_C}2B$.
    If $C$ and $C'$ are distinct components, then $C^+$ and $C{'^+}$
    are disjoint. Moreover, if $C$ is confined, then
    $\#(C)=\#(C^+)$ (see Appendix B).
\ignore{
    \longShort{
		In \cite{becker+3:cisolate:18},
		we do not have such a separation in general.
		if $C$ and $C'$ are two components
		such that both intersect $\partial B_0$ in
		\cite{becker+3:cisolate:18}, then the roots
		in $C^+$ and in $C'^+$ may be arbitrarily close.
		Moreover, it is not possible to restrict to
		components that are ``confined'' in the sense of
		not intersecting $\partial B_0$.  This also
		motivates our use of $\frac{5}{4}B_0$ as the root
		for subdivision.
	      }{}
}
	\item
	  	Define the \dt{component box} $B_C$ to be any smallest
		square containing $C$ subject to $B_C\ib (5/4)B_0$.
       Define $W_C$ as the width of $B_C$ and the disc
       $\Delta_C\as \Delta(B_C)$. Define $R_C$ as the radius of
       $\Delta_C$; note that $R_C=\frac{3}{4}W_C$.
\ignore{
		If $M_C$ and $W_C$ is the center and width of $B_C$,
		then define the disc $\Delta_C\as \Delta(M_C,R_C)$
		where $R_C\as \frac{3}{4}W_C$.
}
\longShort{
		Note that $B_C$ and $\Delta_C$ are computational
		data, and so our definition ensures that they are dyadic.
}{}
	\item
	  	Each component is associated with a ``Newton speed''
\longShort{which is a number
		$N_C$ such that $\lg\lg N_C$ is a positive integer.
		Here, $\lg=\log_2$.  Thus $N_C\ge 2^{2^1}=4$ is also integer.
}{denoted by $N_C$ with $N_C\ge4$.}
		A key idea in the Abbot- Sagraloff technique for
		Newton-Bisection is to automatically update $N_C$:
		if Newton fails, the children of $C$ have speed $\max\set{4,\sqrt{N_C}}$
		else they have speed $N_C^2$.
\longShort{
		We use $N_C$ when applying Newton iteration to $C$.
		Computationally, it suffices to maintain $\lg\lg N_C$
		instead of $N_C$.
}{}
	\item
		Let $k_C\as \#(\Delta_C)$,
        	the number of roots of $\calZ(\Delta_C)$,
		{\em counted with multiplicity}.
		Note that $k_C$ is not always available, but it is needed
		for the Newton step.
\longShort{Moreover, $k_C\ge \#(C)$.

        	In our main loop, we will try to determine
		$k_C$ just before the Newton step.
		It is easy to maintain
		an upper bound on $k_C$ to speed up the $\wtTG_*$ tests.
}{We try to \\determine $k_C$ before the Newton Step in the main loop.}

	\item
	  	A component $C$ is \dt{compact} if $W_C\le 3w_C$.
		Such components have many nice properties, and we will
		require output components to be compact.
	\eenum

	In recap, each component $C$ is associated with the data:
		\begin{equation*}\label{assoc}
		\longShort{\calS_C, }w_C, W_C,
		B_C, \Delta_C, R_C,
		k_C,
		N_C.
		\end{equation*}
\longShort{	Except for $N_C$, these are intrinsic properties of $C$.}{}

	%

	    	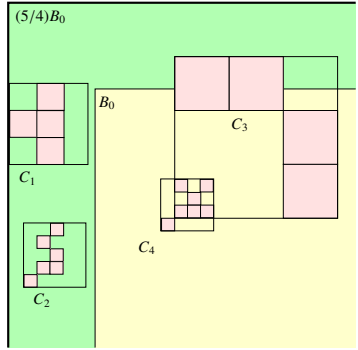
\begin{figure}[htb]
	    	  \begin{center}
		   \scalebox{0.3}{
	    	     \input{./confined.tex}}
	    	   \caption{Four types of components:
		  $C_1$ is not confined, the rest are confined;
		  $C_1$ and $C_2$ are adventitious;
		  $C_3$ may contain adventitious roots;
		  $C_4$ has no adventitious roots.
		}
	    	   \label{fig:confined}
	    	  \end{center}
	    	\end{figure} 	

\section{The Clustering Algorithm}
	As outlined above, our clustering algorithm is a process
	for constructing and maintaining components,
	globally controlled by queues containing components.
	Each component $C$ represents a non-empty set
	of roots.  In addition to the queues $Q_1, \qout, \qdis$
	above, we also need a \dt{preprocessing queue} $Q_0$.
	Furthermore, $Q_1$ is a priority queue such that the operation
	$C\ass Q_1.pop()$ returns the component with the largest width $W_C$.

	We first provide a high level description of the two main subroutines,
	Newton Step and Bisection Step.

\longShort{\bitem\setlength{\itemsep}{0pt}}{}
\longShort{      \item}{}
	The \dt{Newton Step} $\newton(C)$
	is directly taken from \cite{becker+3:cisolate:18}.
	This procedure
	takes several arguments,
      $\newton(C,N_C,k_C,\\x_C)$.
	The intent is to perform an order $k_C$ Newton step:
		$$x'_C \ass x_C - k_C \frac{F(x_C)}{F'(x_C)}.$$
		We then check whether $\calZ(C)$ is actually
	contained in the small disc
	$\Delta'\as \Delta(x'_C, r')$ where
		\beql{r'}
		r'\as \max\set{\vareps, w_C/(8N_C)}.
		\eeql
	This amounts to checking whether $\wtTG_{k_C}(\Delta')$ succeeds.
If it does, Newton test succeeds, and we return a
	new component $C'$ that contains $\Delta'\cap C$ with speed $N_{C'}\as (N_C)^2$
	and constituent width $w_{C'}\as \frac{w_C}{2N_C}$.
	The new component $C'$ consists of at most $4$ boxes
    	and $W_{C'}\le 2w_{C'}$.
	In the original paper \citep{becker+3:cisolate:18},
		$r'$ was simply set to $\frac{w_C}{8N_C}$;
	but \refeQ{r'} ensures that $r'\ge \vareps$.
	This avoids the overshot of Newton Step and simplifies our complexity analysis.
\longShort{ \item
	We now think of $C$ as an object (in Programming Language
	jargon) that stores data such as $N_C$ and $k_C$.
	Moreover, from $C$, we can
	compute the additional argument $x_C$ within the $\newton$
	subroutine.  Therefore we can simply denote this routine
	as ``$\newton(C)$''.  Moreover, $\newton(C)$ returns
	either an empty set (in case of failure)
	or the component $C'$.
}{If $\wtTG_{k_C}(\Delta')$ fails, then Newton test fails, and
it returns an empty set.
In the following context, we simply denote this routine as
``$Newton(C)$''.}

	\ignore{
	We now give the details of the Newton Step:
	\progb{
	\lline $\newton(C,k)$
	\lline[5] OUTPUT: success or failure.  In case of success,
	\lline[15]  a component $C'\ib C$, $\#(C)=\#(C')$, is added to $Q_1$.
	\lline[5] Pick any $x_C\notin C$ such that
	\lline[15]		$x_C$ is at most $w_C/2$ from $C$.
	\lline[5] If $SoftCompare( 4r_C |F'(x_C)|, |F(x_C)|)$ fails
	\lline[10]  Return Fail.
	\lline[5] For $L=1,2,4,\ldots$
	\lline[10]  Compute an $L$-bit approximation of $F(x_C)$ and $F'(x_C)$
	...
	}
      }

\longShort{	\item}{} The \dt{Bisection Step}
	  $\bisect(C)$ returns a set of components.
	  Since it is different from that in \cite{becker+3:cisolate:18},
	  we list the modified bisection algorithm in \refFig{bisect}.

\longShort{	\eitem}

	\begin{figure}[hbt]
	\progb{
	  \lline[-3] $\bisect(C)$
	\lline[0] OUTPUT: a set of components containing all
	\lline[7] the non-adventitious roots in $C$
	\lline[7] (but possibly some adventitious ones)
	\lline[0] Initialize a Union-Find data structure $U$
	\lline[15] for boxes.
	\lline[0] For each constituent box $B$ of $C$
	\lline[5]  For each child $B'$ of $B$
	\lline[10]  	If ($\wtTG_0(\Delta(B'))$ fails)
	\lline[15] 		$U.add(B')$
	\lline[15]  		For each box $B''\in U$ adjacent to $B'$
	\lline[20]  		  $U.union(B',B'')$
	\lline[0] Initialize $Q$ to be empty.
		\ignore{
	\lline[0] $specialFlag \ass \true$
	\lline[0] If ($U$ has only one connected component)
	\lline[5]	$specialFlag \ass \false$
		}%
	\lline[0] For each connected component $C'$ of $U$
	\lline[5]	If {\color[rgb]{0.00,0.70,0.00}($C'$ intersects $B_0$)}
							// $C'$ not adventitious
		\ignore{
	\lline[10]		If ($specialFlag$)
    \lline[15]            $N_{C'}=4$
	\lline[10]		Else
    \lline[15]           $N_{C'}=\max\set{4, \sqrt{N_C}}$
		}%
		\lline[10]		$N_{C'}=\clauses{4
							& \rmif\ U \textrm{ has only 1 component,}\\
						\max\set{4,\sqrt{N_C}} & \ELSe.}$
	\lline[10]      $Q.add(C')$
    \lline[5]	Else	
    \lline[10]      $\qdis.add(C')$
	\lline[0] Return $Q$
	}
	\caption{Bisection Step}
	\label{fig:bisect}
      \end{figure}

	We list the clustering algorithm in \refFig{algo}.

	\begin{figure}[hbt]
	\progb{
	  \lline[-5] {\sc Root Clustering Algorithm}
	  \lline[-5] \INPUt: Polynomial $F(z)$, box $B_0\ib\CC$ and $\vareps>0$
	  \lline[-5] \OUTPUt: Components in $\qout$ representing
	  \lline[5] natural $\vareps$-clusters of $F(z)$ in $2B_0$.
	  \lline[0] \commenT{Initialization}
	  \lline[0] $\qout\ass Q_1 \ass \qdis \ass \es$.
	  \lline[0] $Q_0\ass \set{(5/4)B_0}$ // initial component
	  \lline[0] \commenT{Preprocessing}
	  \lline[0] While $Q_0$ is non-empty
	  \lline[5]	$C\ass Q_0.pop()$
	  \lline[5]	If ($C$ is confined and $W_C\le w(B_0)/2$)
	  \lline[15]		$Q_1.add(C)$
	  \lline[5]	Else
      \lline[10]	$Q_0.add(\bisect(C))$
	  \lline[0] \commenT{Main Loop}
	  \lline[0] While $Q_1$ is non-empty
	  \lline[5]	$C\ass Q_1.pop()$
	  		// $C$ has the largest $W_C$ in $Q_1$
	  \lline[5]	If ($4\Delta_C\cap C'=\es$ for all $C'\in Q_1\cup \qdis$)
	  		// (*)
	  \lline[10]	   $k_C \ass \wtTG_*(\Delta_C)$
	  \lline[10]	   If $(k_C> 0)$ // Note: $k_C\neq 0$.
	  \lline[15]		If ($W_C\ge \vareps$)
	  \lline[20]			$C' \ass \newton(C, other args!)$
	  \lline[20]			If ($C'\neq\es$)
	  \lline[25]		  		$Q_1.add(C')$;\quad Continue
	  \lline[15]		Else if (\cored{$W_C\le 3w_C$}) // $C$ is compact
	  \lline[20]		  	$\qout.add(C)$;\quad Continue
	  \lline[5]    $Q_1.add(\bisect(C))$
	  \lline[0]    Return $\qout$
	}
	\caption{Clustering Algorithm}
	\label{fig:algo}
        \end{figure}

	Remarks on Root Clustering Algorithm:
	\ \\1.
    The steps in this algorithm should appear well-motivated
	(after \cite{becker+3:cisolate:18}).  The only
	non-obvious step is the test ``$W_C\le 3w_C$''
	(colored in red). We may say $C$ is
			\dt{compact} if this condition holds.
	This part is only needed for the analysis; the correctness of the
	algorithm is not impacted if we simply
	replace this test by the Boolean constant $\true$
	(i.e., allowing the output components to have $W_C>3w_C$).
	\\ 2. We ensure that $W_C\ge \vareps$ before we attempt to
	do the Newton Step.  This is not essential, but simplifies
	the complexity analysis.

      Based on the stated properties,
      we prove the correctness of our algorithm.

      \bthmT{Correctness}{corr}
      The Root Clustering Algorithm halts and outputs
      a collection
      $\set{(\Delta_C,k_C): C\in \qout}$
       of pairwise disjoint $\vareps$-isolators
      such that
      $\calZ(B_0)\ib \bigcup_{C\in \qout} \calZ(\Delta_C) \ib \calZ(2B_0)$.
      \ethmT
      \bpf
    First we prove halting.  By way of contradiction, assume $\compTree$
    has an infinite path $\frac{5}{4}B_0=C_0\to C_1\to C_2\to\cdots$.
    After $O(\log n)$ steps, the $C_i$'s are in the main loop
    and satisfies $\#(C_i)=\#(C^+_i)\ge 1$.  Thus the $C_i$ converges
    to a point $\xi$ which is a root of $F(z)$.
    For $i$ large
    enough, $C_i$ satisfies $W_{C_i}\le 3w_{C_i}$ and $w_{C_i}<\vareps$.
    Moreover, if $C_i$ is small enough,
	$4\Delta_{C_i}$ will not intersect other components.
    Under all these conditions,
	the algorithm would have output such a $C_i$.
    This is a contradiction.

    Upon halting, we have a set of output components.
    We need to prove that they represent a set of pairwise disjoint natural
    $\vareps$-clusters.  Here, it is important to use the fact that
    $Q_1$ is a priority queue that returns
    components $C$ in non-increasing width $W_C$.
    Suppose inductively, each component in the $\qout$
    represents a natural $\vareps$-cluster, and they are pairwise disjoint.
    Consider the next component $C$ that we output:
	from line (*), we know that $4\Delta_C$ does not intersect any
    components in $Q_1\cup\qdis$.   But we also know that
    $C\cap 4\Delta_{C'}=\es$ for any $C'$ in $\qout$.
    We claim that this implies that
    		$3\Delta_C \cap C'$ must be empty.
    To see this, observe that $W_C\le W_{C'}$ because of the
    priority queue nature of $Q_1$.  Draw the disc $4\Delta_{C'}$,
    and notice that the center of $\Delta_C$ cannot intersect $3\Delta_{C'}$.
    Therefore, $3\Delta_C$ cannot intersect $\Delta_C'$.
    This proves that $C$ can be added to $\qout$ and preserve the
    inductive hypothesis.

    It is easily verified that the roots represented by the confined components
    belong to $\frac{15}{8}B_0\ibp 2B_0$.
    But we must argue that we cover all the roots in $B_0$.
    How can boxes be discarded?  They might be discarded in the Bisection Step
    because they succeed the exclusion test, or because they belong
    to an adventitious component.   Or we might replace an entire
    component by a subcomponent in a Newton Step, but in this case,
    the subcomponent is verified to hold all the original roots.
    Thus, no roots in $B_0$ are lost.
    \epf

    We now show some basic properties of the components produced in
    the algorithm.
    \bleml{CompBasic}
  \ \\ Let $C$ be a component.
  \ \\(a)
If $C$ is confined with $k=\#(C)$, then
$C$ has at most $9k$ constituent boxes.  Moreover,
$W_C\le 3k \cdot w_C$.
\ \\(b)
If $\calZ(C)$ is strictly contained in a box of width $w_C$,
then $C$ is compact: $W_C\le 3w_C$.
\ \\(c)
If there is a non-special path $(C_1\to\cdots\to C)$ where $C_1$ is special,
then $w_C \le \frac{4w_{C_1}}{N_C}$.
\elem

\bpf
Parts (a) and (b) are easy to verify.
Part (c) is essentially
from \cite[Theorem 4]{becker+3:cisolate:18} with a slight difference:
we do not need to $C_1$ to be equal to the root $\frac{5}{4}B_0$.
That is because our algorithm resets the Newton speed of
the special component $C_1$ to $4$.
\epf

The next lemma addresses the question of lower
bounds on the width $w_C$ of boxes in components.
If $C$ is a leaf, then $w_C<\vareps$, but how much smaller
than $\vareps$ can it be?  Moreover, we want to lower bound $w_C$
as a function of $\vareps$.

\bleml{BoxSize}
  Denote $k=\#(2B_0)$.
  \ \\(a)
  If $C$ is a component in the pre-processing stage, then
  $w_C\ge \frac{w(B_0)}{48k}$.
  \ \\(b)
  Suppose $C_1\to \cdots\to C_2$ is a non-special path
  with $ W_{C_1}<\vareps$.
  Then it holds
  $$ \frac{w_{C_1}}{w_{C_2}}< 57k.$$
  \ \\(c)
  Let $C$ be a confined leaf in $\compTreeHat$
  then
       $$w_C>\frac{\vareps}{2}\Big(\efrac{114k}\Big)^{k}.$$
\elem

A proof of \refLem{BoxSize} is given in Appendix B.

We will need what we call the \dt{small $\vareps$ assumption},
    namely, $\vareps \le \min\set{1,w(B_0)/(96n)}$.
    If this assumption fails, we can simply replace $\vareps$ by $\vareps=\min\set{1,w(B_0)/(96n)}$
    to get a valid bound from our analysis.
	This assumption is to ensure that no $\vareps$-cluster is split
    in the preprocessing stage.

\section{Bound on Number of Boxes}
\longShort{	The rest of this paper is concerned with complexity analysis.}
	In this section,
	we bound the number of boxes produced by our algorithm.
	All the proofs for this section are found in Appendix B.

\ignore{	The key is to bound the maximum length $\smax$ of a
	path in $\compTree$ that is ``non-special''.
}
    The goal is to bound the number of all the constituent boxes
    of the components in $\compTree$.
	But, in anticipation of the following complexity analysis,
	we want to
    consider
	an \dt{augmented component tree} $\compTreeHat$ instead of
    $\compTree$.
	%
	%

\longShort{
		\bitem
        \setlength{\itemsep}{0pt}
	      \item
}{}
\ignore{
		\refFig{confined} shows four components $C_i$ ($i=1\dd 4$)
		illustrating the following remarks.
		The leaves of $\compTree$ may be adventitious (e.g., $C_1, C_2$)
		or non-adventitious (e.g., $C_3, C_4$).
		For the adventitious components, we further classify them
		as confined (e.g., $C_2$) or non-confined (e.g., $C_1$).
}

		Let $\compTreeHat$ be the extension of $\compTree$
		in which, for each confined adventitious components
		in $\compTree$, we (conceptually) continue to run our algorithm
		until they finally produce output components, i.e., leaves
		of $\compTreeHat$.  As before,
		these leaves have at most 9 constituent boxes.

\longShort{		\item}{}
		  Since $C'\to C$ denote the parent-child relation,
		  a path in $\compTree$ may be written
		  \beql{P}
		  P=(C_1\to C_2\to\cdots\to C_s).
		  \eeql
		  We write $w_i, R_i, N_i$, etc,
		  instead of $w_{C_i}, R_{C_i}, N_{C_i}$, etc.

\longShort{		\item}{}
		  A component $C$ is \dt{special} if $C$ is the root or a leaf
		  of $\compTreeHat$, or if $\#(C)<\#(C')$
		  with $C'$ the parent of $C$ in $\compTreeHat$;
		  otherwise it is \dt{non-special}.
		  This is a slight variant of \cite{becker+3:cisolate:18}.

\longShort{		 \item}{}
		   We call $P$ a
              \dt{non-special path led by $C_1$},
            if each $C_i$ ($i=2\dd s$) is non-special,
	    i.e., $\#(C_i)=\#(C_{i-1})$.
		  The \dt{special component tree} $\compTree^*$ is obtained from
		  $\compTreeHat$ by eliminating any non-special components
		  while preserving the descendant/ancestor relationship among
		  special nodes.
\longShort{		\eitem}{}

    We now consider the length of an arbitrary non-special path as in \refeQ{P}.
    In \cite[Lemma 10]{becker+3:cisolate:18}, it was shown that
	$s =O\Big(\log n + \log(\LOG (w(B_0))\cdot \LOG(\sigma_F(2B)^{-1}))\Big)$.
    We provide an improved bound
    which is based on local data, namely, the ratio $w_1/ w_s$ only.	

    We define $\smax$ to be the maximum length
	of a non-special path in $\compTreeHat$.

	\bthml{s_max}
The length of the non-special path \refeQ{P} satisfies
    $$s=O(\log\log\frac{w_{1}}{w_{s}}+\log n).$$
 Particularly,
	   $$
\smax =O\Big(\log n + \log\log\frac{w(B_0)}{\varepsilon}\Big).$$
	\ethm

The proof of \refThm{s_max} is found in Appendix B.


\longShort{	\bigskip}{}
	{\bf Charging function $\phi_0(B)$.}
	For each component $C$,
	define the \dt{root radius} of $C$ to be $r_C\as \rad(\calZ(C))$,
	that is the radius of the smallest disc enclosing all the roots in $C$.
	We are ready to define a charging function $\phi_0$
    for each box $B$ in the components of $\compTreeHat$:
	Let $C_B\in\compTreeHat$ be the component of which $B$ is a constituent box.
	Let $\xi_B$ be any root in $2B$.
	There are two cases:
	(i) If $C_B$ is a confined component,
	there is a unique maximum path in $\compTreeHat$ from $C_B$ to a
	confined leaf $E_B$ in $\compTreeHat$ containing $\xi_B$.
	Define $\phi_0(B)$ to be the first special component $C$
	along this path such that
	   \begin{equation}\label{charge}
	     r_C < 3 w_B.
	   \end{equation}
	   where $w_B$ is the width of $B$.
	(ii) If $C_B$ is not confined, it means that
	$C$ is a component in the preprocessing stage.
	In this case, define $\phi_0(B)$ to be the largest natural
    $\vareps$-cluster containing $\xi_B$.
	Notice that $\phi_0(B)$ is a special component in (i)
	but a cluster in (ii).

	\bleml{phi_0}
	The map $\phi_0$ is well-defined.
	\elem

\bpf
    Consider the component $C_B$ of which $B$ is a constituent box.
    There are two cases in our definition of $\phi_0$:

       (i) If $C_B$ is a confined component, it is easy to see that
           we can find a root $\xi_B\in 2B$,
           and fix a unique maximum path in $\compTreeHat$ from $C_B$ to a
           confined leaf $E_B$ in $\compTreeHat$ containing $\xi_B$.
           It suffices to prove that we can always find a
           special component $C$ in this path such that $r_C<3w_B$.
           This is true because $r_{E_B}<3 w_{E_B}$;
           to see this, note that $E_B$ is a confined leaf of $\compTreeHat$.
	   Thus $W_{E_B}\le 3w_{E_B}$ (this is the condition for output
	   in the main loop of the Root Clustering Algorithm).
	   It follows
              $r_{E_B}\le\frac{\sqrt2}{2}\cdot 3w_{E_B}<3w_{E_B}$.
           Hence $r_{E_B}<3w_{E_B}<3w_B$.
           We can always find a first special component along the path from
           $C_B$ to $E_B$ such that (\ref{charge}) is satisfied. 

           (ii) If $C_B$ is a non confined component, we can also find a root
           $\xi_B$ in $2B$, and we can always charge $B$ to the largest
           natural $\vareps$-cluster containing $\xi_B$.
%
\epf

	Using this map, we can now bound the number of boxes.

	\bleml{TotalBox}
	   The total number of boxes in all the components in $\compTreeHat$ is
	   	$$O(t\cdot s_{\max}) = O(\#(2B_0)\cdot s_{\max})$$
	   with $t=|\{\phi_0(B):B $ is any box in $\compTreeHat\}|$.
	\elem

	This improves the bound in \cite{becker+3:cisolate:18}
	by a factor of $\log n$. A proof for \refLem{TotalBox} is found in Appendix B.
\section{Bit Complexity}
	Our goal is to prove the bit-complexity theorem stated
	in the Introduction.
    From the discussion in \cite[Theorem 7]{becker+3:cisolate:18},
    the total cost of all the $\wtTG$ tests is the main cost of
    the whole algorithm.
    Thus we need to account for the cost of $\wtTG$ tests on all the concerned
boxes and components.

	The road map is as follows:
	we will charge the work of each box $B$
	(resp., component $C$) to some natural
	$\vareps$-cluster denoted $\phi(X)$ (resp., $\phi(C)$).
	We show that each cluster $\phi(X)$ ($X$ is a box or
	a component) is charged $\wtO(1)$ times.
	Summing up over these clusters, we obtain our bound.

	We may assume $\LOG(B_0)=O(\tau_F)$ since
	Cauchy's root bound implies that any root $z_i$
    satisfies $|z_i|\le 1+4\cdot 2^\tau_F$, thus
    we can replace $B_0$ by $B_0\cap B(0,2+8\cdot 2^\tau_F)$.

	%

	{\bf Cost of $\wtTG$-tests and Charging function $\phi(X)$:}
	Our algorithm performs 3 kinds of $\wtTG$-tests:
		\beql{cost}
			\wtTG_*(\Delta_C),\quad
			\wtTG_{k_C}(\Delta'),\quad
			\wtTG_0(\Delta(B))
		\eeql
    respectively appearing in the main loop, the Newton Step and
    the Bisection Step.
	We define the
	\dt{cost} of processing component $C$ to be
	the costs in doing the first 2 tests in \refeQ{cost},
	and
	the \dt{cost} of processing a box $B$ to be the cost of
	doing the last test.
	Note that the first 2 tests do not apply to the non-confined components (which appear in the preprocessing stage only), so there is no corresponding cost.

	We next ``charge'' the above costs to
	natural $\vareps$-clusters.  More precisely, if
	$X$ is a confined component or any box produced
	in the algorithm, we will charge its cost to
	a natural $\vareps$-cluster denoted $\phi(X)$:
	(a) For a special component $C$, let $\phi(C)$ be the
    natural $\vareps$-cluster $\calZ(C')$ where
	$C'$ is the confined
	leaf of $\compTree^*$ below $C$ which minimizes the
	length of path from $C$ to $C'$ in $\compTree^*$.
	%
	%
	(b) For a non-special component $C$,
	we define $\phi(C)$ to be equal to $\phi(C')$ where
	$C'$ is the first special component below $C$.
	%
	(c)
	For a box $B$, we had previously defined $\phi_0(B)$ (see Section 5).
	There are two possibilities:
	If $\phi_0(B)$ is defined as a special component, then
	$\phi(\phi_0(B))$ was already
	defined in (a) above, so we let $\phi(B)\as \phi(\phi_0(B))$.
	Otherwise, $\phi_0(B)$ is defined as
	a natural $\vareps$-cluster, and we let $\phi(B) = \phi_0(B)$.

	\bleml{phi}
		The map $\phi$ is well-defined.
	\elem

\bpf
For a special component $C$, to define $\phi(C)$ we first consider $C'$,
defined as the confined leaf such that path $(C\to\cdots\to C')$ is the
   shortest in $\compTree^*$.  This path has length at most $\log n$
   since there exists a path of length at most $\log n$ in which
   we choose the special node with the least $\#(C_i)$ at each branching
   (this was the path chosen in \cite{becker+3:cisolate:18}).
   Hence, $\phi(C)$ is well-defined. The map $\phi$ for a
   non-special component and a box are defined based on that for a special
   component, it is easy to check that they are well-defined.

   It remains to prove that in the case where $\phi_0(B)$ is a natural
   $\vareps$-cluster, the map $\phi$
   is well-defined.
   This follows from \refLem{phi_0}.
   \ignore{
   For this, we need to prove that $\phi_0(B)$ is a union of
   strong $\vareps$-cluster.
   That is to say: if a strong $\vareps$-cluster $D$ intersects $\phi_0(B)$,
   then $D\ib \phi_0(B)$.

   Note that both $\phi_0(B)$ and $D$ are natural clusters, thus either
   $D\ib \phi_0(B)$ or $\phi_0(B)\ib D$.
   For a constituent box $B$ of a non-confined component, by \refLem{BoxSize}(a) we have
   	$w_B\ge \frac{w(B_0)}{48n}$;
   and by definition of a strong $\varepsilon$-cluster $D$, we have
   $\rad(D)<\frac{\vareps}{48}$.
   From the definition of $\phi_0$ map,
   $\phi_0(B)$ is the largest natural cluster satisfying
   $\rad(\phi_0(B))\le w_B$.
   Therefore, if $\rad(D)\le w_B$, we can conclude
   that $\rad(D)\le\rad(\phi_0(B))$ and thus $D\ib\phi_0(B)$.

   It is easy to see that if $\vareps\le\frac{w(B_0)}{n}$ (by our assumption),
   then $\rad(D)\le\frac{w(B_0)}{n\cdot 48}<w_B$.
   We can further prove that $D\subset 2B_0$:
   since $D\ib\phi_0(B)$, we have $r<\rad(\phi_0(B))\le w_B$
   and by the remark in Section $4$, $w_B\le \frac{w(B_0)}{16}$,
   thus $r<\frac{w(B_0)}{16}$. It is also easy to see that
   for any box $B$, the distance from $2B$ to the boundary of $2B_0$
   is at least $\frac{w(B_0)}{16}$.
   }%
\epf

	Define $\wh{S}$ to be the range of $\phi$, so it is a set of natural
	$\vareps$- clusters.
	The clusters in $\wh{S}$ are of two types: those defined
	by the confined leaves of $\compTreeHat$, and those largest
	$\vareps$-clusters of the form $\phi(B)$ with $B$
	in non-confined components.
	%


    We use the notation $\wtO(1)$
to refer to a quantity that is $O((\LOG (n\tau\log(\vareps^{-1})))^i)$ for some
constant $i$.  To indicate the complexity parameters explicitly,
we could have written ``$\wtO_{n,\tau,\LOG(\vareps^{-1})}(1)$''.
	\bleml{ClustCharge}
	Each natural $\vareps$-cluster in $\wh{S}$ is
	charged $O(\smax\log n)$ times, i.e., $\wtO(1)$ times.
	\elem
\bpf
   First consider the number of components mapped to
   a same natural $\vareps$-cluster.
   From the definition of $\phi(C)$ for a special component,
   it is easy to see that the number of special components mapped to a same natural $\varepsilon$-cluster is at most $\log n$.
   Thus the number of non-special components mapped to
   a same natural $\varepsilon$-cluster is bounded by
      $O(\smax\log n)$.
   Hence the number of components mapped to a same
   natural $\vareps$-cluster is bounded by $O(\smax\log n)$.

   Then we consider the number of boxes mapped to a same
   natural $\vareps$-cluster.
   By \refLem{TotalBox}, the number of boxes charged to
   a same special component by $\phi_0$ is bounded by
   $O(s_{\max})$, and the number of special components mapped to
   a same natural $\vareps$-cluster is bounded by $O(\log n)$,
   thus the number of boxes mapped to a same
   natural $\varepsilon$-cluster is bounded
   by $O(\smax\log n)=\wtO(1)$.
   Also by \refLem{TotalBox}, the number of boxes charged to
   a same natural $\vareps$-cluster by $\phi_0$ is
   bounded by $O(\log n)\wtO(1)$.

   In summary, each natural $\vareps$-cluster is mapped
   $O(\smax\log n) =\wtO(1)$ times.
\epf

   Based on the charging map $\phi$, we can derive a bound for the cost of
   processing each component and box.

   \bleml{NoniceCostX}
    Denote $k=\#(2B_0)$.
   \ \\(a)
       Let $B$ be a box produced in the algorithm.
       The cost of processing $B$ is bounded by
           \beql{BoxCost}
               \wtO\left(n\cdot[\tau_F+n\LOG(B)
                     + k_D\cdot (\LOG(\vareps^{-1})+k)+T_D]\right)
           \eeql
       with $D=\phi(B)$, $k_D=\#(D)$ and
           \beql{T_D}
               T_D\as \LOG\prod_{z_j\notin D} |\xi_D-z_j|^{-n_j}.
           \eeql
       where $\xi_D$ is an arbitrary root contained in $D$.
   \ \\(b)
       Let $C$ be a component produced in the main-loop,
       and let $C_0$ be the last special component above $C$,
       then the cost of processing a component $C$
       is bounded by
       \beql{CompCost}
       \begin{aligned}
          \wtO\Big(n\cdot  [\tau_F+n\LOG(C)+n\LOG(w_{C_0})
               + k_D\cdot (\LOG(\vareps^{-1})+k)+T_D]\Big)
       \end{aligned}
       \eeql
       where $D$ is an arbitrary cluster contained in $C$, $k_D=\#(D)$ and $T_D$ is as defined in \refeQ{T_D}.
\elem
    A proof of \refLem{NoniceCostX} is found in Appendix C.

	We are almost ready to prove the theorems announced
	in Section 1.1.   Theorem A is easier to prove
	if we assume that
	the initial box $B_0$ is \dt{nice} in the following sense:
	  \beql{condition2}
	  \max\nolimits_{z\in 2B_0}\LOG(z) = O( \min\nolimits_{z\in 2B_0}\LOG(z)).
	  \eeql
    Here we only prove the case where the initial box is nice, and
    a complete proof of Theorem A is provided in the end of Appendix C.

	In the nice case, the following lemma bounds the cost of processing $X$ where $X$ is a
	box or a component.

	\bleml{costX}
	   If the initial box is nice, the
	   cost of processing $X$ (where $X$ is a box or a component)
	   is bounded by
	     \begin{equation*}\label{costTest}
	        \wtO\left(n\cdot L_D\right)
	     \end{equation*}
	     bit operations with $D=\phi(X)$ and
	   with $L_D$ defined in \refeQ{L_D}.
	   Moreover, an $L_D$-bit approximation of $F$ is required.
	\elem
  \bpf
     Note that if the initial box satisfies \refeQ{condition2},
   then it holds that
   $\LOG(B)=O(\LOG(\xi))$
   and
   $\LOG(C)= O(\LOG(\xi))$
   for any box $B$ and component $C$ and any root $\xi\in 2B_0$.
   And we know that $\phi(C)\subset C$.
  \epf

   Thus this Lemma is a direct result form \refLem{NoniceCostX}.
	Using this lemma, we could prove Theorem A of Section 1.1
	under the assumption that $B_0$ is nice.

   Before we prove the Theorem A in Section 1.1,
   we want to address a trivial case excluded by the statement in that theorem.  In Theorem A, we assumed that
  the number of roots $k$ in $2B_0$ is at least $1$.  If $k=0$,
   then the algorithm makes only one test, $\wtTG_0(\frac{5}{4}B_0)$.
   We want to bound the complexity of this test.
   Denoting the center of $B_0$ as $M_0$,
   the distance from $M_0$ to any root
   is at least $\frac{w(B_0)}{2}$.  Thus
  $|F(M_0)|>|\lcoeff(F)|\cdot(\frac{w(B_0)}{2})^n$.
   Thus by \cite[Lemma 7]{becker+3:cisolate:18},
   the cost of this $\wtTGk$ test is bounded by
        $\wtO\left(
       n\tau_F+n^2\LOG(B_0)+n\LOG(w(B_0)^{-1})
       \right)$.
   Now we return to the Theorem A in the introduction.

   	\bgenDIY{Theorem A}{
	  Let $S$ be the solution computed by our algorithm for
	  a normal instance $(F(z),B_0,\vareps)$.
	  Then there is an augmentation
        $\wh{S}=\set{D_i: i\in I}$ of $S$ such that
	  the bit complexity of the algorithm is
	  \begin{equation*}
		\wtO\Big(  n \sum\nolimits_{D\in \wh{S}} L_D\Big)
      \end{equation*}
      with
	     \begin{equation*}
	      \begin{aligned}
		      L_D = \wtO \Big( \tau_F
	                 +n\cdot\LOG(\xi_D)
	                 +k_D\cdot(\LOG(k+\varepsilon^{-1})) + \LOG (\prod\nolimits_{z_j\notin D}
                      |\xi_D-z_j|^{-n_j})\Big)
	      \end{aligned}
	   \end{equation*}
	   where
	   $k_D=\#(D)$, and $\xi_D$ is an arbitrary root in $D$.
       Moreover, an $L_D^*$-bit approximation of the coefficients of $F$
       is required with
          $L_D^*\as\max_{D\in\wh{S}}L_D$.
}

   The set $\wh{S}$ in this theorem is precisely the
range of our charge function $\phi$, as defined in the text.

   \blem
   If $B_0$ satisfies \refeQ{condition2}, then
   the Theorem A holds.
\elem

\bpf
   Recall that the number of components
   and that of boxes mapped to any
   natural $\varepsilon$-cluster is bounded by
        $\log n\cdot s_{\max}$.
   Thus from \refLem{costX},
   the cost of processing all the components and boxes
   mapped to a natural cluster $D\in \wh{S}$ is bounded by
        $\wtO(\log n\cdot\smax \cdot nL_{D})$.
   But $\log n\cdot s_{\max}$ is negligible in the sense of
   being $\wtO(1)$.  Thus the
   total cost of all the $\wtTG$ tests
   in the algorithm can be bounded by
   	$$\wtO\Big(n\sum\nolimits_{D\in \wh{S}}L_{D}\Big)$$
   with $L_{D}$ defined in \refeQ{L_D}
   and $\wh{S}$ is the range of $\phi$.
   And it is easy to see that
   $\wtO\Big(n\sum\nolimits_{D\in \wh{S}}L_{D}\Big)$
   is bounded by \refeQ{TotalCost}.

   There is another issue concerning total cost
   (as in \cite[Theorem 7]{becker+3:cisolate:18}):
   There is a non-constant complexity operation
   in the main loop: in each iteration,
   we check if $4\Delta_C\cap C'$ is empty.
   This cost is $O(n)$ since $C'$ has at most $9n$ boxes.
   This $O(n)$ is already bounded by the cost of the iteration,
   and so may be ignored.
\epf

	The appendix will prove Theorem A holds even if $B_0$ is not nice.

	In \cite{becker+3:cisolate:18},
	the complexity bound for global root isolation
	is reduced to the case where $B_0$ is centered at the origin.
	This requires a global pre-processing step.
	It is unclear that we can adapt that pre-processing to
	our local complexity analysis.

    The bit complexity in Theorem A is based on geometric parameters,
    we can also write it in terms of synthetic parameters, although
    the latter bound is not as sharper as the former one.

    \bgenDIY{\sc Corollary to Theorem A}{
\\   The bit complexity of the algorithm is bounded by
     $$\wtO\Big(n^2(\tau_F + k+m)
			    + nk\LOG(\vareps^{-1})
			    + n\LOG|\gdisc(F)|^{-1}
			    \Big).$$
   In case $F$ is an integer polynomial, this bound becomes
     $$\wtO\Big(n^2(\tau_F + k +m)
			    + nk\LOG(\vareps^{-1}) \Big).$$
}

    The proof is found in Appendices A and C.

    Theorem A gives a bit complexity bound in terms of $\wh{S}$.
We now investigate the natural $\vareps$-clusters
in $\wh{S}$.  From the definition of $\wh{S}$, we could write
	\beql{sss}
	\wh{S}=S\cup S'
	\eeql
where $S$ is the set of natural $\vareps$-clusters defined by the confined
leaves of $\compTreeHat$, and
$S'$ is the set of all the natural $\vareps$-cluster
$\phi(B)$ with $B$ being any constituent box of any non-confined
component in the preprocessing stage.
Now we want to show an intrinsic property of the output components
and also of the set $\wh{S}$,
using the concept of strong $\vareps$-clusters
as is defined in the introduction.
	%

    \bgenDIY{{Theorem B}}{
     Each natural $\vareps$-cluster in $\wh{S}$ is a union of strong $\vareps$-clusters.
}

The proof of Theorem B is found in Appendix C.



	\ignore{
	  The general case of an arbitrary $B_0$ can
	be reduced (via a pre-processing step)
	to root clustering of a collection of nice boxes as in
	\cite{becker+3:cisolate:18}.  The overall complexity is unchanged.
	%
	$B_0$. The analysis is analogue to that in the original paper
	\cite{becker+3:cisolate:18}, we adapt the analysis in
	\cite{becker+3:cisolate:18} for root isolation to root clustering.
	}%

\sect{Conclusion}
	This paper initiates the investigation of the local complexity
	of root clustering.  It modifies
	the basic analysis and techniques of
	\cite{becker+3:cisolate:18} to achieve this.
	Moreover, it solves a problem left open in
	\cite{becker+3:cisolate:18}, which is to show
	that our complexity bounds can be achieved without
	adding a preprocessing step to
	search for ``nice boxes'' containing roots.

	We mention some open problems.
	Our Theorem A expresses the complexity in terms of
	local geometric parameters; how tight is this?
	Another challenge is to extend our
	complexity analysis to analytic root clustering
	\citep{yap-sagraloff-sharma:cluster:13}.

	\ignore{
	This paper represents three directions in
	the general topic of ``exact root finding'':
	\\1. ``Root finding'' as a generic topic has a variety of
	subproblems.  Traditionally, in the domain of exact algorithms,
	root finding is viewed as two main subproblem:
	root isolation followed by root refinement (i.e., compute
	an arbitrarily good approximation of an isolated root).
	Alternatively, we can follow Sch\"onhage's view approach
	of root approximation (i.e., produce approximating disks
	for roots, but the disks need not be disjoint.
	Our clustering approach is an alternative to both these
	two views with many useful applications.
	\\2. Ever since Sch\"onhage's landmark, the main criteria for
	evaluating exact root finding algorithms has
	been its performance on the ``benchmark problem'',
	and so biases the
	This paper begins the task of
	providing complexity bounds for the local root finding problem.
	It led to the development of tools based on root geometry.
	\\
	3.  As the benchmark problem is based on integer polynomials,
	it gives rise to assumptions that are not generally
	meaningful.  For integer polynomials, it
	may be inferred that the inherent complexity of root finding
	reduces to the square-free case.   But this is both is
	bad idea in practice (the step of making polynomials square-free
	is a no-op for the typical input), very difficulty for general
	polynomials with algebraic coefficients, and impossible
	with general analytic polynomials.
	To overcome this, we introduce the oracular number model.
	Note that our oracular model is essentially the ``bit-stream model''.
	The bit-stream terminology is unfortunate because it suggests that
	numbers are essentially a unique infinite binary string.
	We know from results in computable analysis that
	this is not a robust model.
	}

\dt{Postscript}: Our clustering algorithm has been implemented in 
~\cite{imbach-pan-yap:ccluster:18}.
    
\end{s4}
\begin{bib}

\input{specialIssue.bbl}

\end{bib}
\begin{s5} 

\appendix
We have omitted the three appendices which may be found
in our full paper: Appendix A contain proofs for Section 2.
Similarly, Appendix B and C are for Sections 5 and 6.

\end{s5}
\begin{s6}
\appendix
\section{Root Bounds}
	To prove Theorem 2,
	we follow \cite{becker+3:cisolate:18}
	by proving three lemmas.
	We then use these bounds to convert the bound in our
	Theorem A into a bound in terms of algebraic parameters
	as in \refeQ{complexity} in Section 1.1.
\ssect{LEMMA A1}
		In the following,
	we will define $G(z)$ and $H(z)$
	relative to any $\Delta$ as follows:
		\beql{fgh2}
		F(z)=G(z) H(z)
		\eeql
	where $G(z)=\prod_{i=1}(z-z_i)$ such
	that $Z_F(\Delta)=\zero(G)=\set{z_1\dd z_k}$ and
	$\zero(H)=\set{z_{k+1}\dd z_n}$.
	Note that the leading coefficients of $F(z)$ and $H(z)$ are the same.
	By induction on $i$, we may verify that
		$$F^{(i)}(z) = \sum_{j=0}^i \binom{i}{j} G^{(i-j)}(z)H^{(j)}(z)$$
	and
		$$\frac{F^{(i)}(z)}{i!} = \sum_{J\in {[n]\choose n-i}}
			\prod_{j\in J}(z-z_j).$$

	\bgenDIY{{\sc Lemma A1}}{
	  	Let $\Delta=\Delta(m,r)$ and $\lambda=\lambda_0\as 4k(n-k)$.
		\\
		If $\#(\Delta)=\#(\lambda\cdot \Delta) =k\ge 0$ then
                for all $z\in \Delta$
		\[
                \Big|\frac{F^{(k)}(z)}{k! H(z)}\Big| > 0.
		\]
                For $z=m$, the lower bound can be improved to half.
	}
	\bpf
		Using the notation \refeQ{fgh2}, we see that
		  $$\frac{F^{(k)}(z)}{k! H(z)}
		  	= \sum_{J\in\binom{[n]}{n-k}}
				\frac{\prod_{j\in J} (z-z_j)}{\prod_{i=k+1}^n
				(z-z_i)}$$
		First suppose $\lambda_0=0$, i.e., $k=0$ or $k=n$.
		If $k=n$, then $H(z)$ is the constant polynomial $a_0$ where
		$a_0$ is the leading coefficient of $F(z)$, and
		clearly, $\frac{F^{(k)}(z)}{k! H(z)}=1$.
		If $k=0$, then $F(z)=H(z)$ and again
		$\frac{F^{(k)}(z)}{k! H(z)}=1$.  In either case the lemma is verified.

		Hence we next assume $\lambda_0>0$.
		We partition any $J\in\binom{[n]}{n-k}$ into
		$J' \as J\cap [k]$ and $J'' \as J\setminus [k]$.
		Then $j'\as |J'|$ ranges from $0$ to $\min(k,n-k)$.
		Also, $j'=0$ iff $J=\set{k+1\dd n}$.
		{\small
		\begin{align*}
		  \lefteqn{\frac{F^{(k)}(z)}{k! H(z)}
		  	= \sum_{J\in\binom{[n]}{n-k}}
				\frac{\prod_{j\in J} (z-z_j)}{\prod_{i=k+1}^n
			      (z-z_i)}}\\
			&= \sum_{j'=0}^{\min(k, n-k)}
				\sum_{J'\in\binom{[k]}{j'}}
				\sum_{J''\in\binom{[n]\setminus [k]}{n-k-j'}}
				\frac{\prod_{i'\in J'} (z-z_{i'})
					\prod_{i''\in J''}(z-z_{i''})}
				{\prod_{i=k+1}^n (z-z_i)}\\
			&= 1 + \sum_{j=1}^{\min(k, n-k)}
				\sum_{J'\in\binom{[k]}{j}}
				\sum_{J''\in\binom{[n]\setminus [k]}{n-k-j}}
				\frac{\prod_{i'\in J'} (z-z_{i'})
					\prod_{i''\in J''}(z-z_{i''})}
				{\prod_{i=k+1}^n (z-z_i)}
                   \end{align*}}
                   We next show that the absolute value of the summation on the RHS is at most $\frac{20}{21}$
                   which completes the proof. Since $z, z_{i'} \in \Delta$, and $z_{i''} \nin 4k(n-k)\Delta$
                   it follows that $|z-z_{i'}| \le 2r$ and $|z- z_{i''}| \ge 3k(n-k)r$. From these inequalities, we get
                   \small{
                   \begin{align*}
                     &\sum_{j=1}^{\min(k,n-k)}
                     \sum_{J'\in\binom{[k]}{j}}
                     \sum_{J''\in\binom{[n]\setminus [k]}{n-k-j}}
                     \frac{\prod_{i'\in J'} |z-z_{i'}|
					\prod_{i''\in J''}|z-z_{i''}|}
				{\prod_{i=k+1}^n |z-z_i|}\\
		  	&\le \sum_{j=1}^{\min(k,n-k)}
				\binom{k}{j}
				\binom{n-k}{n-k-j}
				\Big(\frac{2r}{3k(n-k) r}\Big)^{j} \\
		  	&\le \sum_{j=1}^{\min(k,n-k)}
				\frac{k^{j}}{j !}
				\binom{n-k}{j}
				\Big(\frac{2}{3k(n-k)}\Big)^{j}\\
		  	&< \sum_{j=1}^{k}
				\frac{1}{j !}
				\Big(\frac{2}{3}\Big)^{j} \\
			&= e^{2/3}-1 < \frac{20}{21}.
		\end{align*}
                For $z=m$, the term is upper bounded by $e^{1/4}-1 < \efrac{2}$.
	      }

	\epf

        Since for all $z\in \Delta$, $F^{(k)}(z) \neq 0$, we get the following:

	\bgenDIY{{\sc Corollary A1}}{
	  Let $\lambda = \lambda_0\as 4k(n-k)$.
	  If $\#(\Delta)=\#(\lambda \Delta)=k\ge 0$
	  then $F^{(k)}$ has no zeros in $\Delta$.
	}

\ssect{Lemma A2}
	\bgenDIY{{\sc Lemma A2}}{
	  	Let $\Delta=\Delta(m,r)$,
		$\lambda =4 k(n-k)$ and $c_1 = 7kK$.
		\\
		If $\#(\Delta)=\#(\lambda\Delta) =k$ then
		\[
		\sum_{i<k} \frac{| F^{(i)}(m)|}{| F^{(k)}(m) |} \frac{k!}{i!} (c_1r)^{i-k}
			< \frac{1}{2K}.
		\]
	}
	\bpf
		The result is trivial if $k=0$.
		We may assume that $k\ge 1$.	
		With the notation of 
		\refeQ{fgh2}, we may write
                $$
		  \frac{|G^{(i)}(m)|}{i!}
		  	 \le \sum_{J\in {[k]\choose k-i}} \prod_{j\in J}|m-z_j|
		  	\le \binom{k}{i} r^{k-i},$$
                 since $z_j \in \Delta$. Similarly, we obtain
                 $$
		  \left| \frac{H^{(i)}(m)}{i! H(m)} \right|
			\le
			  \sum_{J\in{[n]\setminus [k]\choose i}}
			  \prod_{j\in J} \frac{1}{|m-z_j|}
			 \le \binom{n-k}{i}\frac{1}{(\lambda r)^{i}}.$$
                From these two results, we derive that
		\begin{align*}
		  \Big|\frac{G^{(i-j)}(m)H^{(j)}(m)}{(i-j)!j! H(m)}\Big|
			&\le
			\binom{k}{i-j}r^{k-(i-j)}
			\cdot
			\binom{n-k}{j} \frac{1}{(\lambda r)^{j}}\\
			&=
			\binom{k}{i-j}\binom{n-k}{j}
			\cdot  \frac{r^{k-i}}{\lambda ^{j}}.\\
		\binom{i}{j} \Big|\frac{G^{(i-j)}(m)H^{(j)}(m)}{i! H(m)}\Big|
			&\le
			\binom{k}{i-j}\binom{n-k}{j}
			 \frac{r^{k-i}}{\lambda ^{j}}.
		\end{align*}
	Thus we get

		{\small
		\begin{align*}
		  \lefteqn{	\sum_{i=0}^{k-1} \frac{| F^{(i)}(m)|}{| F^{(k)}(m) |}
		  \frac{k!}{i!} (c_1r)^{i-k}}\\
			&\le
			\sum_{i=0}^{k-1} \sum_{j=0}^i
			\frac{\binom{i}{j}| G^{(i-j)}(m)H^{(j)}(m)|}{| F^{(k)}(m) |}
				\frac{k!}{i!} (c_1r)^{i-k}\\
			&\le
			\sum_{i=0}^{k-1} \sum_{j=0}^i
					\frac{|H(m)|}{| F^{(k)}(m) |}
					\binom{k}{i-j} \binom{n-k}{j}
				\frac{k! c_1^{i-k}}{\lambda ^j}\\
			&\le
			2 \sum_{i=0}^{k-1} \sum_{j=0}^i
					\binom{k}{k-i+j} \cdot \binom{n-k}{j}
					\frac{c_1^{i-k}}{\lambda ^j}
					\qquad (\text{by Lemma A1 for $z=m$})\\
			&\le
			2 \sum_{i=0}^{k-1} \sum_{j=0}^i
			\frac{(k^j)(k^{k-i})}{(k-i+j)!}\cdot  \frac{(n-k)^j}{j!}
					\frac{c_1^{i-k}}{(4k(n-k))^j}\\
			&=
			2 \sum_{i=0}^{k-1}
				\frac{k^{k-i} c_1^{i-k}}{(k-i)!}
					\sum_{j=0}^i
						\frac{1}{j! 4^j}\\
			&<
			2 \sum_{i=0}^{k-1}
				\frac{k^{k-i} c_1^{i-k}}{(k-i)!}
					e^{1/4}\\
			&<
			2 e^{1/4} \sum_{j=1}^{k}
				\frac{(k/c_1)^j}{j!} \\
			&<
				2 e^{1/4} (e^{1/7K}-1)\\
			&<
				2 e^{1/4} \frac{1}{7K-1}\\
			&\le
				2 e^{1/4} \frac{1}{6K}
			<
			\frac{1}{2K}
			.
		\end{align*}
		}

	\epf

\ssect{Lemma A3}
	\bgenDIY{{\sc Lemma A3}}{
	  Let $\lambda_1 = 3
	  K(n-k)\cdot\max\set{1,4k(n-k)}=3K(n-1)\cdot\max\set{1,\lambda_0}$.
	  \\
	  If $\#(\Delta)=\#(\lambda_1\cdot \Delta)=k\ge 0$ then
		\[
		\sum_{i=k+1}^{n} \left|
		\frac{ F^{(i)}(m) r^{i-k} k! } { F^{(k)}(m) i!} \right|
			< \frac{1}{2K}.
		\]
		where $\Delta=\Delta(m,r)$.
	  }

	\bpf
		First, assume $\lambda_0 = 4k(n-k)>0$ (i.e., $0<k<n$).
		Let $\zero(F^{(k)}) = \set{z^{(k)}_1\dd z^{(k)}_{n-k}}$ be the roots of $F^{(k)}$.
		Since
			$$\#(3K(n-k) \Delta)
			=\#(3K(n-k)\cdot \lambda_0\Delta),$$
		Corollary A1 implies that $F^{(k)}$ has no roots in $3K(n-k)\cdot\Delta$.
		Thus, $|m-z^{(k)}_j|\ge 3K(n-k)r$ and
		\begin{align*}
			\left| \frac{F^{(k+i)}(m)}{F^{(k)}(m)} \right|
			&\le i!
			\sum_{J\in {[n-k]\choose i}}
				\prod_{j\in J} \efrac{|m-z^{(k)}_{j}|} \\
			&\le
				\frac{i! \binom{n-k}{i}}{(3K(n-k)r)^i}\\
			&\le
				\frac{(n-k)^i}{(3K(n-k)r)^i}\\
			&\le
				\frac{1}{(3Kr)^i}.
		\end{align*}

		It follows that
		\begin{align*}
		  \lefteqn{\sum_{j=k+1}^{n}
				\left| \frac{ F^{(j)}(m) r^{j-k} k! }
			      { F^{(k)}(m) j!} \right|}\\
			&\le
			\sum_{i=1}^{n-k}
				\left| \frac{F^{(k+i)}(m)}{F^{(k)}(m)} \right|
				\frac{r^i}{i!}
			\qquad \Big(\text{ since } \frac{k!}{(k+i)!}\le \frac{1}{i!}\Big)\\
			&\le
				\sum_{i=1}^{n-k} \frac{1}{(3 Kr)^i}
				\frac{ r^i}{i!}\\
			&\le
				\sum_{i=1}^{n-k} \Big( \frac{1}{3K} \Big)^i \frac{1}{i!}\\
			&<
			e^{1/3K} - 1 < \frac{1}{3K-1}<\frac{1}{2K}.
		\end{align*}

		It remains to consider the case $k=0$ or $k=n$.
		The lemma is trivial for $k=n$. When $k=0$,
		we have $\lambda_1=3 Kn$ and the roots $z^{(k)}_j$ are the roots of $F$.
		Then $|m-z^{(k)}_j|\ge 3 K n r$ follows from our
		assumption that $\#(\lambda_1\Delta)=\#(\Delta)=0$.
		The preceding derivation remains valid.
	\epf

	\bgenDIY{{\sc Corollary A3}}{
	  Let $c_1\ge 1$.
	  If $\#(\Delta)=\#(c_1\lambda_1\cdot \Delta)=k\ge 0$ then
		\[
		\sum_{i=k+1}^{n} \left|
		\frac{ F^{(i)}(m) (c_1r)^{i-k} k! } { F^{(k)}(m) i!} \right|
			< \frac{1}{2K}.
		\]
		where $\Delta=\Delta(m,r)$.
	}

	\bpf
	Let $\Delta_1= c_1\Delta$.  Then $\#(\Delta_1)=\#(\lambda_1\Delta_1)=k$,
	and the previous lemma yields our conclusion (replacing $r$ by $c_1r$).
	\epf

\ssect{Theorem 2}
	\bgenDIY{{\sc Theorem 2}}{
		\label{new: tktest cor}
		Let $k$ be an integer with $0\le k\le n=\deg(F)$ and $K\ge 1$.
		\\Let $c_1=7kK$, and
			$\lambda_1=3K(n-k)\cdot \MAX\set{4k(n-k)}$.
		\\If
			$$\#(\Delta) = \#(c_1\lambda_1\Delta)=k,$$
		then
		$$T_k(c_1\Delta,K,F)\quad \text{holds.}$$
	}
	\bpf \ \\
	By definition, $T_k(c_1\Delta,K,F)$ holds iff
		$$\sum_{i\neq k} \frac{|F^{(i)}(m)(c_1r)^{i-k}
			k!}{|F^{(k)}(m)|} < \efrac{K}$$
	But the LHS is equal to $A+B$ where
		\begin{align*}
		  A: \qquad& \sum_{i>k} \frac{|F^{(i)}(m)(c_1r)^{i-k} k!}{|F^{(k)}(m)|}\\
		  B: \qquad& \sum_{i<k} \frac{|F^{(i)}(m)(c_1r)^{i-k} k!}{|F^{(k)}(m)|}
		\end{align*}
		By Corollary A3, $A$ is at most $\efrac{2K}$ and by Lemma A2,
		$B$ is at most $\efrac{2K}$.  This proves our theorem.
	\epf

\ssect{Bound on $T_D$ in the Theorem A}
	We will need the following result to derive the bound.
	
	\bgenDIY{{\sc Lemma A4}}{
        Let $g(x)$ be a complex polynomial of degree $n$ with distinct
	roots $\alpha_1 \dd \alpha_m$ where $\alpha_i$ has multiplicity
        $n_i$.  Thus $n=\sum_{i=1}^m n_i$.  Let $I \ib [m]$ and
	$\nu=\min\set{n_i: i\in I}$. Then
        $$\prod_{i\in I} |g_{n_i}(\alpha_i)|
		\ge |\gdisc(g)|
		\paren{\|g\|_\infty^{m} n^{n+1} {\rm Mea}(g)^{n+1-\nu}}^{-1},$$
        where
        $g_{n_i}(\alpha_i) \as g^{(n_i)}(\alpha_i)/n_i!$.
      }

      \bpf
        From the observation     that
		$$g_{n_i}(\alpha_i) = \lcoeff(g)\prod_{1 \le j \le m,\;j\neq i}
		(\alpha_i - \alpha_j)^{n_j},$$
         we obtain the following relation:
         $$
        \prod_{i=1}^m {g_{n_i}(\alpha_i)} = {\lcoeff(g)}^{m}\prod_{1 \le i< j\le m} \paren{\alpha_i - \alpha_j}^{n_i+n_j} = \gdisc(g).
        $$
        From this it follows that
        \beql{bound2}
        \prod_{i\in I} |g_{n_i}(\alpha_i)| = |\gdisc(g)|   \paren{\prod_{i\in [m]\setminus I} |g_{n_i}(\alpha_i)|}^{-1}.
        \eeql

        We next derive an upper bound on $|g_{n_i}(\alpha_i)|$.
        Let $g(x)=\sum_{j=0}^n b_jx^j$.
        By standard arguments we know that
        $$g_{n_i}(\alpha_i) = \sum_{j =n_i}^n {j \choose n_i} b_j \alpha_i^{j-n_i}.$$
        Taking the absolute value and applying triangular inequality, we get
        $$|g_{n_i}(\alpha_i)| \le \|g\|_\infty \sum_{j =n_i}^n {j \choose n_i} \max\set{1,|\alpha_i|}^{j-n_i}.$$
        Applying Cauchy-Schwarz inequality to the RHS we obtain
        $$|g_{n_i}(\alpha_i)| \le \|g\|_\infty \paren{\sum_{j =n_i}^n {j \choose n_i}^2}^{\half} \paren{\sum_{j= n_i}^n\max\set{1,|\alpha_i|}^{2(j-n_i)}}^{\half}.$$
        The second term in brackets on the RHS is smaller than $\max\set{1, |\alpha_i|}^{n-n_i+1}$, and the first
        is bounded by $\sum_{j=n_i}^n{j \choose n_i} = {n+n_i+1 \choose n}\le n^{n_i+1}$. Thus we obtain
        $$|g_{n_i}(\alpha_i)| \le \|g\|_\infty n^{n_i+1} \max\set{1, |\alpha_i|}^{n-n_i+1}.$$
        Taking the product over all $i \in [m] \setminus I$, we get that
        $$\prod_{i \in [k]\setminus I}|g_{n_i}(\alpha_i)| \le \|g\|_\infty^{m} n^{n+1} \rm{Mea}(g)^{n+1-\min_{i \in I}n_i}.$$
        Substituting this upper bound in \refeQ{bound2} yields us the desired bound.
        \epf

        Let $I \ib [m]$.
        We next derive an upper bound on $\sum_{D\in\wh{S}} T_D$, where
                $$T_D=\LOG\prod_{z_j\nin D}|\xi_i-z_j|^{-n_j},$$
        here $\xi_i$ is a representative root in the natural $\vareps$-cluster $D$.
        In this section, we use the convenient shorthand $\xi_D$ to denote the representative for cluster $D$,
        and $k_D$ the number of roots in $D$.
        Moreover, we choose the representative
        $\xi_D$ as a root that has the smallest absolute value among all roots in $D$.
	Let $\calD$ denote a set of disjoint {\em natural} $\vareps$-clusters
    of $F$ such that the union of these clusters
    contains all the roots of $F$.
	%
	Define $F_\vareps$ as the polynomial obtained by replacing
	each natural $\vareps$-cluster $D$ of $F$
        by its representative $\xi_D$ with multiplicity $k_D$, i.e.,
                $$F_\vareps(z) \as \lcoeff(F)
		{\prod_{D\in \calD} (z-\xi_D)^{k_D} }
		$$
	More importantly, the choice of the representative ensures that the
	Mahler measure does not increase,
        i.e., $\rm{Mea}(F_\vareps) \le \rm{Mea}(F)$.
        Since $\xi_D$ is a root of multiplicity $k_D$,  
		it can be verified that
                $$\frac{F_{\vareps}^{(k_D)}(\xi_D)}{k_D!}
		= \lcoeff(F) \prod_{D'\in \calD, D'\neq D}(\xi_D -
		\xi_{D'})^{k_{D'}}.$$
	We first relate the product $\prod_{z \nin D}|\xi_D - z_j|^{n_j}$
	appearing in $T_i$ with the term on the RHS above.
	The two are not the same, since we have replaced all natural
	$\vareps$-clusters with their representative,
	and hence for another cluster $D'$ the distance $|\xi_D - z_j|$,
	for $z_j\in D'$,
        is not the same as $|\xi_D-\xi_{D'}|$.
        Nevertheless, for an isolator $\Delta'$ of $D'$, we have
                $$2\min_{w\in \Delta'}|\xi_D- w| \ge \max_{w\in \Delta'}|\xi_D-w|$$
        and hence
		$$|\xi_D - z_j| \ge \frac{|\xi_D - \xi_{D'}|}{2}.$$
	From this inequality, we obtain that
                $$\prod_{z_j \nin D}|\xi_D - z_j|^{n_j}
			\ge 2^{-n} \frac{\abs{F_\vareps^{(k_D)}(\xi_D)}}{k_D!}.$$
	So to derive an upper bound on $\sum_{D\in\wh{S}} T_D$, it suffices to
	derive a lower bound on
	$\prod_{D\in\wh{S}} |F_{\vareps}^{(k)}(\xi_{D})|/k!$.
        Applying the bound in Lemma A4 above to $F_\vareps$, along with the observations that
        $\|F_\vareps\|_\infty \le 2^n \rm{Mea}(F_\vareps)$,
	and $\rm{Mea}(F_\vareps) \le \rm{Mea}(F)$,
        we get the following result:

	\bgenDIY{{\sc Theorem A5}}{
                $$\sum_{D \in \wh{S}} T_D=
                \wtO(\LOG|\gdisc(F_\vareps)|^{-1}+ nm+ n \LOG \rm{Mea}(F)).$$
	      }
         Note, however, that
                $$|\gdisc(F_\vareps)| > \frac{|\gdisc(F)|}{\vareps^{\sum_{D \in \wh{S}}k_{D}^2}}.$$
         If we assume that $\vareps < 1$, i.e., $|\gdisc(F_\vareps)|$ is larger than $|\gdisc(F)|$,
         then the term $(\sum_{D\in \wh{S}}k_{D}^2) \log \vareps < 0$ and so we can
         replace $|\gdisc(F_\vareps)|^{-1}$ by $|\gdisc(F)|^{-1}$ in Theorem A5 to obtain a larger bound.
         Moreover, if $F$ is an integer polynomial, not necessarily square-free,
         from \cite[p.~52]{mehlhorn-sagraloff-wang:15} we know that $\LOG|\gdisc(F)|^{-1}=O(n\tau_F+n \log n)$.
        Hence we obtain the following bound (using Landau's inequality ${\rm{Mea}}(F) \le \|F\|_2 \le n2^{\tau_F}$):

	\bgenDIY{{\sc Corollary A6}}{
          Let $\{D_i; i \in I\ib [m]\}$ be any set of disjoint nature $\vareps$-clusters of an integer
          polynomial $F$ with $m$ distinct roots. Then
                $$\sum_{i \in I} T_{D_i}=  \wtO(n\tau_F+nm).$$
         }
\end{s6}
\begin{s7}
\section{Bound on Number of Boxes}
	\ignore{
	    In Section 3, properties (C4) and (C8) refers to
	    the following fact about confined components:
	
	\blemBl{cplus}
	If $C$ is confined, then $\#(C)=\#(C^+)$.
	\elemB
	\bpf
	Since $C$ is confined, the separation of $C$ from
	$\partial((5/4)B_0)$ is at least $w_C$.
	Suppose $z$ is a root in $C^+\setminus C$.
	That means that $z\in B^+$ for some constituent box $B$ in $C$.
	Thus $\Sep(C,z)\le w_C/2$.
	This proves that $z$ is in $(5/4)B_0$.
	Since $z\in (5/4)B_0$, there is a box $B$ containing $z$
	that is in some component $C'$ in $Q_0\cup Q_1\cup\qdis$.
	But $\Sep(C,C')\ge w_C$.  This is a contradiction.
	\epf
	}%
\ssect{Lemma 8}
	\bgenDIY{{\sc Lemma 8}}{
	  Denote $k=\#(2B_0)$.
	  \ \\(a)
	  If $C$ is a component in the pre-processing stage, then
	  $w_C\ge \frac{w(B_0)}{48k}$.
	  \ \\(b)
	  Suppose $C_1\to \cdots\to C_2$ is a non-special path
	  with $ W_{C_1}<\vareps$.
	  Then it holds
	  $$ \frac{w_{C_1}}{w_{C_2}}< 57k.$$
	  \ \\(c)
	  Let $C$ be a confined leaf in $\compTreeHat$
	  then
	       $$w_C>\frac{\vareps}{2}\Big(\efrac{114k}\Big)^{k}.$$
	}
	
	\bpf
	  (a) By way of contradiction, assume $w_C < \frac{w(B_0)}{48k}$.
	  Then the parent component $C'$ satisfies
		$w_{C'}<\frac{w(B_0)}{24k}$ since
		$C$ is obtained from $C'$ in a Bisection Step.
		Then $W_{C'}\le 3k w_{C'} < \frac{w(B_0)}{8}$.
		Thus $C'\cap B_0$ is empty or $C'$ is confined.
		In either case, we would not bisect $C'$ in the pre-processing stage,
		contradicting the existence of $C$.
	
	  (b)
	  In this proof and in the proof of part (c) of this Lemma,
	  we write $w_i, R_i, N_i$, etc,
		  instead of $w_{C_i}, R_{C_i}, N_{C_i}$, etc.
	  By way of contradiction, assume that
	     $\frac{w_1}{w_2}\ge57k$.
	  Since $W_1\le \vareps$, from the algorithm, we know that
	  each step in the path $C_1\to \cdots \to C_2$ is a Bisection step.
	  Thus there exists a component $C'$ such that
	       $3k\cdot w_{2}< w_{C'}\le 6k\cdot w_{2}$.
	  The following argument shows that $C'$ is a leaf
	  of $\compTreeHat$.
	  By \refLem{CompBasic}(a), we have
	       $W_{2}\le 3kw_{2}$,
	  thus $W_{2}< w_{C'}$.
	  Thus the roots in $C'$
	  are contained in a square of width less than $w_{C'}$.
	  By \refLem{CompBasic}(b), we conclude that $C'$ is compact.
	  To show that $C'$ is a leaf, it remains to
	  show that $4\Delta_{C'}$ has no intersection with
	  other components.
	  We have
	       $4R_{C'}=4\cdot \frac{3}{4}W_{C'}\le 9w_{C'}
	       $.
	  Meanwhile, since $C'$ is compact, it is easy to see that
	  the distance from the center of $\Delta_{C'}$ to $C'$ is at most
	  $\efrac{2}w_{C'}$.
	  Thus the separation between $C'$ and any point in $4\Delta_{C'}$
	  is less
	  than $9w_{C'}+\efrac{2}w_{C'}=\frac{19}{2}w_{C'}$
	       $\le \frac{19}{2} \cdot 6k\cdot w_2\le
	       		\frac{19}{2}\cdot 6k\cdot\frac{w_{1}}{57k} =w_1$.
	  By Property (C3) in Section 3,
	  we know that $C'$ is separated from other components by
	  at least $w_1$,
	  thus $4\Delta_{C'}$ has no intersection with
	  other components.
	  We can conclude that $C'$ is a leaf of $\compTreeHat$.
	  Contradiction.
	
	  (c)
	  Let $C_0$ be the first component above $C$ such that
	  $w_0<\vareps$.
	  From the algorithm, we have $w_0\ge \frac{\vareps}{2}$.
	  Consider the path $P=C_0\to \cdots\to C$.
	  There exists a consecutive sequence of special components
	  below $C_0$, denoted as $\{C_1\dd C_t\}$ with $C_t=C$.
	  Split $P$ into a concatenation $P=P_0; P_1;\cdots; P_{t-1}$
	  of $t$ subpaths where subpath
	  	$P_i= (C_i\to \cdots C_{i+1})$ for $i\in\{0\dd t-1\}$.
	  Let $C_i'$ be the parent of $C_i$ in $\compTreeHat$ for
	  $i\in\{1\dd t\}$.
	  Consider the subpath of $P_i$ where we drop the last special configuration: $(C_i\to\cdots\to C'_{i+1}).$
	  By part (b) of this lemma, we have
	  	$$\frac{w_{C_i}}{w_{C'_{i+1}}} < 57k$$
	  for $i\in\{0\dd t-1\}$.
	  The step $C_{i+1}'\to C_{i+1}$ is evidently a Bisection step and so
	  	$$\frac{w_{i}}{w_{i+1}} < 114k.$$
	  Hence
	      $\frac{w_{0}}{{w_t}}< (114k)^{k}$.
	  It follows
	      $w_C>\frac{\vareps}{2}(\efrac{114k})^{k}$.
	\epf
\ssect{Lemma 9}
	Before proving \refThm{s_max}, we state the following lemma,
	which is an adaptation of
	\cite[Lemma 8]{becker+3:cisolate:18}, giving
	a sufficient condition for the success of the Newton step.
	
	\vspace{3pt}
	\blemBl{SucNew}
	  Let $C$ be a confined component with $W_C\ge \vareps$.
	  Then $\newton(C)$ succeeds provided that
	    \benum[(i)]
	      \item
		$\#(\Delta_C) =\#((2^{20}\cdot n^2\cdot N_C)\cdot\Delta_C)$.
	      \item
		$\rad(\calZ(C))\le(2^{20}\cdot n)^{-1}\cdot \frac{R_C}{N_C}$.
	    \eenum
	\elemB

	
	\bgenDIY{{\sc Theorem 9}}{
	The length of the non-special path \refeQ{P} satisfies
	    $$s=O(\log\log\frac{w_{1}}{w_{s}}+\log n).$$
	 Particularly,
		   $$
	\smax =O\Big(\log n + \log\log\frac{w(B_0)}{\varepsilon}\Big).$$
	}
	\bpf
	   From \refLem{BoxSize}(a), we can see that the length of path in the
	   preprocessing stage is bounded by $O(\log n)$.
	   From \refLem{BoxSize}(b), the length of non-special path is bounded
	   by $O(\log n)$ if the width of components is smaller than
	   $\vareps$.
	   Hence it remains to bound the length of
	   non-special path in the main loop
	   such that any component $C$ in the path satisfies $W_C\ge \vareps$.
	   Lemma B2 gives us the sufficient conditions to perform Newton step
	   in this path.
	
	   As in \cite{becker+3:cisolate:18}, the basic idea
	   is to divide the path $P=(C_1\to\cdots\to C_s)$ (using the notation
	   of \refeQ{P}) into 2 subpaths
	   $P_1=(C_1\to\cdots\to C_{i_1})$ and
	   $P_2=(C_{i_1}\to\cdots\to C_s)$ such that
	   the performance of the Newton steps in $P_2$
	   can be controlled by \refLem{SucNew}.
	   This lemma has two requirements ((i) and (ii)):
	   we show that the components in $P_2$ automatically satisfies requirement (i).
	   Thus if component $C_i$ in $P_2$ satisfies requirement (ii), we
	   know that $C_i\to C_{i+1}$ is a Newton step.   This allows us to
	   bound the length of $P_2$ using the Abbot-Sagraloff Lemma
	   \cite[Lemma 9]{becker+3:cisolate:18}.
	
	   We write $w_i, R_i, N_i$, etc,
		  instead of $w_{C_i}, R_{C_i}, N_{C_i}$, etc.
	
	   Define $i_1$ as to be the first index satisfying
	   $N_{i_1}\cdot w_{i_1}< 2^{-24}\cdot n^{-3}\cdot w_{1}$.
	   If no such index exists, take $i_1$ as $s$.
	
	   First we show that the length of $P_1$ is $O(\log n)$.
	   Note that $N_i\cdot w_{i}$ decreases by a factor of at least 2 in each step
       (see \cite{becker+3:cisolate:18}).
	   There are two cases: if step $C_i\to C_{i+1}$ is a Bisection
	   step, $w_{i+2}=w_i/2$ and $N_i$ does not increase; if it is a Newton
	   step, then $w_{i+1}=\frac{w_{i}}{2N_i}$ and $N_{i+1}=N_i^2$, so
	   $N_{i+1}\cdot w_{i+1}=N_i^2\cdot\frac{w_i}{2N_i}=\frac{1}{2}\cdot N_i\cdot w_{i}$.
	   It follows that at most
	   $\log(2^{24}\cdot n^3)$ steps are performed to reach an $i'$ such that
	   $N_{i'}\cdot w_{i'}\le 2^{-24}\cdot n^{-3}\cdot N_1\cdot w_{1}$.
	   This proves $i'\le 1+\log(2^{24}\cdot n^3)$.
	   Since $C_1$ is a special component, our algorithm reset
	   $N_1=4$ (cf.~proof of \refLem{CompBasic}).  So it
	   takes 2 further steps from $i'$ to satisfy the
	   condition of $i_1$.  Thus $i_1\le 3+\log(2^{24}\cdot n^3)=O(\log n)$.
	   Note that this bound holds automatically if $i_1=s$.
	
	   We now show that requirement (i) of Lemma B2 is satisfied in $P_2$:
	   from the definition of $i_1$,
	   for any $i\ge i_1$, $2^{20}\cdot n^2 \cdot N_{i}\cdot r_{i}\le 2^{20}\cdot
	   n^2\cdot N_{i}\cdot \frac{3}{4}\cdot9n\cdot w_{i}<w_{1}$, and the separation
	   of $C_1$ from any other component is at least $w_{1}$, so
	   $(2^{20}\cdot n^2\cdot N_i)\cdot\Delta_i$
	   contains only the roots in $\calZ(C_1)$, fulfilling requirement (i).
	
	   Next consider the path $P_2$.
	   Each step either takes a bisection step or a Newton step.
	   However, it is guaranteed to take the Newton step if requirement (ii)
	   holds (note that it may take a Newton step even if requirement (ii) fails).
	   Let $\#(\Delta_{s})=k$.
	   If component $C_i$ satisfies
		   \beql{fracri}
		   \frac{R_{i}}{N_i}\ge2^{20}\cdot n\cdot R_{s},
		   \eeql
	   the requirement (ii) is satisfied.  But
	   	$R_{s}<\frac{3}{4}\cdot 9n \cdot w_{s}<2^4 \cdot n\cdot w_{s}$
	   and
	   	$R_{i}\ge w_{i}$
	   so if
	   	\beql{fracwi}
		\frac{w_{i}}{N_{i}}\ge (2^{20}\cdot n)\cdot
			(2^4 \cdot n)\cdot w_{s}=2^{24}\cdot n^2\cdot w_{s}
		\eeql
	   holds, it would imply \refeQ{fracri}.
	   On the other hand, \refeQ{fracwi} is precisely the
	   requirement that allows us to invoke
	   \cite[Lemma 9]{becker+3:cisolate:18}.
	   Applying that lemma bounds the length of $P_2$ by
	   \\ $A\as (\log\log N_{i_1}+2\log\log(w_{i_1}\cdot (2^{24}\cdot n^2)^{-1}\cdot
	   \frac{1}{w_{C_s}})+2)+(2\log n +24)$.
	   Since $N_{i_1}\le \frac{w_{i_1}}{w_{s}}$, we conclude that
	   $A =O(\log\log\frac{w_{i_1}}{w_{s}}+\log n)$.
		This concludes our proof.
	
	    The second part of this theorem  is a direct result from the first part.
	\epf
	
	%
	%
	\ignore{
	We say that a component $C$ has \dt{small root radius}
	if $r_C<3w_C$; otherwise it has \dt{big root radius}.
	It is easy to see that if $C$ has small root radius, then it
	has at most $64$ constituent boxes.
	}
\ssect{Lemma 11}
	We first prove two lemmas that is useful for later proof.
	
	\blemBl{C_1C_2}
	    Let $C_1$ be the parent of $C_2$ in $\compTree^*$,
	    then
	      $$r_{C_1} \le 3\sqrt2 n\cdot w_{C_2} $$
	\elemB
	\bpf
	    Suppose $C_2'$ is the parent of $C_2$ in the component tree
	    $\compTreeHat$.
	    Then all the roots in $C_1$ remain in $C_2'$, meaning that
	    $r_{C_2'}=r_{C_1}$.
	    It is easy to see that the step $C_2' \to C_2$ is a Bisection Step,
	    thus $w_{C_2'}=2w_{C_2}$.
	    By \refLem{CompBasic}(a), we have $W_{C_2'}\le 3n\cdot w_{C_2'}=6n\cdot w_{C_2}$.
	    It follows $r_{C_2'}\le \frac{1}{2}\cdot \sqrt2W_{C_2'}\le 3\sqrt2n\cdot w_{C_2}$.
	    Hence $r_{C_1}=r_{C_2'}\le 3\sqrt2n\cdot w_{C_2}$.
	\epf

	\blemBl{in2B_0}
	\ \\(a)
	    For any box $B$ produced in the preprocessing stage,
	    if $\phi_0(B)$ is a natural $\vareps$-cluster,
	    then we have $w_B\ge 2\cdot\rad(\phi_0(B))$.
	(b)
	For any $B\neq \frac{5}{4}B_0$ produced in the algorithm, $\phi_0(B)\ib 2B_0$.
	\elemB
	
	\bpf
	  (a)
	  \beqarrys
	  w_B &\ge& \frac{w(B_0)}{48n}
	                    & \textrm{(by \refLem{BoxSize}(a))}\\
		  &\ge& 2\cdot\vareps
	                    & \textrm{(by small $\vareps$ assumption)}\\
		  &\ge& 2\cdot\rad(\phi_0(B))
	                   &\textrm{(by definition of $\vareps$-cluster)}
	  \nonumber
	  \eeqarrys
	
	  (b)
	  If $\phi_0(B)$ is a special component, it is easy to see that
	  $\phi_0(B)\ib 2B_0$.
	
	  We now discuss the case where $\phi_0(B)$ is a
	  natural $\vareps$-cluster.
	  To show that $\phi_0(B)\ib 2B_0$, note that since
	  $B$ is a proper subbox of $\frac{5}{4}B_0$, it follows that
	  $2B\ib \frac{15}{8}B_0$.
	  Thus there is a gap of $\frac{w(B_0)}{16}$ between the boundaries
	  of $2B_0$ and $\frac{15}{8}B_0$.
	  Since $\phi_0(B)$ is a $\vareps$-cluster, thus
	     $\rad(\phi_0(B))< \vareps
	     \le \frac{w(B_0)}{96n}$,
	  and $\phi_0(B)\cap 2B$ is non-empty, we conclude that
	  $\phi_0(B)$ is properly contained in $2B_0$.
	\epf

	\bgenDIY{{\sc Lemma 11}}{
	   The total number of boxes in all the components in $\compTreeHat$ is
	   	$$O(t\cdot s_{\max}) = O(\#(2B_0)\cdot s_{\max})$$
	   with $t=|\{\phi_0(B):B $ is any box in $\compTreeHat\}|$.
	}
	
	\bpf
	By the discussion above, we charge each box $B$ to $\phi_0(B)$
	which can be a special component or a cluster.
	
	First consider the case where $\phi_0(B)$ is special component. Note
	   that $\frac{1}{3}r_{\phi_0(B)}<w_B$.  We claim that
	   the number of boxes congruent with
	   $B$ that are charged to $\phi_0(B)$ is at most 64:
	   to see this, note that $2B\cap \calZ(\phi_0(B))$.
	   If $\Delta$ is the minimum disc containing $\calZ(\phi_0(B))$,
	   then $2B$ must intersect $\Delta$.  By some simple calculations,
	   we see that at most $64$ aligned boxes congruent to $B$ can
	   be charged to $\phi_0(B)$.
	
	   We now analyze the number of different sizes of the boxes that
	   are
	   charged to the same special component $C$.
	
	    Denote the parent of $C$ in the special component tree $\compTree^*$
	    as $C'$.
	    Let $B$ be a box such that
	       $\phi_0(B)=C$
	    and suppose $B$ is the constituent boxes of the component $C_B$,
	    evidently, $w_B=w_{C_B}$.
	    From the definition of $\phi_0$,
	    $B$ satisfies one of the two following conditions:
	    (i) $C_B$ is an component in the path $C'\to\cdots\to C$
	        and $w_B>\efrac{3}r_C$;
	    (ii) $C_B$ is a component above $C'$
	        and $\efrac{3}r_{C'}\ge w_B > \efrac{3}r_C$.
	    It is easy to see that there number of components
	    $C_B$ satisfying condition (i) is bounded by $\smax$
	    from \refThm{s_max}.
	    It remains to count the number of components $C_B$
	    that satisfy condition(ii).
	    By \refLem{C_1C_2}, we have
	    $r_{C'}\le 3\sqrt2n\cdot w_C$.
	    Since $B$ is charged to $C$ but not $C'$,
	    we have
	         $w_B\le \frac{1}{3}\cdot r_{C'}
	             \le \sqrt2n\cdot w_C$.
	    The box $B$ is constitute an ancestor of $C$, thus $w_C\le w_B$.
	    Therefore, we have
	         $w_C\le w_B \le \sqrt2 n\cdot w_C$,
	    and note that $w_B$
	    decreases by a factor of at least 2 at each step,
	    so $w_B$ may take $\log(\sqrt2n)$ different values.
	    Hence, the number of boxes
	    charged to each special component is bounded by $64s_{\max}$.
	\ignore{
	    The special component $C'$ could be a component
	    with small root radius or a component with big root radius.
	    Consider the case where $C'$ has small root radius,
	    which means $w_{C'}>\frac{1}{3}r_{C'}$,
	    then the constituent boxes of $C'$ is charged
	    to $C'$ itself.
	    Thus only the constituent boxes of the components between $C'$ and $C$ are charged to $C$.
	    By \refThm{s_max}, the length of this sequence is bounded by
	    $s_{\max}$.
	    Now consider the case where $C'$ is a component with big root radius, thus \refLem{C_1C_2} applies,
	    and it gives $r_{C'}\le 3\sqrt2n\cdot w_C$.
	}
	
	    Now consider the case where a box is charged to
	    a natural $\vareps$-cluster, this case only happens
	    in preprocessing step where the number of steps is
	    bounded by $O(\log n)$.
	    On the other hand, by \refLem{in2B_0}(a), we have $2\rad(\phi_0(B))\le w_B$
	    if $\phi_0(B)$ is a $\vareps$-cluster.
	    Thus the number of boxes of the same size charged to a
	    natural $\vareps$-cluster by $\phi_0$ is at most $9$.
	    Therefore, the number of boxes charged to a
	    natural $\vareps$-cluster by $\phi_0$ is bounded by $O(\log n)$.
	
	    Thus we can
	    conclude that the total number of boxes is bounded by
	    $O(t\cdot \smax)$ with $t=|\{\phi_0(B):B $ is any box in $\compTreeHat\}|$.
	%
	\epf
\end{s7}
\begin{s8}
\section{Bit Complexity}
\ssect{Lemma 14}
	We first prove two lemmas for later use.
	\blemCl{maxF}
		Let $\Delta=\Delta(m,R)$ and $\wh{\Delta}\as K \Delta$ for some $K\ge 1$.
		Let $D$ be any subset of $\calZ(\wh{\Delta})$ and $\zeta\in D$.
		If $\wh{\mu}=\#(\wh{\Delta})$ and $k_D=\#(D)$ then
	{\small
	  	$$
	         \max_{z\in\Delta}|F(z)|> R^{k_D}
		    \cdot n^{-\wh{\mu}}
		    \cdot K^{-\wh{\mu}+k_D}
			\cdot 2^{-3n+1}
			\cdot \prod_{z_j\notin D}|\zeta-z_j|^{n_j}.
			$$
	}where $z_j$ ranges over all the roots of $F$ outside $D$
	   and $\#(z_j)=n_j$.
	\elemC
	
	\bpf
	   Let $\{z_1, z_2,\ldots, z_r\}$ be the set of all the distinct roots of $F$.
	   Wlog, assume that $\zeta$ appearing in the lemma is $z_1$.
	   There exists a point $p\in\Delta(m,\frac{R}{2})$ such that the distance from
	   $p$ to any root of $F$ is at least $\frac{R}{2n}$, this is because the union
	   of all discs $\Delta(z_i,\frac{R}{2n})$ covers an area of at most $n\cdot
	   \pi(\frac{R}{2n})^2=\pi\frac{R^2}{4n}<\pi(\frac{R}{2})^2$.
	   Then for a root $z_i\in\wh{\Delta}$, it holds $\frac{|p-z_i|}{|z_{1}-z_i|}\ge \frac{R/(2n)}{2KR}=\frac{1}{4nK}$, and for a root $z_j\notin\wh{\Delta}$, it holds $\frac{|p-z_j|}{|z_{1}-z_j|}\ge \frac{|p-z_j|}{|p-z_j|+|p-z_{1}|}= \frac{1}{1+\frac{|p-z_{1}|}{|p-z_j|}} \ge \frac{1}{1+\frac{2KR}{KR-R/2}}=\frac{1}{5}$. Note that $|F(p)|=\lcoeff(F)\cdot \prod\nolimits_{i=1}^r|p-z_i|^{n_i}$, it follows
	  {\small
	     \begin{align*}
	     & \frac{|F(p)|}{\prod_{z_j\notin D}|z_{1}-z_j|^{n_j}}\\
	     &=  \lcoeff(F) \prod_{z_i\in D}|p-z_i|^{n_i}
	          \prod_{z_j\in \wh{\Delta}
		   , z_j\notin D}\left|\frac{p-z_j}{z_{1}-z_j}\right|^{n_j}
	          \prod_{z_k\notin\wh{\Delta}}
	         \left|\frac{p-z_k}{z_{1}-z_k}\right|^{n_k}\\
	     &\ge \frac{1}{4}\cdot\left(\frac{R}{2n}\right)^{k_D}
		 \cdot \left(\frac{1}{4nK}\right)^{\wh{\mu}-k_D}\cdot \left(\frac{1}{5}\right)^{n-\wh{\mu}}\\
		 &> R^{{k_D}}\cdot n^{-\wh{\mu}}\cdot K^{-\wh{\mu}+k_D}\cdot
	           2^{-3n-1},
		  \end{align*}
	    }
	   which proves the Lemma.
	\epf
	
	\blemCl{in14B}
	    For any box $B$, $\phi(B)$ is contained in $14B$.
	\elemC
	
	\bpf
	    Consider $\phi_0(B)$.  If $\phi_0(B)$ is a cluster, then
	    $2B$ intersects $\phi_0(B)$, and $2\rad(\phi_0(B))\le w_B$ (\refLem{in2B_0}(a)).
	    Thus $\phi_0(B)\ib 4B$.
	
	    Next suppose $\phi_0(B)$ is a special component.
	    Then $w_B>\efrac{3}r_C$ where $r_C=\rad(\calZ(C))$.
	    Since $2B\cap \calZ(C)$ is non-empty, we conclude that
	       $\calZ(C)\ib 14B$.
	\epf
	
	Now we derive a bound for the cost of
	processing each component and box.
	
	\bgenDIY{{\sc Lemma 14}}{
	    Denote $k=\#(2B_0)$.
	   \ \\(a)
	       Let $B$ be a box produced in the algorithm.
	       The cost of processing $B$ is bounded by
	           \beql{BoxCost}
	               \wtO\left(n\cdot[\tau_F+n\LOG(B)
	                     + k_D\cdot (\LOG(\vareps^{-1})+k)+T_D]\right)
	           \eeql
	       with $D=\phi(B)$, $k_D=\#(D)$ and
	           \beql{T_D}
	               T_D\as \LOG\prod_{z_j\notin D} |\xi_D-z_j|^{-n_j}.
	           \eeql
	       where $\xi_D$ is an arbitrary root contained in $D$.
	   \ \\(b)
	       Let $C$ be a component produced in the main-loop,
	       and let $C_0$ be the last special component above $C$,
	       then the cost of processing a component $C$
	       is bounded by
	       \beql{CompCost}
	       \begin{aligned}
	          \wtO\Big(n\cdot & [\tau_F+n\LOG(C)+n\LOG(w_{C_0})\Big.\\
	                & \Big.+ k_D\cdot (\LOG(\vareps^{-1})+k)+T_D]\Big)
	       \end{aligned}
	       \eeql
	       where $D$ is an arbitrary cluster contained in $C$, $k_D=\#(D)$ and $T_D$ is as defined in \refeQ{T_D}.
	}
	
	\bpf
	     (a) According to \cite[Lemma 7]{becker+3:cisolate:18}:
	    the cost for carrying out a $\wtTG(\Delta)$ test
	    (associated with a box $B$ or component $C$)
	    is bounded by
	    	\beql{ldf}
		\wtO \Big(n\cdot[\tau_F+n\cdot \LOG(m,r)+L(\Delta,F)]\Big).
	    	\eeql
	     Thus for each call of $\wtTG(\Delta)$ test, we need to bound
	     $\LOG(m,r)$ and $L(\Delta,F)$.
	
	     For $\wtTG_0(\Delta(B))$, we need to perform $\wtTG_0$ test for
	     each subbox $B_i$ into which $B$ is divided.
	     We have $\Delta_{B_i}=\Delta(m,r)$,
	     it is easy to see that
	          $\LOG(m,r)\le \LOG(B)$.
	\ignore{
	     (where $X$ is a box or a component), hence
	     condition \eqref{condition2} gives
	          $\LOG(m,r)=O(\LOG(\xi_D))$ with $\xi_D$
	     an arbitrary root in an arbitrary strong
	     $\varepsilon$-cluster $D$ that is contained in $C$.
	}%
	    So it remains to bound the term $L(\Delta,F)$ in \refeQ{ldf}.
	    By definition, $L(\Delta,F)=2\cdot(4+\LOG(||F_{\Delta}||_{\infty}^{-1}))$
	    And for any $z\in \Delta$, it holds $|F(z)|\le n\cdot
	    ||F_{\Delta}||_{\infty}$. Hence, we need to prove that
	        $\LOG((\max_{z\in \Delta_{B_i}}|F(z)|)^{-1})$
	    can be bounded by \refeQ{BoxCost}.
	
	    We apply \refLem{maxF} to obtain the bound of
	         $\log((\max_{z\in \Delta_{B_i}}
	         \\|F(z)|)^{-1})$.
	    Since $\phi(B)\ib \calZ(14B\cap 2B_0)$ (\refLem{in14B}),
	    it suffices to take
	    $\wh{\Delta}=42\cdot \Delta_{B_i}$ since $42\Delta_{B_i}$
	    contains $14\cdot\Delta_B$ which (by \refLem{in14B}) contains $\phi(B)$.
	    Hence with $K'=42$, \refLem{maxF} yields that
	       $\max_{z\in\Delta_{B}}|F(z)|>
	             (\frac{3}{4}\cdot\frac{w_{B}}{2})^{k_D}
	             \cdot n^{-\#(\wh{\Delta})}
	             \cdot(K')^{-\#(\wh{\Delta})+k_D}
	             \cdot 2^{-3n-1}
	             \prod\nolimits_{z_j\notin D}|\xi_D-z_j|^{n_j}$
	    where $D=\phi(B)$,
	    $k_D=\#(D)$, and $\xi_D$ is an arbitrary root contained in $D$.
	    From \refLem{BoxSize}(c), we have
	       $w_B>\frac{\vareps}{2}(\efrac{114k})^{k}.$
	    It is easy to check that
	       $\LOG((\max_{z\in\Delta_{B}}|F(z)|)^{-1})$
	    is bounded by \refeQ{BoxCost}.
	
	    (b)
	    To bound the cost of processing a component $C$,
	    we need to bound the cost of performing
	       $\wtTG(\Delta_C)$ and $\wtTG(\Delta')$.
	    It is easy to see that in both cases
	    where  $\Delta(m,r)=\Delta_C$ and $\Delta(m,r)=\Delta'$,
	    we have
	       $\LOG(m,r)=O(\LOG(C))$.
	    With the same arguments in the proof of (a),
	    it remains to prove that  both
	        $\log{\max_{z\in \Delta_C|F(z)|^{-1}}}$
	    and
	        $\log{\max_{z\in \Delta'|F(z)|^{-1}}}$
	    are bounded by \refeQ{CompCost}.
	
	    First consider the $\wtT_{*}^{G}(\Delta_C)$ test,
	    by applying \refLem{maxF} with $K=1$, we have
	        $\max_{z\in\Delta}|F(z)|
	          > R_C^{k_D}\cdot n^{-k_C}\cdot 2^{-3n-1}\cdot \prod\nolimits_{z_j\notin D}|\xi_D-z_j|^{n_j}$
	    with $D$ an arbitrary cluster in $C$, $k_D=\#(D)$ and $\xi_D$ an arbitrary root in $D$.
	    We know that $R_C\ge \frac{4}{3}w_C$.
	\ignore{
	    From \refLem{BoxSize},     
	     $R_C>\frac{3}{4}\cdot w_C
	     >\frac{3}{4}\cdot\frac{1}{2^{8n+7}\cdot n^{2n+1}}\cdot\frac{\varepsilon^2}{w_{C_0}}$.
	    Thus
	     $\LOG((R_C^{k_D})^{-1})
	     <\wtO(k_D\cdot(\LOG(\varepsilon^{-1})+\log(w_{C_0})+n))$.
	}%
	    With the same arguments as in part (a),
	    we can conclude that the cost of $\wtTG_*(\Delta_C)$ test is bounded
	    by \refeQ{CompCost}.
	
	    Now consider $\wtT_{k_C}^G(\Delta')$ test with $\Delta'=\Delta(m', \frac{w_C}{8N_C})$ and $m'$ as defined in the algorithm of Newton test.
	    Here we take
	        $\wh{\Delta}=2\cdot 3n \cdot 8N_C\cdot \Delta'
	          =48nN_C\cdot\Delta'$
	    since $48nN_C\Delta'$ will contain $C$ and thus contain all the
	    roots in $C$.
	    By applying \refLem{maxF} with $K=48nN_C$,
	    we have
	        $\max_{\Delta'}|F(z)|> (\frac{w_C}{8N_C})^{k_D}
	             \cdot n^{-\#(\wh{\Delta})}
	             \cdot K^{-\#(\wh{\Delta})+k_D}
	             \cdot 2^{-3n-1}
	             \cdot \prod\nolimits_{z_j\notin D}|\xi_D-z_j|^{n_j}$
	    with $D$ an arbitrary cluster in $C$, $k_D=\#(D)$ and
	    $\xi_D$ an arbitrary root in $D$.
	    First consider the lower bound for
	         $(\frac{w_C}{8N_C})^{k_D}$.
	    By lemma B0(b), we have
	         $N_C\le \frac{4w_{C_0}}{w_C}$,
	    thus
	         $\frac{w_C}{8N_C}\ge \frac{w_C^2}{32w_{C_0}}$.
	    It follows
	         $\LOG(((\frac{w_C}{8N_C})^{k_D})^{-1})
	         =k_D(2\LOG(w_C^{-1})+\LOG(w_{C_0})+5)$.
	    As is proved, $k_D(2\LOG(w_C)+\LOG(w_{C_0})+5)$ is bounded by
	    \refeQ{CompCost}.
	
	    The bound for the other terms except
	    $K^{\#(\wh{\Delta})-k_D}$ are similar
	    to the case discussed above.
	    Hence it remains to bound $K^{\#(\wh{\Delta})-k_D}$.
	    Denote the radius of $\wh{\Delta}$ as ${\wh{R}}$, then
	    $\wh{R}=18n w_{C}$ from the definition of $\wh{\Delta}$.
	    Note that
	         $K=48nN_C \le 48n \cdot \frac{w_{C_0}}{w_C}
	         =48n\cdot 18n \cdot\frac{w_{C_0}}{\wh{R}}$
	    and
	         $\LOG\paren{(48n\cdot18n\cdot w_{C_0})^{\#(\wh{\Delta})-k_D}}
	             =O(n\log n+n\LOG(w_{C_0}))$,
	    thus it suffices to bound $\wh{R}^{-\#(\wh{\Delta})+k_D}$.
	    For any root $\xi_D$ of $F$ in any $\varepsilon$-cluster
	    $D\ib C$ which contains $k_D$ roots counted with multiplicities, we have
	    {{
	      \beqarrays
	          \prod\nolimits_{z_i\notin D}|\xi_D-z_i|^{n_i}
	          &=& \prod_{z_j\in \wh{\Delta},
	                 z_j\notin D}|\xi_D-z_j|^{n_j}
			 \prod_{z_k\notin \wh{\Delta}}|\xi_D-z_k|^{n_k} \\
	          &\le&
	               (2\wh{R})^{\#(\wh{\Delta})-k_D}
	                  \cdot \frac{{\rm{Mea}}(F(\xi_D+z))}{|\lcoeff(F)|}\\ 
	            &\le&
			  (2(\wh{R})^{\#(\wh{\Delta})-k_D}
	                 \cdot 2^{\tau_F} 2^{n+3}\max\nolimits_1(\xi_D)^n \\
	            &\le&  2^{\tau_F+2n+3}\cdot
	                   \max\nolimits_1(\xi_D)^n\cdot
	                   \wh{R}^{\#(\wh{\Delta})-k_D}
	     \eeqarrays
	     }
	     }
	     So $\log(\wh{R}^{-\#(\wh{\Delta})+\xi_D})$ is bounded by
	     \refeQ{CompCost}.
	    Hence the cost for processing component $C$,
	    that is the two kind of
	    $\wtTG$ tests discussed above can be bounded by \refeQ{CompCost}.
	\epf
\ssect{Corollary to Theorem A}
	\bgenDIY{\sc Corollary to Theorem A}{
	\\   The bit complexity of the algorithm is bounded by
	     $$\wtO\Big(n^2(\tau_F + k+m)
				    + nk\LOG(\vareps^{-1})
				    + n\LOG|\gdisc(F_\vareps)|^{-1}
				    \Big).$$
	   In case $F$ is an integer polynomial, this bound becomes
	     $$\wtO\Big(n^2(\tau_F + k +m)
				    + nk\LOG(\vareps^{-1}) \Big).$$
	}
	\bpf
	   From our assumption in Section 6, $\LOG(B_0)=O(\tau_F)$.
	   We can also see that
	       $\sum_{D \in \wh{S}} L_D\le n\tau_F+k(k+\LOG(\vareps^{-1}))+\sum_{D \in \wh{S}} T_D+\sum_{i=1}^{k}\LOG(z_i)$.
	
	   By Theorem A5, $\sum_{D \in \wh{S}} T_D=
	                \wtO(\LOG|\gdisc(F_\vareps)|^{-1}+ nm+ n \LOG \rm{Mea}(F)).$
	   And
	        $\sum_{D \in \wh{S}} \LOG(\xi_D)
	        \le \sum_{i=1}^{k}\LOG(z_i)
	        \le\LOG{\rm{Mea}(F)}+k=O(\tau+k+\log n)$ (using Landau's inequality).
	   From the equations above, we can deduce the first part of this lemma.
	
	   The second part comes from Corollary A6.
	\epf
\ssect{Theorem B}
	We first show two useful lemmas:
	\refLem{RootSep} is about root separation in components,
	and \refLem{StrongNatu} says that
	strong $\vareps$-clusters are actually natural clusters.
	
	\blemCl{RootSep}
		If $C$ is any confined component,
		and its multiset of roots $\calZ(C)$ is partitioned into
		two subsets $G, H$.  Then there exists $z_g\in G$
		and $z_h\in H$ such that $|z_g-z_h|\le (2+\sqrt{2})w_C$.
	\elemC
	
	\bpf
	   We can define the $\calS_G\as \set{B\in \calS_C: 2B\cap G\neq \es}$
	   and $\calS_H\as \set{B\in \calS_C: 2B\cap H\neq \es}$.
	   Note that $\calS_G\cup \calS_H=\calS_C$.
	   Since the union of the supports of $\calS_G$ and $\calS_H$ is connected,
	   there must a box $B_g\in \calS_G$ and
	   $B_h\in \calS_H$ such that $B_g\cap B_h$ is non-empty.
	   This means that the centers of $B_g$ and
	   $B_h$ are at most $\sqrt{2}w_C$ apart.
	   From \refCor{exclusion},
	   there is root $z_g$ (resp., $z_h$) at distance $\le w_C$ from the
	   centers of $B_g$ (resp., $B_h$).
	   Hence $|z_g-z_h|\le (2+\sqrt{2})w_C$.
	\epf
	
	\blemCl{StrongNatu}
	  	Each strong $\vareps$-cluster is a natural $\vareps$-cluster.
	\elemC
	
	\bpf
	In the definition of $\vareps$-equivalence,
	if $z\equiV z'$ then there is a witness isolator $\Delta$
	containing $z$ and $z'$.
	If $z'\equiV z''$ we have another witness $\Delta'$ containing $z'$ and $z''$.
	It follows from basic properties of isolators that if
	$\Delta$ and $\Delta'$ intersect, then there is
	inclusion relation between $\calZ(\Delta)$ and $\calZ(\Delta')$.
	Thus $\Delta$ or $\Delta'$ is a witness for $z\equiV z''$.
	Proceeding in this way, we eventually get a witness isolator for
	the entire equivalence class.
	\epf

	\bgenDIY{{Theorem B}}{
	   \\  Each natural $\vareps$-cluster in $\wh{S}$ is a union of strong $\vareps$-clusters.
	}
	
	\bpf
	    First we make an observation:
	    For any strong $\vareps$-cluster $D'$ and confined component $C'$,
	    if $D'\cap \calZ(C') \neq \es$ and $w_{C'}>2\cdot \rad(D')$,
	    then $D'\subset \calZ(C')$.
	    To see this: suppose, $z_1\in D'\cap\calZ(C')$
	    and $z_2\in\calZ(D)$ belong to a component other than $C'$.
	    By Property (C3),
	     $|z_1-z_2|\ge w_{C'}>2r$, contradicting the fact that any 2 roots
	    in $D'$ are separated by distance at most $2r$.
	
	    Let $D\in \wh{S}$.   There are two cases: $D$ is either in $S$
	    or in $S'$ where $\wh{S}=S\cup S'$ as defined in \refeQ{sss}.
	
	    First, assume that $D\in S'$.  This case is relatively easy.
	    Suppose $E$ is a strong $\vareps$-cluster and
	    $D\cap E\neq \es$.
	    From \refLem{StrongNatu}, $E$ is also a natural cluster;
	    thus either $D\subset E$ or $E\subset D$.
	    By the definition of $\phi_0(B)$, $D$ is a largest natural
	    $\vareps$-cluster, meaning that there is no natural
	    $\vareps$-cluster strictly containing $D$. Hence it
	    follows $E\subset D$, which is what we wanted to prove.
	
	    In the remainder of this proof,
	    we show that each natural $\vareps$-cluster in $D$ is $S$
	    is a union of strong $\vareps$-cluster.
	   The observation above and \refLem{in2B_0}(a) imply that
	   for each component $C'$ in the preprocessing stage,
	   $C'$ is a union of strong $\vareps$-clusters.
	   Thus, when the mains loop starts, for each component $C$ in
	   $Q_1$, $\calZ(C)$ is a union of strong $\vareps$-clusters.
	
	   Suppose $D$ is a strong $\varepsilon$-cluster and $C$ is a confined leaf of
	    $\compTreeHat$.  It is sufficient to prove that
	    if $D\cap \calZ(C)\neq \emptyset$, then $D\ib \calZ(C)$.
	     Let $r=\rad(D)$. Suppose $z_1\in D\cap \calZ(C)$. There is an unique maximal path
	     in $\compTreeHat$ such that all the components in this path contain $z_1$.
	\ignore{
	     Observation: for any component $C'$ in this path, if $w_{C'}>2r$, then
	     $D\ib\calZ(C')$ [To see this: suppose, $z_2\in D$ belongs to a different
	       component than $C'$.  By Property (C3),
	     $|z_1-z_2|\ge w_{C'}>2r$, contradicting the fact that any 2 roots
	   in $D$ is separated by distance $\le 2r$.]
	}%
	
	    Consider the first component $C_1$ in the path above such that $C_1$ contains
	    the root $z_1$ and $w_{C_1}\le4r$. If $C_1$ does not exist, it means that
	    the leaf $C_t$ in this path satisfies $w_{C_t}\ge 4r$, and by the observation
	    above, it follows that $D\ib\calZ(C_t)$.
	    Henceforth assume $C_1$ exists;
	    we will prove that it is actually a leaf of $\compTreeHat$.
	
	    Consider $C'_1$, the parent of $C_1$ in $\compTreeHat$.
	    Note that $w_{C'_1}\ge 4r$, and by the observation above, $D\ib \calZ(C'_1)$.
	    We show that $w_{C_1}>2r$.
	    To show this, we discuss two cases.
	    If the step $C'_1\to C_1$ is a Newton Step, then all the roots in $C_1$ are contained in a disc of
	    radius $r'=\frac{w_{C'_1}}{8N_{C'_1}}$.
	    Note that $r'\ge r$ since the Newton disc contains all the roots
	    in $C'_1$ and hence contains $D$.
	    Newton step gives us
	    $w_{C_1}=\frac{w_{C'_1}}{2N_{C'_1}}=4r'\ge 4r$.
	    If $C'_1\to C_1$ is a Bisection Step, then $w_{C_1}= w_{C'_1}/2> 2r$.
	    To summarize, we now know that $2r < w_{C_1}\le 4r$.
	    Again, from our above observation, we conclude that $D\ib \calZ(C_1)$.
	
	    First a notation: let $\Delta_D$ be the smallest disc containing $D$.
	    We now prove that $\calZ(C_1)\ib D$.
	    By way of contradiction, suppose there is a root $z\in \calZ(C_1)\setminus D$.
	    Since $D$ is a strong $\vareps$-cluster,
	    $\#(\Delta_D)=\#(114\Delta_D)$.
	    It follows that for any $z'\in D$, we must have have $|z-z'|>113r$.
	    On the other hand, by \refLem{RootSep}, there exists $z$ and $z'$
	    fulfilling the above assumptions with the property
	    that $|z-z'|\le (2+\sqrt{2})w_{C_1} \le (2+\sqrt{2})4r <113r$.
	    Thus we arrived at a contradiction.
	
	    \ignore{
	    The component $C$ satisfies $w_C\ge2r$, thus $C$ has at most $9$ constituent boxes and $W_C\le3\cdot w_C< 48r\le \varepsilon$.
	    Moreover, $R_C= \frac{3}{4}W_C<9r$,
	    thus $\Delta_C$ is contained in the disc $9\cdot\Delta$,
	    and $\frac{4}{3}\cdot3\Delta_C$ is contained in the disc $36\Delta$.
	    Since $\#(\Delta)=\#(36\cdot\Delta)$, the disc $\frac{4}{3}\cdot 3\Delta_C$
	    contains only the roots in $D$. Hence both $\wtTG_k(3\Delta_C)$ and
	    $\wtTG_k(\Delta_C)$ will succeed.
	  }%
	
		From the above discussion, we conclude that
	        $\calZ(C_1)=D$ and $2r<w_{C_1}\le4r$,
	    it is easy to see that $W_{C_1}\le3w_{C_1}$.
	    Hence we can conclude that
	       $W_{C_1}\le 12r< 12\cdot \frac{\vareps}{12}\le \vareps$.
	    Therefore, to show that $C_1$ is a leaf, it remains to
	    prove that $4\Delta_{C_1}\cap C_2=\es$ for all $C_2$
	    in $Q_1\cup \qdis$.
	
	    Since $2r<w_{C_1}\le4r$,
	\ignore{
	    it is easy to see that $C_1$ consists of at most $9$ boxes
	    and $W_{C_1}\le3w_{C_1}\le12r$.
	    Hence $R_{C_1}=\frac{3}{4}\cdot W_{C_1}\le 9r$ and
	}%
	    by some simple calculations,
	    we can obtain that $C_1\subset 8\Delta_{D}$
	    thus $\Delta_{C_1}$ is
	    contained in $9\Delta_D$, it follows $4\Delta_{C_1}\subset36\Delta_D$.
	    It suffices to prove that $36\Delta_D\cap C_2=\es$ for all $C_2$.
	    Note that for any root $z_1\in C_1$ and any component $C_2$,
	    we have $\Sep(z_1,C_2)\ge w_{C_2}$ by property (C3).
	    Assume that $\Sep(z_1,C_2)=|z_1-p|$ for some $p\in C_2$.
	    We claim that there exists a root $z_2\in C_2$ such that
	    $|z_2-p|\le \frac{3\sqrt2}{2}w_{C_2}$.
	    [To see this, suppose that $p$ is contained in
	    a constituent box $B_2$ of $C_2$, note that $2B_2$ must contain a root,
	    assume that $z_2\in2B_2$, it follows $|z_2-p|\le\frac{3\sqrt2}{2}w_{C_2}$.]
	    Hence
	        $|z_1-p|+|z_2-p|\le\Sep(z_1,C_2)+\frac{3\sqrt2}{2}\cdot\Sep(z_1,C_2)$.
	    Note that $\#(\Delta_D)=\#(114\Delta_D)$, thus $|z_1-z_2|\le 113r$ .
	    By triangular inequality, we have
	        $|z_1-z_2|\le|z_1-p|+|z_2-p|<(1+\frac{3\sqrt2}{2})\cdot\Sep(z_1,C_2)$.
	    Hence $\Sep(z_1,C_2)\ge\frac{1}{1+3\sqrt2/2}|z_1-z_2|>36r$,
	    implying $36\Delta_D\cap C_2=\es$.
	
	    This proves that our algorithm will output $C_1$, i.e., $C_1$ is a confined
	    leaf of $\compTreeHat$.
	
	    In summary, each natural $\vareps$-cluster in $\wh{S}$
	    is a union of strong $\vareps$-cluster.
	\epf

\ssect{A complete proof of Theorem A}
    Based on \refLem{NoniceCostX}, we can now derive the total cost of carrying out all the
    $\wtTG$ tests in the algorithm.

    A direct result from \refLem{NoniceCostX} is that the cost of processing all the
    boxes can be bounded by
        \begin{align*}
         \wtO\Big( \sum_{B\in \mathcal{B}}\left(n\cdot[\tau_F+n\LOG(B)
                     + k_{\phi(B)}\cdot (\LOG(\vareps^{-1})+k)+T_{\phi(B)}]\right)
                     \Big)
        \end{align*}
    where $\mathcal{B}$ is
    the set of all the boxes produced in the algorithm.

   Taking into account the fact that the number of boxes
   charged to a natural $\epsilon$-cluster by the map $\phi$ is bounded by $O(\smax\log n) =\wtO(1)$, we can write the above bound as
     \beql{costTotalB}
         \wtO\Big( \sum_{B\in \mathcal{B}}n^2\LOG(B)+
                    \sum_{D\in \wh{S}}\left( n\cdot[\tau_F + k_{D}\cdot (\LOG(\vareps^{-1})+k)+T_{D}]\right)
                     \Big).
     \eeql
   Analogously, we can obtain that the cost of processing all the
   components can be bounded by
    \beql{costTotalC}
         \wtO\Big( \sum_{C\in \mathcal{C}}n^2\left(\LOG(C)+\LOG(w_{C_0})
         \right)+
                    \sum_{D\in \wh{S}}\left( n\cdot[\tau_F + k_{D}\cdot (\LOG(\vareps^{-1})+k)+T_{D}]\right)
                     \Big)
    \eeql
    where $\mathcal{C}$ is the set of all the components produced in
    the algorithm and $C_0$ is the last special component above $C$.

    The bounds \refeQ{costTotalB} and \refeQ{costTotalC} add up to the
    cost of processing all the boxes and components produced in the algorithm.
    To prove Theorem A, we want to show that both
    \beql{partB}
    \wtO\Big( \sum_{B\in \mathcal{B}}n^2\LOG(B)
         \Big)
    \eeql
    and
    \beql{partC}
    \wtO\Big(\sum_{C\in \mathcal{C}}n^2\left(\LOG(C)+\LOG(w_{C_0})
         \right)\Big)
    \eeql
    can be bounded by
    \beql{partA}
    \wtO\Big( n^2 \LOG(B_0) + n^2 \sum_{D\in \wh{S}} \LOG(\xi_D)
     \Big)
    \eeql
    where $\xi_D$ is an arbitrary root contained in $D$.

    First we show that the bound \refeQ{partC} can be bounded by
    \refeQ{partA}.
    Notice that for each component $C$, we have
    $$\LOG(C)+\LOG(w_{C_0})\le \LOG(C_0)+\LOG(C_0)$$
    where $C_0$ is the last special component above $C$.
    Since the length of each non-special path is at most $\smax$,
    we can bound \refeQ{partC} by
    \beql{costSpecial}
    \wtO\Big(\smax\cdot\sum_{C_0\in \mathcal{SC}}n^2\LOG(C_0)
         \Big)
    =  \wtO\Big(\sum_{C_0\in \mathcal{SC}}n^2\LOG(C_0)
        \Big)
    \eeql
    where $\mathcal{SC}$ is the set of all the special component produced in the algorithm. Thus it suffices to prove the following lemma.

    \bleml{costSpecial}
    The bound
    \refeQ{costSpecial} can be bounded by \refeQ{partA}.
    \elem

    Before proving this lemma, we first consider a simple case where
    each special component $C$ satisfies the following condition:
    \beql{niceC}
    \max\nolimits_{z\in C}\LOG(z) = O( \min\nolimits_{z\in C}\LOG(z)).
    \eeql
    Since $\phi(C)\in C$ and \refeQ{niceC} holds, it follows that
   $$ \wtO(\sum_{C\in \mathcal{SC}}n^2\LOG(C))
    = \wtO(\sum_{C\in \mathcal{SC}}n^2\LOG(\xi_{\phi(C)}))
        $$
    where $\xi_{\phi(C)}$ is an arbitrary root contained in $\phi(C)$.
    Thus, it is easy to see that \refLem{costSpecial} holds.

    In general case, condition \refeQ{niceC} may not hold for all the
    special components. And we call a special component \dt{nice} if
    it satisfies \refeQ{niceC}, otherwise it is \dt{non-nice}.

    Now we define a set of square annuli for later use.
    Denote by $w_0$ the width of the smallest box centered at the origin
	containing $\frac{5}{4}w(B_0)$ and denote
	     $t_0\as \lfloor\log(w_0)\rfloor$
	for short.
	Note that if $B_0$ is centered at the origin,
	we have $w_0=\frac{5}{4}w(B_0)$.
	We now define $I_{t_0+1}\as\es$ and
	   \begin{equation*}
	      \begin{aligned}
	         I_i &\as [-\efrac{2^i}, \efrac{2^i}]w_0,\\
	         A_i &\as (I_i\times I_i)\setminus (I_{i+1},I_{i+1}),
	      \end{aligned}
	   \end{equation*}
	for $i\in \{1\dd t_0\}$.
	Denote $w(A_i)\as \frac{1}{2}\cdot\frac{w_0}{2^i}$ as the width
	of the square annulus $A_i$.

	    	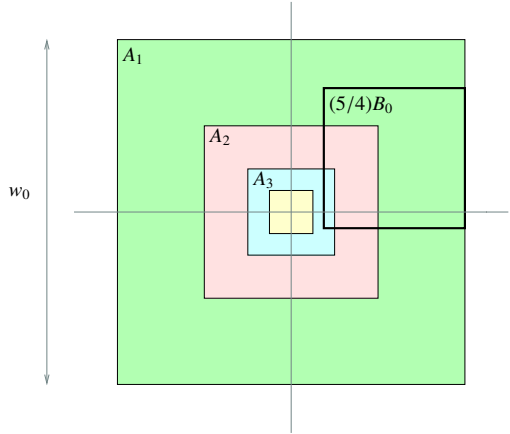
\begin{figure}[htb]
	    	  \begin{center}
		   \scalebox{0.45}{
	    	     \input{./annulus2.tex}}
	    	   \caption{Annulus $A_1$, $A_2$, $A_3$ and box
	$\frac{5}{4}B_0$.}
	    	   \label{fig:annulus2}
	    	  \end{center}
	    	\end{figure} 	

	An observation is that: for a component $C$,
	if there exists an integer $i\in\{1\dd t_0-1\}$ such that
	       $C\ib A_i\cup A_{i+1}$,
	then $C$ satisfies \refeQ{niceC}.

    Now we are prepared to prove \refLem{costSpecial}.

    \bpf
    Denote by $\mathcal{SC}_1$ the set of all the nice special components
    and $\mathcal{SC}_2$ the set of all the non-nice special components.
    From the discussions above, we can see that $ \wtO(\sum_{C\in \mathcal{SC}_1}n^2\LOG(C))$ is bounded by \refeQ{partA}. Thus it remains to prove that $ \wtO(\sum_{C\in \mathcal{SC}_2}n^2\LOG(C))$
    can be bounded by \refeQ{partA}.

    We define the unique set $I$ such that $i\in I$ if and only if
	   $A_i$ contains at least one root in $\calZ(Q)$. Suppose $I={i_1\dd i_m}$
	   with $i_1<\cdots<i_m$.

    We consider the components in $\mathcal{SC}_2$ that contain at
    least one root in $A_{i_1}$. Denote by $\mathcal{SC}_2(A_{i_1})$
    the set of all such components and $\calZ(A_{i_1})$ the union
    of the roots contained in $\mathcal{SC}_2(A_{i_1})$.
    We classify these components into $2$
    categories: the special component that contains all the roots in $\calZ(A_{i_1})$ and the special components part of the roots in $\calZ(A_{i_1})$.
    The first category consists of at most one components since any
    two special components contain different roots. If the first
    category is not empty, suppose $C$ is the component in it.
    We can bound $ \wtO(n^2\LOG(C))$ with $ \wtO(n^2\LOG(B_0))$.

    Now we consider the second category.

    We claim that for any component $C$ in the second category,
	   it holds that
	       $\LOG(C)=O(\LOG(w(A_{i_1})))$.
	   The proof is as follows.
	   We can easily see that
	         $\calZ(A_{i_1})\subset B(0,4w(A_{i_1}))$
       with
	    $ B(0,4w(A_{i_1}))$ the square centered at the
       origin and of width $4w(A_{i_1})$.
	    Thus    $\rad(\calZ(A_{i_1}))\le 2\sqrt2 w(A_{i_1})$.
	   Since the second category consists of at least $2$
       components,
       thus for any component $C\in \mathcal{SC}_2(A_{i_1})$, we have
	        $w_C\le 2\cdot\rad(\calZ(A_{i_1}))
	        \le 4\sqrt2w(A_{i_1})$
       (See the observation in the proof of Theorem B).
	   Now for any $C\in P_{i_1}$, we have
	        $\calZ(C)\subset B(0, 4w(A_{i_1}))$
	   and
	        $w_C\le 4\sqrt2 w(A_{i_1})$.
       By Corollary 5(b), the distance from any point in $C$
       to a closest root in $C$ is at most $2\sqrt2w_C$.
	   Hence it is easy to see that
	        $C\subset B(0, 4w(A_{i_1})+2\sqrt2\cdot 4\sqrt2 w(A_{i_1}))
	        =B(0, 20w(A_{i_1}))$.
	   It follows
	        $\LOG(C)=O(\LOG(w(A_{i_1})))$.

       By the definition of $\mathcal{SC}_2(A_{i_1})$, for each component $C$ in the second category, there exists a root contained in $A_{i_1}$.
       And since each natural $\vareps$-cluster has width less than $1$,
       there exists a natural $\vareps$-cluster $D_C$ in $C$ such that $D_C\in A_{i_1}\cup A_{i_1+1}$. With the claim above, we have $\LOG(C)=O(\LOG(\xi_{D_C}))$ where $\xi_{D_C}$ is an arbitrary root contained in $D_C$. And in this case, we charge the component $C$
       to the natural $\vareps$-cluster $D_C$ that is contained in $C$.
       Now we prove that each $D_C$ is charged at most $O(\log n)$ times.
       Suppose $C'$ is a component in $\mathcal{SC}_2(A_{i_1})$ that is charged to
       $D_C$.
       Since $C'$ is not a nice component, $C'$ must contain a root
       inside $A_{i_1+2}$. Otherwise, $C'$ would have
       satisfied the condition \refeQ{niceC} since we have $\LOG(C')=O(\LOG(w(A_{i_1})))$.
       From the fact that $C'$ contains both a root in $A_{i_1}$ and
       a root inside $A_{i_1+2}$, we conclude that $W_{C'}\ge\frac{w(A_{i_1})}{2}$. Hence we have
       $w_{C'}\ge\efrac{3n}\cdot\frac{w(A_{i_1})}{2}$.
        Meanwhile, since $C' \subset B(0, 20w(A_{i_1}))$, thus
        $w_{C'}\le 40w(A_{i_1})$. It is easy to see that the number
        of different sizes of $C'$ is bounded by $O(\log n)$.
       Thus we come to the conclusion that
       $ \wtO(\sum_{C\in \mathcal{SC}_2}n^2\LOG(C))$ is bounded
       by $ O(\log n)\cdot \wtO(\sum_{D\in A_{i_1}\cup A_{i_1+1}} n^2\LOG(\xi_D))= \wtO(\sum_{D\in A_{i_1}\cup A_{i_1+1}} n^2\LOG(\xi_D)).$

       Hence we have $ \wtO(\sum_{C\in \mathcal{SC}_2(A_{i_1})}n^2\LOG(C))= \wtO(n^2 \LOG(B_0)+
       \sum_{D\in A_{i_1}\cup A_{i_1+1}} n^2\LOG(\xi_D))$.

       Analogously, if we consider the components in
       $\mathcal{SC}_2\setminus \mathcal{SC}_2(A_{i_1})$
       that contain at least one root in $A_{i_2}$,
       we will obtain that
       $ \wtO(\sum_{C\in \mathcal{SC}_2(A_{i_2})}n^2\LOG(C))= \wtO(n^2 \cdot w(A_{i+1})+
       \sum_{D\in A_{i_2}\cup A_{i_2+1}} n^2\LOG(\xi_D))$.

       By recursive analysis, we can eventually obtain that
       the bound \refeQ{partC} is bounded by \refeQ{partA}.
    \epf

    It remains to prove the following lemma.
    \bleml{partB}
     The bound \refeQ{partB} can be
    bounded by \refeQ{partA}.
    \eleml
    Likewise, we first consider a simple case where each
    box $B$ satisfies the following condition:
    \beql{niceB}
	  \max\nolimits_{z\in 14B}\LOG(z) = O( \min\nolimits_{z\in 14B}\LOG(z)).
	  \eeql
     Since $\phi(B)\in 14B$ and \refeQ{niceB} holds, it follows that
   $$ \wtO(\sum_{B\in \mathcal{B}}n^2\LOG(B))
    = \wtO(\sum_{B\in \mathcal{B}}n^2\LOG(\xi_{\phi(B)}))
        $$
    where $\xi_{\phi(B)}$ is an arbitrary root contained in $\phi(B)$.
    Thus, it is easy to see that \refLem{costSpecial} holds.

    In general case, condition \refeQ{niceB} may not hold for all the
    boxes. And we call a box \dt{nice} if
    it satisfies \refeQ{niceB}, otherwise it is \dt{non-nice}.

    Before we proving \refLem{partB}, we need to give a useful result.

    \bleml{noniceBox}
     There exists at most $400$ aligned non-nice boxes
            of the same size.
    \eleml
    \bpf
     Denote $M_B$ as the middle of a box $B$.
         We will
         shows that if $M_B\notin B(O,20w_B)$(the box
         centered at the origin and of width $20w_B$), then $B$ is a nice box.

     If $M_B\notin B(O,20w_B)$, then $|M_B|>10 w_B$. We have
     $\min_{z\in 14B}\LOG(z)\ge \LOG(M_B-7\sqrt2w_B)\ge \LOG(\frac{M_B}{100}) $ and
     $\max_{z\in 14B}\LOG(z)\le \LOG(M_B+7\sqrt2w_B)\le
     \LOG(20M_B) $.
     It follows $\max_{z\in 14B}\LOG(z)=O(\min_{z\in 14B}\LOG(z))$.

     We can count that the number of aligned boxes satisfying
          $M_B\in B(0,20w_B)$ is at most $20^2=400$.
         Thus the number of non-nice boxes of width $w_B$
         is at most $400$.
    \epf

    Now we prove \refLem{partB}.

    \bpf
    Denote by $\calB_1$ the set of all the nice boxes produced in
    the algorithm and $\calB_2$ the set of all the non-nice boxes.
    From the discussions above, it follows that
    $\wtO( \sum_{B\in \mathcal{B}_1}n^2\LOG(B))$ can be bounded
    by \refeQ{partA}.

    It remains to prove that
    $\wtO( \sum_{B\in \mathcal{B}_2}n^2\LOG(B))$ can be bounded by
    \refeQ{partA}.
    By \refLem{noniceBox}, the number of non-nice boxes of the same
    size is at most $400$. And for a box $B$, if $B$ is a constituent
    box of a component $C$, it is evident that $\LOG(B)\le \LOG(C)$.
    Hence $\wtO( \sum_{B\in \mathcal{B}_2}n^2\LOG(B))= 400 \wtO( \sum_{C\in \mathcal{C}}n^2\LOG(C))= \wtO( \sum_{C\in \mathcal{C}}n^2\LOG(C))$. By \refLem{costSpecial}, the latter is
    bounded by \refeQ{partA}.
    \epf

\end{s8}

\end{document}

%% file: component.tex
\begin{picture}(0,0)%
\includegraphics{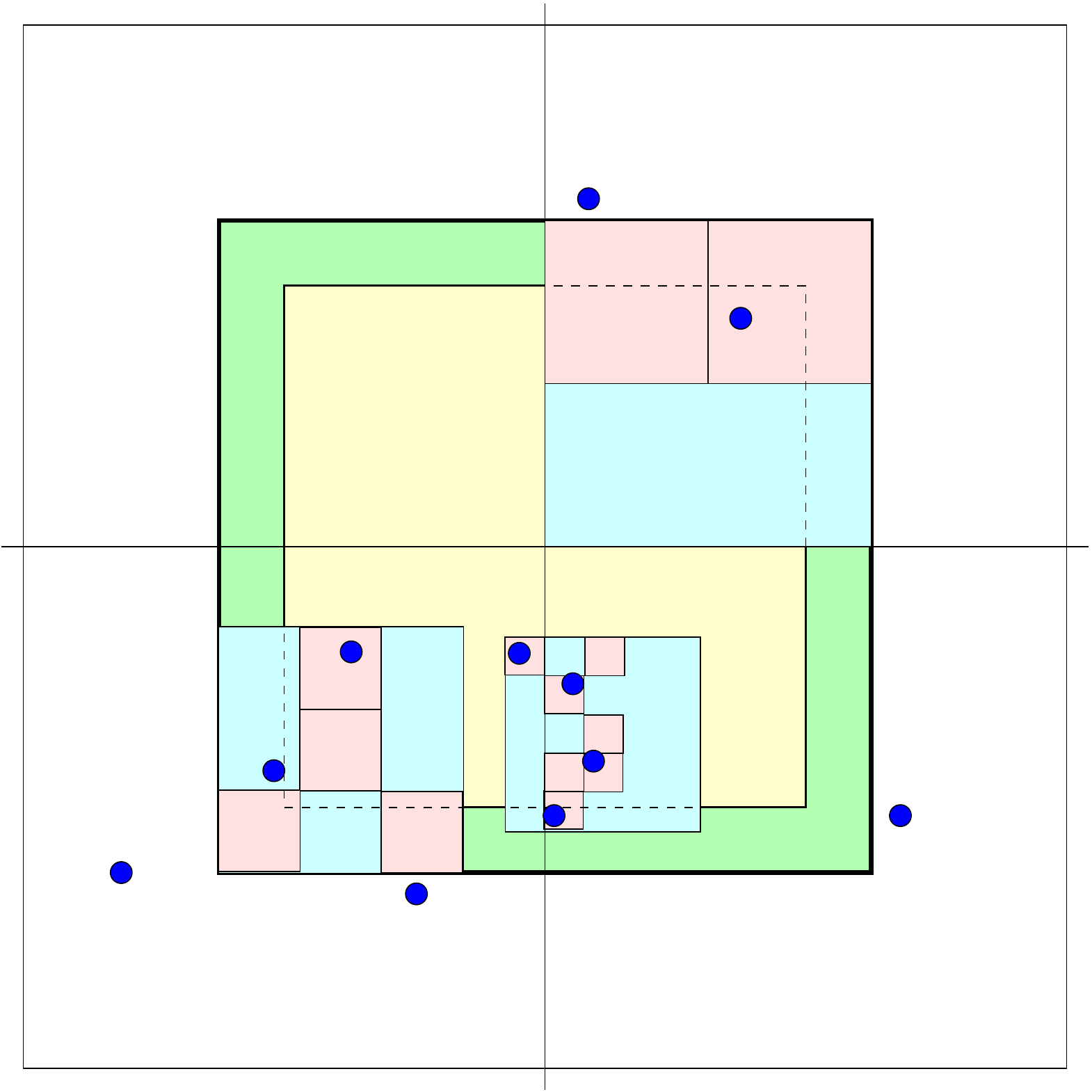}
\end{picture}%
\setlength{\unitlength}{3947sp}%
\begingroup\makeatletter\ifx\SetFigFont\undefined%
\gdef\SetFigFont#1#2#3#4#5{%
  \reset@font\fontsize{#1}{#2pt}%
  \fontfamily{#3}\fontseries{#4}\fontshape{#5}%
  \selectfont}%
\fi\endgroup%
\begin{picture}(7524,7524)(1039,-7723)
\put(3301,-4411){\makebox(0,0)[lb]{\smash{{\SetFigFont{20}{24.0}{\familydefault}{\mddefault}{\updefault}{\color[rgb]{0,0,0}$C_2$}%
}}}}
\put(3076,-2536){\makebox(0,0)[lb]{\smash{{\SetFigFont{20}{24.0}{\familydefault}{\mddefault}{\updefault}{\color[rgb]{0,0,0}$B_0$}%
}}}}
\put(1351,-661){\makebox(0,0)[lb]{\smash{{\SetFigFont{20}{24.0}{\familydefault}{\mddefault}{\updefault}{\color[rgb]{0,0,0}$2B_0$}%
}}}}
\put(2551,-2011){\makebox(0,0)[lb]{\smash{{\SetFigFont{20}{24.0}{\familydefault}{\mddefault}{\updefault}{\color[rgb]{0,0,0}$(5/4)B_0$}%
}}}}
\put(5033,-3242){\makebox(0,0)[lb]{\smash{{\SetFigFont{20}{24.0}{\familydefault}{\mddefault}{\updefault}{\color[rgb]{0,0,0}$C_1$}%
}}}}
\put(5289,-5105){\makebox(0,0)[lb]{\smash{{\SetFigFont{20}{24.0}{\familydefault}{\mddefault}{\updefault}{\color[rgb]{0,0,0}$C_3$}%
}}}}
\end{picture}%

%% file: confined.tex
\begin{picture}(0,0)%
\includegraphics{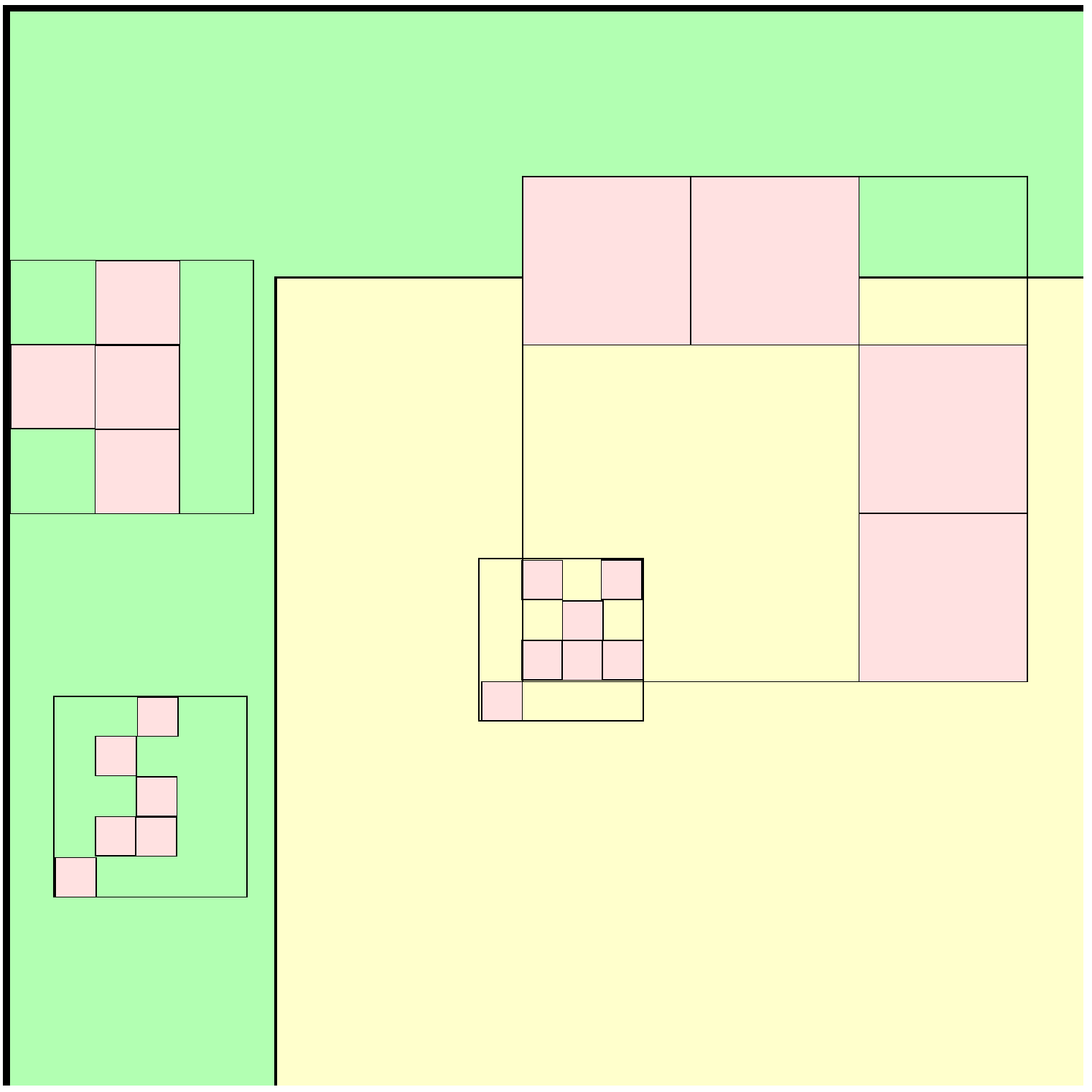}
\end{picture}%
\setlength{\unitlength}{3947sp}%
\begingroup\makeatletter\ifx\SetFigFont\undefined%
\gdef\SetFigFont#1#2#3#4#5{%
  \reset@font\fontsize{#1}{#2pt}%
  \fontfamily{#3}\fontseries{#4}\fontshape{#5}%
  \selectfont}%
\fi\endgroup%
\begin{picture}(7288,7288)(1157,-7605)
\put(3901,-5559){\makebox(0,0)[lb]{\smash{{\SetFigFont{20}{24.0}{\familydefault}{\mddefault}{\updefault}{\color[rgb]{0,0,0}$C_4$}%
}}}}
\put(1726,-6661){\makebox(0,0)[lb]{\smash{{\SetFigFont{20}{24.0}{\familydefault}{\mddefault}{\updefault}{\color[rgb]{0,0,0}$C_2$}%
}}}}
\put(1426,-4111){\makebox(0,0)[lb]{\smash{{\SetFigFont{20}{24.0}{\familydefault}{\mddefault}{\updefault}{\color[rgb]{0,0,0}$C_1$}%
}}}}
\put(3076,-2536){\makebox(0,0)[lb]{\smash{{\SetFigFont{20}{24.0}{\familydefault}{\mddefault}{\updefault}{\color[rgb]{0,0,0}$B_0$}%
}}}}
\put(1351,-736){\makebox(0,0)[lb]{\smash{{\SetFigFont{20}{24.0}{\familydefault}{\mddefault}{\updefault}{\color[rgb]{0,0,0}$(5/4)B_0$}%
}}}}
\put(5828,-2997){\makebox(0,0)[lb]{\smash{{\SetFigFont{20}{24.0}{\familydefault}{\mddefault}{\updefault}{\color[rgb]{0,0,0}$C_3$}%
}}}}
\end{picture}%

%% file: annulus2.tex
\begin{picture}(0,0)%
\includegraphics{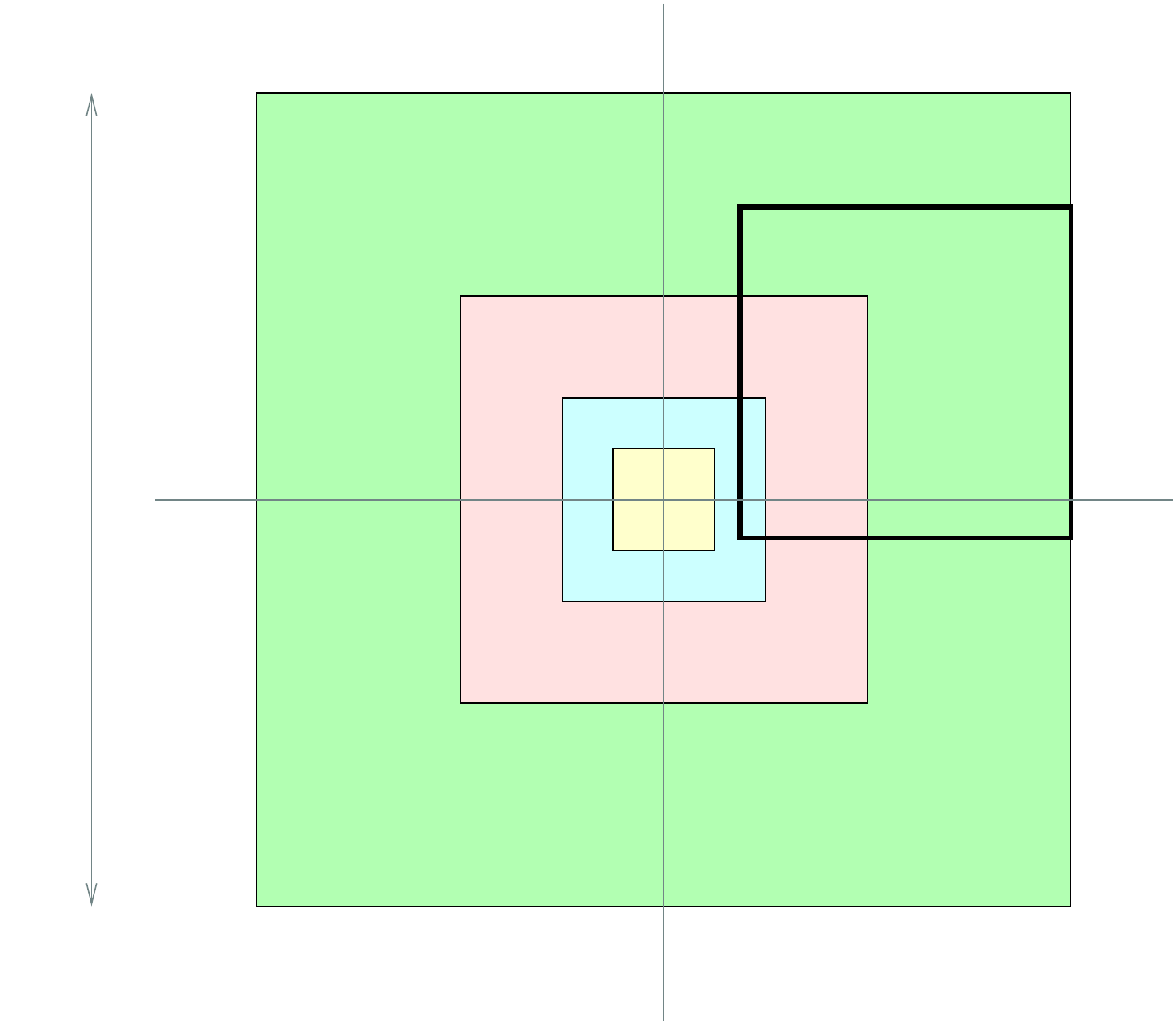}
\end{picture}%
\setlength{\unitlength}{3947sp}%
\begingroup\makeatletter\ifx\SetFigFont\undefined%
\gdef\SetFigFont#1#2#3#4#5{%
  \reset@font\fontsize{#1}{#2pt}%
  \fontfamily{#3}\fontseries{#4}\fontshape{#5}%
  \selectfont}%
\fi\endgroup%
\begin{picture}(6927,6024)(886,-7048)
\put(3676,-2986){\makebox(0,0)[lb]{\smash{{\SetFigFont{17}{20.4}{\rmdefault}{\mddefault}{\updefault}{\color[rgb]{0,0,0}$A_2$}%
}}}}
\put(2476,-1861){\makebox(0,0)[lb]{\smash{{\SetFigFont{17}{20.4}{\rmdefault}{\mddefault}{\updefault}{\color[rgb]{0,0,0}$A_1$}%
}}}}
\put(4276,-3586){\makebox(0,0)[lb]{\smash{{\SetFigFont{17}{20.4}{\familydefault}{\mddefault}{\updefault}{\color[rgb]{0.000,0.000,0.000}$A_3$}%
}}}}
\put(901,-3736){\makebox(0,0)[lb]{\smash{{\SetFigFont{17}{20.4}{\familydefault}{\mddefault}{\updefault}{\color[rgb]{0.000,0.000,0.000}$w_0$}%
}}}}
\put(5326,-2536){\makebox(0,0)[lb]{\smash{{\SetFigFont{17}{20.4}{\familydefault}{\mddefault}{\updefault}{\color[rgb]{0.000,0.000,0.000}$(5/4)B_0$}%
}}}}
\end{picture}%

%% file: specialIssue.bbl
\begin{thebibliography}{35}
\expandafter\ifx\csname natexlab\endcsname\relax\def\natexlab#1{#1}\fi
\expandafter\ifx\csname url\endcsname\relax
  \def\url#1{\texttt{#1}}\fi
\expandafter\ifx\csname urlprefix\endcsname\relax\def\urlprefix{URL }\fi

\bibitem[{Abbott(2014)}]{abbott:qir}
Abbott, J., 2014. Quadratic interval refinement for real roots. ACM
  Communications in Computer Algebra 48~(1), 3--12.

\bibitem[{Becker(2012)}]{becker:thesis:12}
Becker, R., May 2012. The {B}olzano {M}ethod to isolate the real roots of a
  bitstream polynomial. Bachelor thesis, University of Saarland, Saarbruecken,
  Germany.

\bibitem[{Becker et~al.(2016)Becker, Sagraloff, Sharma, Xu, and
  Yap}]{becker+4:cluster:16}
Becker, R., Sagraloff, M., Sharma, V., Xu, J., Yap, C., 2016. Complexity
  analysis of root clustering for a complex polynomial. In: 41st Int'l Symp.
  Symbolic and Alge. Comp. pp. 71--78, {I}SSAC 2016. July 20-22, Wilfrid
  Laurier University, Waterloo, Canada.

\bibitem[{Becker et~al.(2018)Becker, Sagraloff, Sharma, and
  Yap}]{becker+3:cisolate:18}
Becker, R., Sagraloff, M., Sharma, V., Yap, C., May-June 2018. A near-optimal
  subdivision algorithm for complex root isolation based on {Pellet} test and
  {Newton} iteration. J. Symbolic Computation 86, 51--96.

\bibitem[{Burr and Krahmer(2012)}]{burr-krahmer:sqfree:12}
Burr, M., Krahmer, F., 2012. {SqFreeEVAL}: An (almost) optimal real-root
  isolation algorithm. J. Symbolic Computation 47~(2), 153--166.

\bibitem[{Burr et~al.(2009)Burr, Krahmer, and
  Yap}]{burr-krahmer-yap:continuousAmort:09}
Burr, M., Krahmer, F., Yap, C., December 2009. Continuous amortization: A
  non-probabilistic adaptive analysis technique. Electronic Colloquium on
  Computational Complexity (ECCC) TR09~(136).
\newline\urlprefix\url{http://eccc.hpi-web.de/report/2009/136/}

\bibitem[{Burr(2016)}]{burr:contAmortization:16}
Burr, M.~A., 2016. Continuous amortization and extensions: With applications to
  bisection-based root isolation. J. Symb. Comput. 77, 78--126.
\newline\urlprefix\url{http://dx.doi.org/10.1016/j.jsc.2016.01.007}

\bibitem[{Collins and Akritas(1976)}]{collins-akritas:76}
Collins, G.~E., Akritas, A.~G., 1976. Polynomial real root isolation using
  {D}escartes' rule of signs. In: Jenks, R.~D. (Ed.), Proceedings of the 1976
  ACM Symposium on Symbolic and Algebraic Computation. ACM Press, pp. 272--275.

\bibitem[{Davenport(1985)}]{davenport:85}
Davenport, J.~H., 1985. Computer algebra for cylindrical algebraic
  decomposition. {T}ech.\ {R}ep., The Royal Inst. of Technology, Dept. of
  Numerical Analysis and Computing Science, S-100 44, Stockholm, Sweden,
  reprinted as Tech.~Report 88-10 , School of Mathematical Sci., U. of Bath,
  Claverton Down, Bath BA2 7AY, England. URL
  http://www.bath.ac.uk/\textasciitilde masjhd/TRITA.pdf.

\bibitem[{Du et~al.(2007)Du, Sharma, and Yap}]{du-sharma-yap:sturm:07}
Du, Z., Sharma, V., Yap, C., 2007. Amortized bounds for root isolation via
  {S}turm sequences. In: Wang, D., Zhi, L. (Eds.), Symbolic-Numeric
  Computation. Trends in Mathematics. Birkh{\"a}user Verlag AG, Basel, pp.
  113--130, proc. Int'l Workshop on Symbolic-Numeric Computation, Xi'an, China,
  Jul 19--21, 2005.

\bibitem[{Eigenwillig et~al.(2006)Eigenwillig, Sharma, and
  Yap}]{eigenwillig-sharma-yap:descartes:06}
Eigenwillig, A., Sharma, V., Yap, C., 2006. Almost tight complexity bounds for
  the {D}escartes method. In: 31st Int'l Symp. Symbolic and Alge. Comp.
  (ISSAC'06). pp. 71--78, genova, Italy. Jul 9-12, 2006.

\bibitem[{Emiris et~al.(2014)Emiris, Pan, and
  Tsigaridas}]{emiris-pan-tsigaridas:handbk:14}
Emiris, I.~Z., Pan, V.~Y., Tsigaridas, E.~P., 2014. Algebraic algorithms. In:
  Gonzalez, T., Diaz-Herrera, J., Tucker, A. (Eds.), Computing Handbook, 3rd
  Edition: Computer Science and Software Engineering. Chapman and Hall/CRC, pp.
  10: 1--30.

\bibitem[{Imbach et~al.(2018)Imbach, Pan, and Yap}]{imbach-pan-yap:ccluster:18}
Imbach, R., Pan, V., Yap, C., 2018. Implementation of a near-optimal complex
  root clustering algorithm. In: Proc.~Int'l Congress on Mathematical Software.
  Vol. 10931 of LNCS. pp. 235--244, 6th ICMS, Notre Dame University. July
  24-27, 2018.

\bibitem[{Kamath(2010)}]{kamath:thesis}
Kamath, N., Aug. 2010. Subdivision algorithms for complex root isolation: \
  empirical comparisons. Msc thesis, Oxford University, Oxford Computing
  Laboratory.

\bibitem[{Kamath et~al.(2011)Kamath, Voiculescu, and
  Yap}]{kamath-voiculescu-yap:study:11}
Kamath, N., Voiculescu, I., Yap, C., 2011. Empirical study of an
  evaluation-based subdivision algorithm for complex root isolation. In: 4th
  Intl. Workshop on Symbolic-Numeric Computation (SNC). pp. 155--164.

\bibitem[{Kirrinnis(1998)}]{kirrinnis:poly-factor-newton:98}
Kirrinnis, P., 1998. Polynomial factorization and partial fraction
  decomposition by simultaneous {N}ewton's iteration. J. of Complexity 14,
  378--444.

\bibitem[{Kobel et~al.(2016)Kobel, Rouillier, and
  Sagraloff}]{kobel-rouillier-sagraloff:for-real:16}
Kobel, A., Rouillier, F., Sagraloff, M., 2016. Computing real roots of real
  polynomials ... and now for real! In: 41st Int'l Symp. Symbolic and Alge.
  Comp. pp. 303--310, july 19-22, Waterloo, Canada.

\bibitem[{Lane and Riesenfeld(1981)}]{lane-riesenfeld:81}
Lane, J.~M., Riesenfeld, R.~F., 1981. Bounds on a polynomial. BIT 21, 112--117.

\bibitem[{Marden(1949)}]{marden:bk}
Marden, M., 1949. The Geometry of Zeros of a Polynomial in a Complex Variable.
  Math. Surveys. American Math. Soc., New York.

\bibitem[{McNamee and Pan(2013)}]{mcnamee:roots:bk2}
McNamee, J., Pan, V., 2013. Numerical Methods for Roots of Polynomials, Part 2.
  Elsevier, Amsterdam.

\bibitem[{Mehlhorn et~al.(2015)Mehlhorn, Sagraloff, and
  Wang}]{mehlhorn-sagraloff-wang:15}
Mehlhorn, K., Sagraloff, M., Wang, P., 2015. From approximate factorization to
  root isolation with application to cylindrical algebraic decomposition. J.
  Symbolic Computation 66, 34--69.
\newline\urlprefix\url{http://dx.doi.org/10.1016/j.jsc.2014.02.001}

\bibitem[{Neff and Reif(1996)}]{neff-reif:complex-roots:96}
Neff, C.~A., Reif, J.~H., 1996. An efficient algorithm for the complex roots
  problem. J. of Complexity 12, 81--115.

\bibitem[{Pan(1997)}]{pan:history-progress:97}
Pan, V.~Y., 1997. Solving a polynomial equation: some history and recent
  progress. SIAM Review 39~(2), 187--220.

\bibitem[{Pan(2000)}]{pan:weyl:00}
Pan, V.~Y., 2000. Approximating polynomial zeros: Modified quadtree ({W}eyl's)
  construction and improved {Newton's} iteration. J.Complexity 16~(1),
  213--264.

\bibitem[{Pan(2002)}]{pan:poly-roots:02}
Pan, V.~Y., 2002. Univariate polynomials: Nearly optimal algorithms for
  numerical factorization and root-finding. J. Symb. Comput. 33~(5), 701--733.

\bibitem[{Renegar(1987)}]{renegar:approximate-poly-zeros}
Renegar, J., 1987. On the worst-case arithmetic complexity of approximating
  zeros of polynomials. Journal of Complexity 3, 90--113.

\bibitem[{Rouillier and Zimmermann(2004)}]{rouillier-zimmermann:roots:04}
Rouillier, F., Zimmermann, P., 2004. Efficient isolation of [a] polynomial's
  real roots. J. Computational and Applied Mathematics 162, 33--50.

\bibitem[{Sagraloff(2012)}]{sagraloff:newton-descartes:12}
Sagraloff, M., 2012. When {N}ewton meets {D}escartes: A simple and fast
  algorithm to isolate the real roots of a polynomial. In: 37th Int'l Symp.
  Symbolic and Alge. Comp. pp. 297--304.

\bibitem[{Sagraloff and Mehlhorn(2016)}]{mehlhorn-sagraloff:real-roots:16}
Sagraloff, M., Mehlhorn, K., 2016. Computing real roots of real polynomials. J.
  Symb. Comput. 73, 46--86.
\newline\urlprefix\url{http://dx.doi.org/10.1016/j.jsc.2015.03.004}

\bibitem[{Sagraloff and Yap(2011)}]{sagraloff-yap:ceval:11}
Sagraloff, M., Yap, C.~K., 2011. A simple but exact and efficient algorithm for
  complex root isolation. In: Emiris, I.~Z. (Ed.), 36th Int'l Symp. Symbolic
  and Alge. Comp. pp. 353--360, june 8-11, San Jose, California.

\bibitem[{Sch{\"o}nhage(1982)}]{schonhage:fundamental}
Sch{\"o}nhage, A., 1982. The fundamental theorem of algebra in terms of
  computational complexity. Manuscript, Department of Mathematics, University
  of T{\"u}bingen. Updated in 2004 with typo corrections and an appendix of
  related subsequent papers.
\newline\urlprefix\url{www.informatik.uni-bonn.de/~schoe/fdthmrep.ps.gz}

\bibitem[{Sharma and Yap(2012)}]{sharma-yap:near-optimal:12}
Sharma, V., Yap, C., 2012. Near optimal tree size bounds on a simple real root
  isolation algorithm. In: 37th Int'l Symp. Symbolic and Alge. Comp.(ISSAC'12).
  pp. 319 -- 326, jul 22-25, 2012. Grenoble, France.

\bibitem[{Yakoubsohn(2000)}]{yakoubsohn:cluster:00}
Yakoubsohn, J.-C., 2000. Finding a cluster of zeros of univariate polynomials.
  Journal of Complexity 16~(3), 603--638.

\bibitem[{Yap et~al.(2013)Yap, Sagraloff, and
  Sharma}]{yap-sagraloff-sharma:cluster:13}
Yap, C., Sagraloff, M., Sharma, V., 2013. Analytic root clustering: A complete
  algorithm using soft zero tests. In: The Nature of Computation. Logic,
  Algorithms, Applications. Vol. 7921 of LNCS. pp. 434--444.

\bibitem[{Yap(2009)}]{yap:praise:09}
Yap, C.~K., 2009. In praise of numerical computation. In: Albers, S., Alt, H.,
  N{\"a}her, S. (Eds.), Efficient Algorithms. Vol. 5760 of LNCS. Springer, pp.
  308--407.

\end{thebibliography}
